\documentclass{article}

\usepackage{arxiv}

\usepackage{amssymb}
\usepackage{amsmath}

\usepackage[utf8]{inputenc} %
\usepackage[T1]{fontenc}    %
\usepackage{xcolor}
\usepackage{hyperref}       %
\definecolor{BlueLink}{RGB}{15,78, 155}
\definecolor{lightgray}{gray}{0.9}
\hypersetup{
    colorlinks, allcolors=BlueLink
}

\usepackage{url}            %
\usepackage{booktabs}       %
\usepackage{amsfonts}       %
\usepackage{nicefrac}       %
\usepackage{microtype}      %
\usepackage{cleveref}       %
\usepackage{lipsum}         %
\usepackage{graphicx}
\usepackage{doi}
\usepackage{framed}      
\definecolor{shadecolor}{rgb}{0.95,0.95,0.95}

\usepackage{amssymb}
\usepackage{amsmath}
\usepackage{bm} %
\usepackage{stmaryrd} %
\usepackage{enumitem} %
\usepackage{wrapfig}
\usepackage{subcaption}  
\usepackage{float} %
\input{macros.sty}

\usepackage{dsfont}
\usepackage{mathrsfs}
\usepackage{dirtytalk}

\title{Deep Reinforcement Learning\\for Adaptive Mesh Refinement}

\author{Corbin Foucart\thanks{\texttt{foucartc@mit.edu}} \,$^{,1,2}$, \,
	Aaron Charous $^{1,2}$, \,
	Pierre F.J. Lermusiaux\thanks{Corresponding author: \texttt{pierrel@mit.edu}} \,$^{,1}$\\
        $^1$Department of Mechanical Engineering, \,
        $^2$Center for Computational Science and Engineering\\
	Massachusetts Institute of Technology\\
	Cambridge, MA 02139\\
}

\renewcommand{\shorttitle}{Deep Reinforcement Learning for Adaptive Mesh Refinement}

\hypersetup{
pdftitle={Deep Reinforcement Learning for Adaptive Mesh Refinement},
pdfsubject={},
pdfauthor={C.Foucart},
pdfkeywords={Reinforcement learning, finite element methods, adaptive mesh refinement},
}

\begin{document}
\maketitle

\begin{abstract}
Finite element discretizations of problems in computational physics often rely on adaptive mesh refinement (AMR) to preferentially resolve regions containing important features during simulation.
However, these spatial refinement strategies are often heuristic and rely on domain-specific knowledge or trial-and-error.
We treat the process of adaptive mesh refinement as a local, sequential decision-making problem under incomplete information, formulating AMR as a partially observable Markov decision process.
Using a deep reinforcement learning approach, we train policy networks for AMR strategy directly from numerical simulation.
The training process does not require an exact solution or a high-fidelity ground truth to the partial differential equation at hand, nor does it require a pre-computed training dataset.
The local nature of our reinforcement learning formulation allows the policy network to be trained inexpensively on much smaller problems than those on which they are deployed.
The methodology is not specific to any particular partial differential equation, problem dimension, or numerical discretization, and can flexibly incorporate diverse problem physics.
To that end, we apply the approach to a diverse set of partial differential equations, using a variety of high-order discontinuous Galerkin and hybridizable discontinuous Galerkin finite element discretizations.
We show that the resultant deep reinforcement learning policies are competitive with common AMR heuristics, generalize well across problem classes, and strike a favorable balance between accuracy and cost such that they often lead to a higher accuracy per problem degree of freedom.
\end{abstract}


\section{Introduction}
\label{sec:introduction}

In recent decades, the finite element community has developed principled, efficient techniques for solving partial differential equations (PDEs).
Not only are these techniques extremely general methods that may be applied to almost any PDE, they also provide guarantees such as stability, consistency, and convergence \cite{brenner2008mathematical,hesthaven2007nodal}.
On the other hand, the machine learning community has developed a broad set of methods to learn latent patterns from large datasets in the absence of a model
\cite{bishop2006pattern,kaiser2018sparse,kulkarni_et_al_DDDAS2020}.
However, for problems in computational physics, it is often the case that the PDE is an excellent model of the underlying physical phenomena, often down to the molecular level, where the continuum assumption begins to lose validity.
Attempts to use machine learning to learn solutions to PDEs directly have shown promise for relatively trivial problems, but are as of yet subject to several failure modes in terms of generalization to even moderately more complicated problems \cite{krishnapriyan2021characterizing}.
Rather than ignoring the extensive body of work in either field, we propose combining techniques from numerical mathematics and machine learning in order to preserve mathematically hard-earned guarantees while improving accuracy and efficiency by applying machine learning to the peripheral aspects of numerical methods which lack a model and rely purely on heuristics. 
Adaptive mesh refinement is one such aspect.

Uniform meshes are often computationally inefficient for finite element simulations in that the mesh density required to resolve complex physical features such as steep gradients or small-scale processes is used everywhere over the computational domain, even in regions where the numerical solution is smooth and much coarser resolution could be used.
In many problems, such features are dynamic: eddies, meandering jets, or nonlinearly evolving cracks constitute some examples.
To optimize efficiency, adaptive mesh refinement (AMR) techniques are a class of methods that dynamically modify the computational mesh during simulation in an attempt to increase resolution specifically where it is needed \cite{plewa2005adaptive}.

Most AMR techniques follow an iterative procedure of numerically solving the PDE, estimating the error on each element, marking a subset of the mesh elements for refinement or de-refinement (coarsening), and executing the alterations to the mesh \cite{braess2007convergence,daniel2018adaptive}; this well-established paradigm is commonly referred to as the SOLVE$\rightarrow$ESTIMATE$\rightarrow$MARK$\rightarrow$REFINE loop, terminology we will use in the present work. 
Following each numerical solve, the ESTIMATE process involves estimating the discretization error on each cell in the mesh.
In this paper, we draw the distinction between an \textit{error estimator}, which provides objective measures of error in a specific norm, and an \textit{error indicator}, which offers an empirical indication  as to the local magnitude of numerical error but provides no theoretical guarantees. 
The MARK process selects cells for coarsening and refinement based on the estimates of the error. 
Two common approaches are bulk refinement and fixed-number refinement \cite{bangerth2012algorithms,binev2004adaptive,braess2007convergence,desterck2008efficiency,hoppe2006posteriori,hoppe2009convergence,morin2000data}.
The former marks all the cells responsible for a specified percentage of the (estimated) total error for refinement and similarly for coarsening.
The latter refines and coarsens a fixed percentage of the total number of cells.

While AMR methodologies have allowed computational scientists to solve problems which are completely intractable on a uniform mesh \cite{klein1999star}, the application of AMR strategies remains largely heuristic, and best practices are not universally agreed upon \cite{plewa2005adaptive}.
The process of estimating the error on each cell is often complex, dependent on both the PDE and the numerical method used to solve it, and constitutes an active subject of research \cite{ainsworth1997posteriori,babuska2010finite,ern2010guaranteed}.
In general, for nonlinear partial differential equations, it is often difficult, and in some cases, impossible to provide an upper bound on the error.
Even in the situations where rigorous error estimation is possible, the bounds on the error may not be tight, or may apply to a limited subset of problems, rendering the estimator ineffectual.
As a consequence, in practice, these estimators tend to be optimistically applied as \textit{ad hoc} error indicators.
A relevant example is the well-known Kelly error \say{estimator} \cite{kelly1983posteriori}, which is derived from analysis specifically for the Poisson equation, but is widely employed in AMR strategies for the spatial discretizations of many other PDEs \cite{dealII93,bonilla2020monotonicity,joshi2004adaptive,vidal2017hp,young2010strategies} despite its lack of theoretical applicability.
Equally important, but often overlooked in the literature, the MARK process is also fraught with challenges.
In attempting to refine only the cells which constitute a certain percentage of the error, bulk refinement strategies can be expensive in cases with very few cells at singularities, miss important features of the solution, and present difficulties controlling the number of cells in the mesh \cite{desterck2008efficiency,morin2000data}.
With fixed-number refinement, the number of cells in the mesh can be readily controlled, but the approach can wastefully refine too many cells \cite{desterck2008efficiency}.
Furthermore, the parameters in either of these methods (\ie, the algorithmic realization of a refinement strategy with regard to the percentages of the estimated error to refine and coarsen in the case of bulk refinement or exactly how many cells to refine and coarsen in the case of fixed number refinement) are choices typically made \textit{a priori} before a numerical simulation begins.
In doing so, the balance between coarsening and refinement is implicitly assumed to be static in time, rather than dynamically driven by the behavior of the underlying solution and available computational resources.

As a result, even after decades of research, in practice, AMR strategies remain largely unprincipled and often require domain-specific knowledge, trial and error, or manual intervention.
Selecting and combining an efficient set of AMR heuristics for a new problem is a challenging endeavor and open research problem in general \cite{desterck2008efficiency}---there exists a clear need for an automated, flexible, and principled approach to AMR, motivating the present research.

We propose the treatment of AMR as a partially observable Markov decision process (POMDP) that can be interpreted as a local Markov decision process (MDP) on each element and apply a deep reinforcement learning (RL) approach \cite{arulkumaran2017deep} in which we train an agent to increase or decrease mesh resolution, balancing improved accuracy against the computational cost associated with each mesh modification decision. 
Deep reinforcement learning replaces the expected reward function from the well-known classical Q-learning algorithm with a neural network as a function approximator \cite{sutton2018reinforcement}, allowing for the discovery of arbitrarily sophisticated decision-making policies based on unstructured input data  \cite{hornik1989multilayer,sonoda2017neural} and obviating the need to manually engineer aspects of the strategy.
In doing so, we replace the ESTIMATE and MARK processes in AMR with a trained RL policy learned from numerical simulation.

In the present work, we focus primarily on high-order discontinuous Galerkin finite element methods (DG-FEM), in part due to their incredible success in modeling a wide range of phenomena in computational physics (see \cite{hesthaven2007nodal,Kronbichler2021} and references therein) and due to their discontinuous representation of the numerical solution.
It is this discontinuous representation that reliably links the smoothness of the local solution as measured by the interface jumps to local error \cite{cockburn2003discontinuous} in a way that can be well-leveraged by a decision-making agent, as we develop in \S\ref{sec:observation_space}.
The same property gives rise to the simple and effective non-conformity error estimator and its variants examined in the DG-FEM literature \cite{gassner2009explicit,ern2010guaranteed,krivodonova2003error,krivodonova2004shock,naddei2019comparison}.
As demonstrated in what follows, the DG-FEM discretization plays an important role in the construction of the observation spaces for the reinforcement learning problem. 
However, in principle, our methodology could be applied with minor modifications to classical continuous Galerkin finite element methods or even other numerical methods such as finite difference or finite volume methods.

\subsection{Related work and novel contributions}
\label{sec:objectives_and_novel_contributions}

Optimal mesh refinement strategies have been shown to be theoretically learnable in a recurrent neural network setting \cite{bohn2021recurrent}. 
The application of deep RL specifically to problems in computational physics is in its infancy, and recent work at the intersection between the disciplines can be found in \cite{farahmand2017deep,wang2019learning,wei2019general}.
A deep RL approach for the related problem of mesh generation is explored in \cite{pan2022reinforcement}.
The work in \cite{yang2021reinforcement} constitutes the first attempt to formulate AMR as an RL problem and demonstrates the feasibility of the approach. 
The present work was developed concurrently and independently, and we formulate the deep RL problem differently.
To avoid growing and shrinking action and observation spaces, our decision-making problem is inherently local rather than defined over the entire mesh; to that end, our observation space includes measures of local non-conformity of the solution.
We elect not to impose a max refinement depth or provide a hard limit on the refinement budget in the action space; rather, we impose these restrictions implicitly through our reward function. 
Accordingly, we do not aim to maximize total error reduction; our reward function seeks to strike a tunable balance between the accuracy of the numerical solution and available computational resources.

In this work, we present novel theory and schemes that use a POMDP representation to formulate AMR as a deep RL problem under incomplete information and obtain a trained RL policy that can be thought of as a custom error indicator discovered through trial and error.
To the best of our knowledge, this is the first work that includes both refinement and de-refinement actions as part of the resultant policy.
We provide a very general framework for choosing an observation space that can incorporate physically relevant features of the PDE to be solved.
The methodology is non-reliant on an exact solution or ground truth during model training or deployment. The agent learns a cost-effective refinement and coarsening strategy that balances improved solution accuracy with computational cost, as opposed to a hard threshold.

DG-FEM methods have been shown to be competitive in the \textit{under-resolved} regimes of fluid flow simulation, both in the sense of stability and robustness of the schemes \cite{fehn2017stability}, and in their ability to transport high-frequency features over long time-integration horizons without altering the shape of the features or losing amplitude (dispersion and dissipation) \cite{ueckermann_lermusiaux_OD2010,fehn2018robust}. 
Therefore, practically, the ability of DG-FEM schemes to capture and preserve physical features 
is more an important criterion of merit than the norm-measured errors often presented in academic convergence studies, (see \cite{Kronbichler2021}, pp.35-48). 
This is crucial, as it suggests that any performant AMR technique should be measured against its ability to resolve dynamical features while making efficient use of problem degrees of freedom.
This motivates the specific form of our reward function, as well as informs our primary objective of this work; to obtain an AMR policy that accurately resolves features of the numerical solution and efficiently uses computational resources.

The paper is organized as follows. 
In \S\ref{sec:deep_RL_framework}, we introduce our RL formulation of the adaptive mesh refinement problem. 
The partially observable Markov decision process is described in \S\ref{sec:adaptive_mesh_refinement_as_a_markov_decision_process}, with the action and observation spaces detailed in \S\ref{sec:action_space}-\S\ref{sec:observation_space}.
The design and implementation of a new reward function that allows for model training without the need for exact or ground-truth solutions to the underlying problem are provided in \S\ref{sec:reward_fn}.
Procedures for training and deployment are given in \S\ref{sec:training}, and a brief discussion of the deep learning policy architectures employed therein is specified in \S\ref{sec:policy_architectures}.
The discontinuous Galerkin spatial discretizations for the numerical forward models are provided in \S\ref{sec:finite_element_discretization} for different PDEs.
In \S\ref{sec:numerical_experiments}, we demonstrate the efficacy of the methodology on a set of numerical test cases spanning a variety of PDEs and numerical methods. 
We show that the resultant RL policies can outperform widely-used AMR heuristics in terms of accuracy per degree of freedom and that the policy corresponding to a single trained model generalizes well to different boundary conditions, forcing functions, and problem sizes.
We offer concluding remarks in \S\ref{sec:conclusions_and_future_work}.

\section{Deep reinforcement learning framework}
\label{sec:deep_RL_framework}

Deep reinforcement learning 
combines reinforcement learning with deep learning using a neural network to represent the value function, policy, or model considered in a classical RL setting \cite{li2017deep}.
Similar to other deep learning approaches, the network is typically trained by optimization of a loss function over many episodes in which a strategy that maximizes the expected reward (\ref{eq:Q_expected_reward}) is determined through trial and error.
This approach is attractive, as it does not require domain-specific knowledge, nor does it require a large training data set; training data is generated experientially by \say{self-play} as the agent interacts with its environment.

Formulating an RL approach (\S\ref{sec:RL_notation}) to a problem involves describing the underlying decision process and representation of the agent and the state (\S\ref{sec:adaptive_mesh_refinement_as_a_markov_decision_process}).
We must specify a reward function (\S\ref{sec:reward_fn}) which encodes desirable behaviors of the resultant learned policy.
Lastly, we must specify the neural network architectures which are to represent the different parts of the RL problem (\S\ref{sec:policy_architectures}).
The subsections that follow present our novel formulation of adaptive mesh refinement as a deep RL problem.

\subsection{Notation}%
\label{sec:RL_notation}

Reinforcement learning is characterized by a decision-making agent that interacts with an environment described by a state $S$. 
We consider the discrete-time case: at time step $t\in\left\{0,1,2,\ldots \right\}$, the observable state of the environment is $S_t$ and the action selected by the agent is $A_t$, the latter of which gives rise to the reward $R_{t+1}$ and modified state $S_{t+1}$.

A partially observable Markov decision process (POMDP) is a formalism for describing a sequential decision-making process under incomplete information \cite{aastrom1965optimal}.
A POMDP is characterized by $(\mathcal{S},\, \mathcal{A},\, \mathcal{R},\, T,\, \mathcal{O},\, O,  \gamma)$, where the sets $\mathcal{S}$, $\mathcal{A}$, and $\mathcal{R}$ are the sets of possible states, actions, and rewards, respectively. 
The state transition function $T:\mathcal{S}\times\mathcal{A}\times \mathcal{S}\to[0,1]$ encodes the distribution of transition probabilities from a state $S_t = s$ to $S_{t+1}=s^{\prime}$ given the action $a$; that is, $T(s, a, s^{\prime})$ returns the probability of transitioning to state $S_{t+1}=s^{\prime}$ under the action $A_t = a$ executed in state $S_t = s$.
The set $\mathcal{O}$ is the set of possible observations, and the probability distribution $O(o|s,a)$ describes the probabilities over the set of possible new observations $O_{t+1} = o\in\mathcal{O}$ upon taking action $A_t = a$ in state $S_t=s$.
The scalar $\gamma\in[0,1]$ is a time discount factor---a lower discount factor motivates the agent to favor taking actions early.
A POMDP differs from a standard Markov decision process (MDP) in that the agent does not have access to the complete state $S_t$, but rather, indirect observations of it.

The goal of RL is to find a stochastic policy function $\pi:\mathcal{O}\to \mathcal{A}$ that maps the observation space, that we take as a subset of the entire space $\mathcal{O}\subseteq \mathcal{S}$, to the set of actions that maximizes the expected reward 
\begin{equation*}
  Q_t(s, a) = \Expect_{\pi} \left[ \sum_{k=0}^{\infty} \gamma^k R_{t + k + 1} \mid S_t = s,\, A_t = a\right]
  \label{eq:Q_expected_reward}
\end{equation*}
over an infinite time horizon \cite{silver2015}.

\subsection{Elemental refinement as a local Markov decision process}%
\label{sec:adaptive_mesh_refinement_as_a_markov_decision_process}

\begin{figure}[H]
  \centering
  \includegraphics[width=0.48\linewidth]{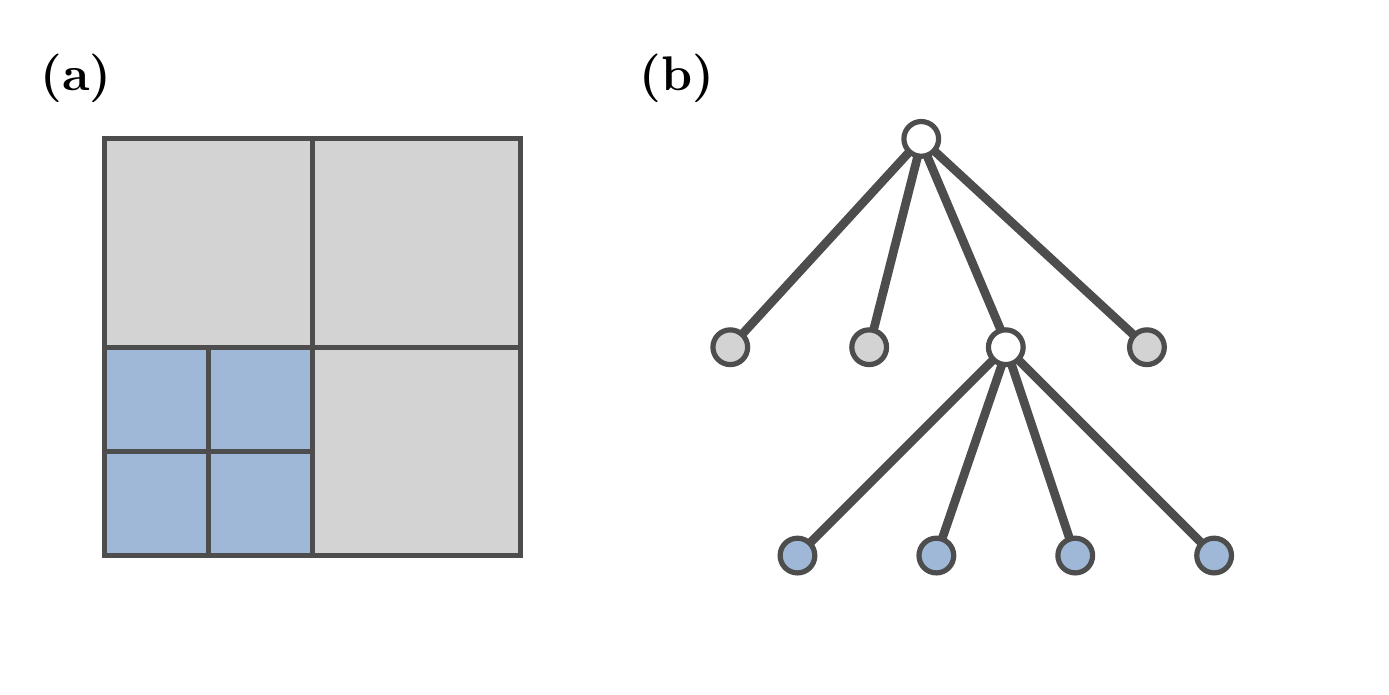}
  \vskip -0.3truecm
  \vskip -0.3truecm
  \caption{\small (a) Adaptively refined elements in a computational mesh $\mathcal{T}_h$ generated from a single element. (b) Underlying data structure representation as a tree, with inactive parent elements and active elements as the leaves of the tree.}
  \label{fig:mesh_representation}
\end{figure}

As per the approach described in \cite{arndt2021deal,bangerth2012algorithms} and references therein, the computational mesh is composed of linear, quadrilateral, or hexahedral elements, and represented by a tree data structure.
Upon refinement of an element, $2^d$ child elements are generated by bisection and marked as active, while the parent element is marked as inactive.
The active cells of the computational mesh at any given time are the leaves of the tree.
This process is schematically illustrated in Figure \ref{fig:mesh_representation}.
In this paper, we refer to elements (active or inactive) which share the same parent element as \say{sibling} elements. 

The state $S_t$ is described by the current computational mesh $\mathcal{T}_h$, its numerical solution, $u_h$, and any data known to the PDE (boundary conditions, forcing functions, \etc), which we refer to as $\mathcal{D}$.
Formally, the POMDP is well-defined, as we can define the global state space as the set of all possible mesh configurations and numerical solutions. 
However, the consideration of the global state and action spaces is computationally intractable, as the agent would encounter a combinatorially growing space of actions to explore.

In light of this fact, we consider a local description of the state as the observation space in the POMDP in order to train the agent.
During training, the agent observes a single cell locally, takes action on the current cell alone, and receives a reward which is the effect of its local action on the entire computational domain.
The agent is subsequently moved to another cell in the mesh randomly, according to a distribution $B_t(O_t, S_{t+1})$ which encodes the POMDP belief in the state of the environment and which cell should be visited next.
In this work, for simplicity, we take the distribution to be uniform over all cells in the mesh.
However, a Bayesian treatment of the belief state is possible within our RL formulation.

Although it may seem counter-intuitive to train the agent on small subsets of the entire state space, formulating the learning problem locally in this way has several advantages.
Considering the entire state of the mesh as an MDP leads to a growing and shrinking action space as the mesh changes, with a combinatorially growing number of action possibilities; this can be addressed \cite{yang2021reinforcement}---however, the complicated action space may present training and generalization difficulties.
Second, many problems in computational physics are governed by PDEs with strong spatial locality, which is why the linear systems arising from the discretization of PDEs are typically sparse. 
Therefore it makes sense to view the RL problem on a single element rather than on the entire mesh, as the numerical solution is typically more sensitive to local phenomena. 
This is not always the case; \eg, in elliptic PDEs, it is well-known that errors are propagated globally \cite{strauss2007partial}; to address this, we specifically include elliptic PDE test cases in our numerical experiments in \S\ref{sec:numerical_experiments}.
Lastly, the resultant trained neural network constituting the RL policy has the intuitive interpretation of a custom cell-wise error indicator learned during training.
With this interpretation, it can be seen that the same trained model can be deployed repeatedly over the entire mesh, in parallel if desired.

\subsubsection{Action space}%
\label{sec:action_space}

The action space of the agent is the discrete set $\mathcal{A} = \left\{\texttt{coarsen},\,  \texttt{do nothing},\, \texttt{refine}\right\}$, with respect to the current cell.
The goal of the RL problem can be stated succinctly as: train the agent such that it refines in areas that are locally under-resolved, coarsens in areas that are locally over-resolved, and takes no action otherwise.
Furthermore, the agent should take these actions flexibly and efficiently with respect to the available computational resources.

While it is always possible to refine, the mesh may be in a state such that it is not topologically possible to coarsen the current cell.
This is because the current cell could be at the coarsest possible level of the mesh, \eg, a computational mesh comprised of a single cell, or because the siblings of the current element could have arbitrarily many children, and coarsening is only possible when all the sibling cells are at the highest possible refinement level (\ie, leaves on the tree).
For example, in the mesh depicted in Figure \ref{fig:mesh_representation}, all of the elements marked as blue can be coarsened, whereas those marked as gray can not.
Our implementation is such that if coarsening is not possible, the agent defaults to doing nothing.
Under this set of actions, the transition function between states is technically deterministic.
On the other hand, conceptually, it is useful to think of the transition as probabilistic, i.e., an attempt to coarsen succeeds with a probability that is a function of the present state of the mesh.

\subsubsection{Observation space}%
\label{sec:observation_space}

The observation space $\mathcal{O}$ is the observable portion of the total state $\mathcal{S}$. The observation space on an element $K\in \mathcal{T}_h$ of the computational mesh consists of: 
\begin{enumerate}[label=\Alph*., itemsep=0pt]
  \item The jumps in the numerical solution integrated over the boundary of the cell, denoted $\Xi_K = \int_{\partial K} \jmp{u_h} \,d\partial K$, as well as $\Xi_{K^\prime}$ for the cell neighbors $K^{\prime}$  \label{it:obs_space:jumps}
  \item The average integrated jump across all mesh elements $\sum_{K} \Xi_K / N_K$ \label{it:obs_space:avg_jump}
  \item The current usage of available computational resources $p$ \label{it:obs_space:p}
  \item Any physically relevant features of the numerical solution or data $\phi(u_h, \mathcal{D})$ local to the cell, where $\phi(\cdot)$ refers to a feature and $\mathcal{D}$ refers to available data \label{it:obs_space:misc}
\end{enumerate}

The rationale for the inclusion of item (\ref{it:obs_space:jumps}) is that discontinuities present at the element interfaces measure the non-conformity of the numerical solution.
This is due to the assumption that, in the absence of a physical discontinuity, the exact solution is continuous across element interfaces. 
This observation is borne out by straightforward analysis, which shows a strong relationship between the residuals of the numerical solution $u_h$ over the element interior and the jumps across the inter-element boundaries \cite{cockburn2003discontinuous}, leading to the development of so-called ``non-conformity'' error estimators \cite{naddei2019comparison} specific to discontinuous Galerkin finite element methods.

Although formulating the observation space of the POMDP as a completely local problem is computationally attractive, it is clear that in the absence of global information, the agent should always perform the \texttt{refine} action. 
Items (\ref{it:obs_space:avg_jump}) and (\ref{it:obs_space:p}) both serve the purpose of communicating global information regarding the relative utility of local refinement to the local decision process.

The inclusion of the average integrated jump over all mesh elements provides information about the relative estimate of the error on the current cell as compared to other cells in the mesh. 
During training, this element of the observation space allows the agent to decide when to forgo the opportunity to refine the current cell in order to spend computational resources elsewhere in the mesh, where the numerical errors may be greater than those observed locally.

The scalar $p\in[0,1]$ indicates the current usage of available computational resources. In this work, we take $p$ to be the fraction of active mesh elements out of a user-specified maximum number of elements. 
However, $p$ need not have this representation: in the case of an HPC application, the value of $p$ could be measured directly by monitoring CPU or memory usage in real time. 
Alternatively, $p$ could be defined with respect to a maximum allowable wall-clock time per solution time step or could be computed from some other \textit{a priori} allocation of computational resources such as RAM or arithmetic throughput.
Regardless of how the computational usage $p$ is defined, its inclusion in the observation space serves to indicate to the agent the availability of computational resources; the reward function \S\ref{sec:reward_fn} specifies the cost-benefit relationship of making use of them.

Due to item (\ref{it:obs_space:misc}), this observation space is very flexible and general. 
This is by design, as it allows the user to include physically or computationally relevant information which may help the agent learn a more effective strategy.
As an example, the advection problem as described in \S\ref{sec:linear_advection_equation} may include the local convective velocity in the observation space, as it determines the flow of information.

\textbf{Remark.} In this paper, we emphasize discontinuous Galerkin methods due to the simplicity and effectiveness of the non-conformity of the local solution as a proxy for the local error.
In the case of classical continuous Galerkin finite element schemes, the jump of the solution along the element boundary is identically zero, due to the continuous nature of the approximation spaces in which the numerical solution is sought \cite{brenner2008mathematical}. 
However, item (\ref{it:obs_space:jumps}) can be readily replaced with other indicators of non-conformity in order to generalize the methodology to other 
finite element discretizations. 
For example, the jump of the solution gradient along the element boundary can be applied as a simple surrogate in the continuous Galerkin case---the core principle is to select a numerical quantity that vanishes as the numerical solution approaches the exact solution of the PDE. 
However, more sophisticated indicators of non-conformity exist \cite{babuska2010finite}.

\subsection{Designing the reward function}%
\label{sec:reward_fn}

\begin{figure}[H]
  \centering
    \includegraphics[width=0.33\textwidth]{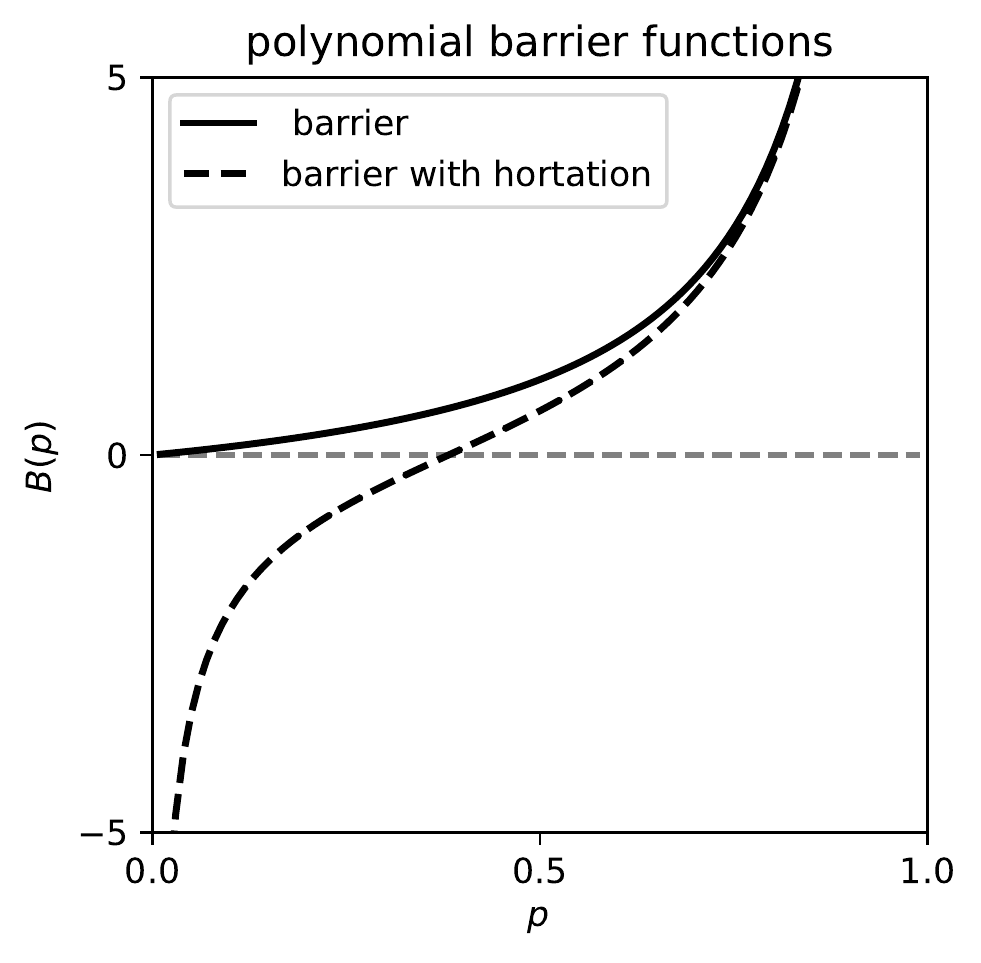}
    \caption{\small Example of polynomial barrier function $B(p) = p/(1-p)$  and hortative barrier function $B(p) = p/(1-p) - [1/\sqrt{p} - 1]$.}
    \label{fig:barrier_fns}
\end{figure}

The reward function is an environment-provided reinforcement signal that determines the immediate reward or penalty due to each decision on the part of the agent \cite{sutton2018reinforcement}.
In that sense, the reward function is central to the ultimate behavior of the agent after training and should encode all the information pertaining to what we wish our agent to accomplish.
For AMR, the reward function used to train the agent should strike a balance between improvements in accuracy and incurring additional computational costs. With these conditions, our proposed reward function takes the general form of
\begin{equation}
  [\text{accuracy}] - \gamma_c [\text{cost}] B(p).
  \label{eq:reward_fn_general}
\end{equation}
The coefficient $\gamma_c$ is a scaling factor expressing the relative importance of the accuracy of the solution versus increasing computational cost in the RL agent's policy; it can either be tuned as a hyperparameter or empirically computed for different computational regimes. 
The function $B(p):[0,1)\to[0,\infty)$ acts as an asymptotic barrier discouraging the agent from exceeding the limit of its computational resources. 
Two choices are shown in in Figure \ref{fig:barrier_fns}.
The barrier function can be modified to encourage the agent to use a certain percentage of resources, for example, to incentivize aggressive refinement in the under-resolved case, which we refer to as a barrier function \say{with hortation}.
In this paper, unless otherwise specified, we use the barrier function $B(p) = \sqrt{p}/(1 - p)$.

It remains to define a metric for improvement in solution and change in computational cost for each of the terms in equation (\ref{eq:reward_fn_general}).

\begin{figure}[h]
  \centering
      \includegraphics[width=0.8\textwidth]{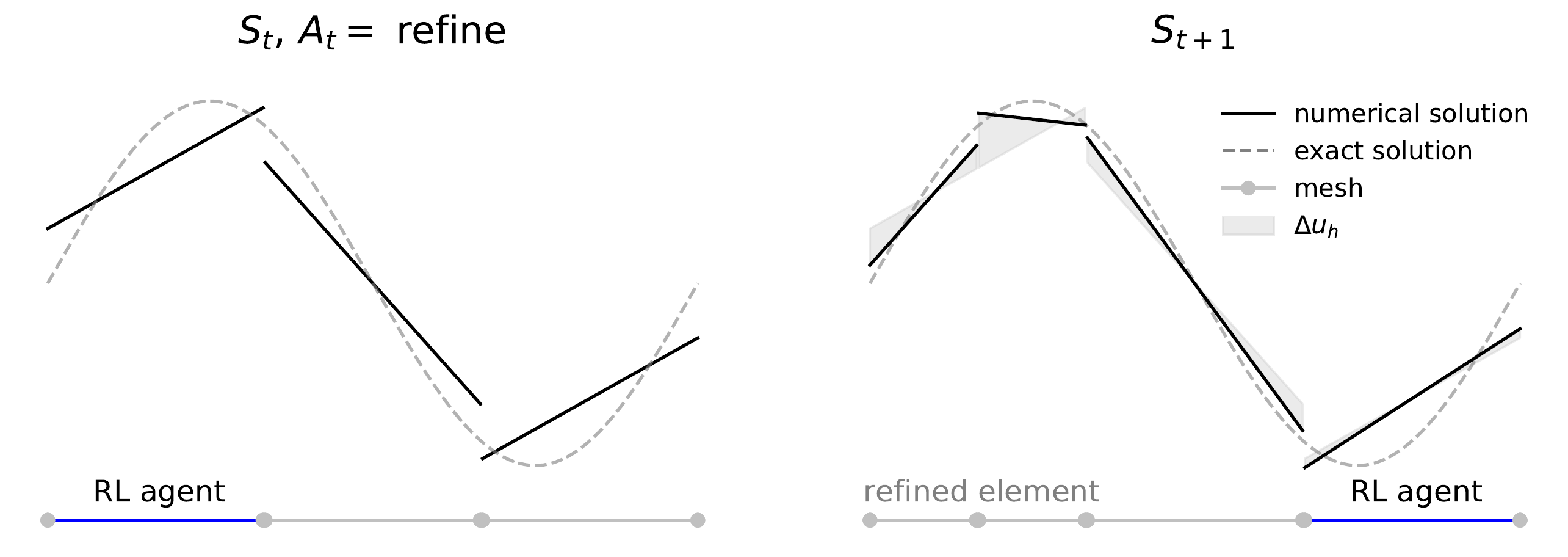}
      \caption{\small Schematic of $\Delta u_h$ upon refinement. The RL agent's new location is sampled from the belief distribution $B_t(O_t, S_{t+1})$.} 
    \label{fig:refinement-schematic}
\end{figure}

\textbf{Accuracy.}
\,The optimality properties of the numerical solution arising from continuous and discontinuous Galerkin finite element discretizations ensure that the error of the solution (measured in an appropriate norm) decreases or remains the same upon any refinement of the computational mesh \cite{brenner2008mathematical,cockburn2003discontinuous,hesthaven2007nodal}.
These properties are crucial, as they guarantee that any change in the error of the numerical solution, $e(u_h) = \left\lVert u_h - u_{\text{exact}}\right\rVert_{L^2(\Omega)}$, that arise as a result of refinement will be monotonically non-increasing. 
Therefore the fundamental idea is to reward any change in the numerical solution upon refinement, and to penalize any change to the numerical solution upon coarsening.  We define the quantity
\begin{equation}
\Delta u_h = \sum_{K\in\mathcal{T}_h} \int_{K} |u_h^{t+1} - u_h^{t}|  \,dK
\label{eq:change_in_soln}
\end{equation}
to represent the change in the solution, upon the action $A_t$, as is schematically illustrated in Figure \ref{fig:refinement-schematic}.
The integers $t$ and $t+1$ refer to a time step with respect to the POMDP, not a time integration step in the case of a time-dependent PDE.
We always perform the computation of $\Delta u_h$ by first interpolating the coarser of the two solutions onto the finer of the two grids, then performing the integration. 
This way, the change in solution $\Delta u_h$ upon refining and coarsening the same element are identical in magnitude, but with opposite signs.

Naively applying the reward $+\Delta u_h$ upon refinement and $- \Delta u_h$ upon coarsening is logical; however, this choice often results in an inconsistent reward signal due to scaling. 
One of the attractive features of high-order finite element methods is that the error in the numerical solution decreases quickly with respect to the mesh element size for problems with smooth solutions, typically as $1/h^\porder$ where $h$ is a representative size of a mesh element, and $\porder$ is the polynomial order of the scheme \cite{brenner2008mathematical,hesthaven2007nodal}.
As a result, the error in numerical solutions often spans many orders of magnitude over a small number of mesh refinement cycles.
To account for these large differences in the scale of the error, and to prevent the initial refinements from dominating the total achievable reward during training, we scale the rewards logarithmically, awarding the agent 
\vspace{0.4em}
\begin{equation*}
R_{\Delta u} = 
\pm\left[\log(\Delta u_h + \epsilon_{\text{machine}}) -\log \epsilon_{\text{machine}} \right],
\vspace{0.4em}
\end{equation*}
where the positive and negative signs apply to the cases of refinement and coarsening, respectively.
Here, $\epsilon_{\text{machine}}$ is a representation of machine precision, which we take to be $10^{-16}$,
and is included for the edge cases in which refinement or coarsening does not change the numerical solution. 
Of course,
it can also be a desired accuracy for the numerical solution.
Lastly, the additive factor is chosen to center the positive and negative rewards around zero, rather than $\log(\epsilon_{\text{machine}})$ for interpretability, and follows from our specific choice of machine precision.

\textbf{Cost.} \, 
We can specify the reward due to the change in computational cost upon action $a_t$ as the quantity
\begin{equation}
  R_{\Delta C }= B\left(p^{t+1}\right) - B\left(p^t\right).
  \label{eq:reward_barrier}
\end{equation}
The sign of $R_{\Delta C}$ is not explicitly dependent on the action $a_t$, as it purely relates to the increase or decrease in computational cost as measured by $p$ before and after the action occurs.

\textbf{The reward function.} \, 
Combining the constituent components, the final explicit form of the new reward function that we employ is
\begin{equation}
  R(s_t, a_t) = 
  \begin{cases}
+\left[\log(\Delta u_h + \epsilon_{\text{machine}}) -\log(\epsilon_{\text{machine}}) \right] - \gamma_c R_{\Delta C}, \qquad &\text{ if } a_t = \texttt{refine} \\
-\left[\log(\Delta u_h + \epsilon_{\text{machine}}) -\log(\epsilon_{\text{machine}}) \right] - \gamma_c R_{\Delta C}, \qquad &\text{ if } a_t = \texttt{coarsen}  \\
0, \qquad &\text{ if } a_t = \texttt{do nothing}
  \end{cases}
  \label{eq:reward_fn}
\end{equation}

\subsection{Training and model deployment}%
\label{sec:training}

To train the model on a static problem, we begin each episode with a very coarse mesh.
The agent is placed randomly on a cell in the mesh; after collecting the reward for each action, the agent is moved to a different element randomly as described in \S\ref{sec:adaptive_mesh_refinement_as_a_markov_decision_process}.
After a large given number of actions have occurred, the episode ends, with early termination possible in the case of repeated \texttt{do nothing} actions.
If at any point the agent exceeds the allotted computational budget, it receives a large negative reward and the episode terminates (see Appendix C).
For time-dependent problems, the training is similar, although at random intervals, the numerical solution is stepped forward by one time step.

For model deployment, the trained model considers every cell in the mesh sequentially, in the order specified by the non-conformity estimator, and makes a decision based on the local observation on that particular element.
The percentage of available resources $p$ is updated in between elements so that the model may update its recommendation.
This constitutes the marking process---once every element has been visited, the recommendations of the RL policy can be executed. This constitutes a single AMR \say{cycle} for the RL model.

\subsection{Policy and deep learning  architectures}
\label{sec:policy_architectures}

We now describe the particular architectures we used for our AMR agent.
Our deep Q-network \cite{dqn} is the simplest policy.
It consists of an input layer, the size of which corresponds to the size of the observation space, two hidden-layers, each with 64 neurons, and an output layer with one neuron for each possible action. Rectified linear units (ReLUs) are chosen as our activation functions.
Taking the argmax of the output yields the action we take (assuming we are using the optimal policy deterministically).

The Advantage Actor-Critic (A2C) policy \cite{a2c} uses the same network architecture as the deep Q-network, but it trains two networks, one to estimate the value function (the critic) and one to determine the optimal policy (the actor).
At each step, the actor chooses an action that the critic evaluates to estimate the state-value and improve the policy of the actor.
Finally, the Proximal Policy Optimization (PPO) policy \cite{ppo} uses the same network architecture as A2C; the main difference is that PPO uses clipping in the objective function so that policy updates remain relatively small.
We compare the performance of these algorithms in \S\ref{sec:lin_adv_training_and_deployment}.

\subsection{Summary: reinforcement learning framework}
\label{sec:RL-summary}

The design of the reward function is pivotal in RL. The idea behind the reward signal (\ref{eq:reward_fn}) is illustrated in Figure \ref{fig:refinement-schematic}. 
When the agent uses resources to refine the solution, it accumulates a reward corresponding to how much the numerical solution changes as a result of the refinement ($R_{\Delta u}$), and is penalized according to the increased usage of its resource budget ($R_{\Delta C})$.
In the limiting case, in which the solution is perfectly resolved, the agent accrues only penalization for additional refinement.
Conversely, when the agent elects to coarsen the resolution, it is rewarded according to the resources it saves, but pays for any change in the numerical solution.
In the limiting case where the solution is perfectly resolved with more elements than necessary, the agent receives only reward for decreasing the number of elements, as the numerical solution does not change upon coarsening.
In essence, the reward function provides a signal which communicates the trade-off between accuracy and computational cost.
To preserve the reward signal, the reward function logarithmically scales and machine-precision-normalizes the quantities $R_{\Delta u}$ and $R_{\Delta C}$, which can span many orders of magnitude.
The hyperparameter $\gamma_c$ is a tunable setting that reflects the degree of displeasure incurred upon using additional resources, effectively weighting the user priority of accuracy versus cost.

By formulating the sequential decision process locally as a POMDP, the observation space provides the link between the local solution data and its conformity to the cost-benefit trade-off of refinement in that region.
Informally speaking, the observation $O_t$ provides a signal relating local solution conformity to that of the global solution as well as the remaining computational resources and other potentially relevant data.

\begin{figure}[h] 
\small
\begin{shaded*}
\begin{minipage}[t]{.4\textwidth}
\begin{tabbing}
\= \texttt{training\_step}$(A_t,\, O_t)$: \\
\> $K_t$ $\leftarrow$ cell corresponding to $O_t$ \\
\> if \= $A_t = $ \texttt{refine} \\
\= \> $S_{t+1}$ $\leftarrow$ refine $K_t$\\
\> if \= $A_t =$ \texttt{coarsen}\\
\= \> $S_{t+1}$ $\leftarrow$ coarsen $K_t$\\
\> update $u_h^{t+1} = \mathcal{M}(S_{t+1})$, \, $p_{t+1}$  \\
\> compute $\Delta u_h$ (on finer mesh) via (\ref{eq:change_in_soln}), $R_{\Delta C}$ via (\ref{eq:reward_barrier})\\
\> compute $R_{t+1}$ via (\ref{eq:reward_fn})\\
\> sample cell $K_{t+1}\leftarrow B_t(O_{t}, S_{t+1})$\\
\> $O_{t+1}\leftarrow$ compute observation space for $K_{t+1}$\\
\> if \= $p_{t+1} > 1$\\
\= \> $R_{t+1} \leftarrow$ large, negative reward\\
\= \> \texttt{done} $\leftarrow$ true \\
\> if \= iteration $\geq$ episode iterations\\
\= \> \texttt{done} $\leftarrow$ true \\
\>\texttt{return} $R_{t+1}$, \texttt{done}, $O_{t+1}$
\end{tabbing}
\end{minipage}
\begin{minipage}[t]{.09\textwidth}  \hspace{1em} \end{minipage}
\begin{minipage}[t]{.4\textwidth}
\begin{tabbing}
\= \texttt{model\_deployment}$(S_0, \, \pi_h)$: \\
\> $S^\prime \leftarrow S_0$\\
\> $\{T\}\leftarrow$\texttt{sort} $K\in\Th$ by $\Xi_K = \int_{\partial K} \jmp{u_h} d\,\partial K$\\
\>For \= $K^\prime \in \{T\}$  \\
\>\> $O_{K^\prime} \leftarrow$  compute observation space for $K^\prime$\\
\>\> $A_{K^\prime}\leftarrow$  $\pi_h(O_{K^\prime})$  query policy network \\
\>\> execute action $A_{K^\prime}$, update $S^{\prime}, \,p$\\
compute new solution $u_h^{\prime}\leftarrow \mathcal{M}(S^{\prime})$  \\
\>\texttt{return} $u_h^{\prime}$, \,$S^\prime$
\end{tabbing}
\end{minipage}
\caption{\small Deep RL-AMR. Algorithms for a single RL training time step (left) and for the deployment of the trained policy over a single refinement cycle (right). The training step takes as input an action $A_t$ and an observation $O_t$, and returns the reward $R_{t+1}$, the Boolean episode termination condition \texttt{done} (default false), and the new observation $O_{t+1}$. The model deployment procedure takes as input a trained policy network $\pi_h$ and starting state $S_0$; it returns the proposed next mesh state $S^\prime$ and the updated numerical solution $u_h^\prime$.}
\label{fig:summary_table}
\end{shaded*}
\end{figure}

We summarize the algorithm for a training time step and model deployment in Figure \ref{fig:summary_table}. 
The notation $u^{t}_h = \mathcal{M}(S_t)$ describes running the forward model on the mesh state $S_t$ to compute the corresponding solution $u_h^{t}$ by numerically solving the PDE.
The trained policy network is denoted $\pi_h$ (\S\ref{sec:adaptive_mesh_refinement_as_a_markov_decision_process}), and we use the Boolean \texttt{done} as a sentinel for episode termination.
Lastly, we refer to the belief distribution over the mesh cells $K\in \Th$ at state $S_{t}$ as $B_t(O_t, S_t)$.
For the architectures we employ (\S\ref{sec:policy_architectures}), we refer to \cite{raffin2019stable}
for the complete specification as to how the weights of the policy network are updated as a result of the training steps taken.

\section{Finite element spatial discretization and problem physics}%
\label{sec:finite_element_discretization}

\subsection{Notation and approximation spaces}%
\label{sec:FEM_notation}
In order to state the weak forms of the subsequent problems that follow, we introduce some requisite notation.
We let $\mathcal{T}_h = \cup_i K_i$ be a finite collection of non-overlapping elements $K_i$ that discretizes the entire problem domain $\Omega\subset\mathbb{R}^d$.
We refer to the boundary of the problem domain as $\Gamma$.
The set $\partial \mathcal{T}_h = \{\partial K : K \in \mathcal{T}_h\}$ refers to all boundary edges and interfaces of the elements, where $\partial K$ is the boundary of element $K$.
For two elements $K^+$ and $K^-$ sharing an edge, we define $e = \partial K^+ \cap \partial K^-$ as the edge between elements $K^+$ and $K^-$.
Each edge can be classified as belonging to either $\varepsilon^\circ$ or $\varepsilon^\partial$, the set of interior and boundary edges, respectively, with $\varepsilon = \varepsilon^\circ \cup \varepsilon^\partial$. 

The elements $K^+$ and $K^-$ have outward pointing unit normals $\bm{n}^+$ and $\bm{n}^-$, respectively.
The quantities $a^\pm$ denote the traces of $a$ on the edge $e$ from the interior of $K^\pm$.
When relevant for element-wise operations, we take as convention that the element $K^-$ refers to the local element, and $K^+$ to the neighboring element.
The jump $\jmp{\cdot}$ operator for scalar quantities are then defined as $ \jmp{a} = a^{-} - a^{+}$ on the interior faces $e \in \varepsilon^\circ$. On the set of boundary faces $e \in \varepsilon^\partial$, we can unambiguously refer to a single-valued trace and outward facing normal $\bm{n}$ on $\Gamma$. 
Lastly, we define the inner products over a set $D \subset \mathbb R^d$ and its boundary $\partial D \subset \mathbb R^{d-1}$ using typical discontinuous Galerkin finite element notation as
\begin{equation*}
  (c, d)_{D} = \int_{D} cd \,\mathrm{d}D, \qquad
 \langle c, d \rangle_{\partial D} = \int_{\partial D} cd \, \mathrm{d}{\partial D}.
\end{equation*}

Let $\mathcal{P}^{\porder}(D)$ denote the set of polynomials of degree $\porder$ on a domain $D$.
We consider the discontinuous finite element spaces
\begin{equation*}
\begin{aligned}
  W^{\porder}_{h} &= \left\{ w \in L^2(\Omega) \colon \restr{w}{K} \in W^{\porder}(K)\: \forall K\in\mathcal{T}_h \right\}, \\
  V^{\porder}_{h} &= \left\{ \bm{v} \in \left[L^2(\Omega)\right]^{d} \colon \restr{\bm{v}}{K} \in
    V^{\porder}(K)\: \forall K\in\mathcal{T}_h \right\}, \\
    M^{\porder}_{h} &= \left\{ \mu \in L^2\left({\varepsilon_{h}}\right) \colon \restr{\mu}{e} \in
    \mathcal{P}^{\porder}(e) \: \forall e\in\varepsilon_h \right\},
\end{aligned}
\label{eq:imp:polynomial_spaces}
\end{equation*}
where $L^2(\Omega)$ is the space of square-integrable functions on the domain $\Omega$. 
Informally, $W_h^\porder$ represents the space of piecewise discontinuous polynomials of degree at most $p_{\text{order}}$ on every element in the mesh.

\subsection{Linear advection equation}%
\label{sec:linear_advection_equation}

We will consider the linear advection equation
\begin{equation}
    \begin{aligned}
        \frac{\partial u}{\partial t} + \nabla \cdot \left(\bm{c}u\right) &= f \qquad & \text{ in } \Omega,\\
        u &= g_D \qquad & \text{ on } \Gamma_{\text{in}},
    \end{aligned}
    \label{eq:linear_advection_PDE}
\end{equation}
on the domain $\Omega$ with boundary $\Gamma$ separated into inflow and outflow regions $\Gamma = \Gamma_{\text{in}} \cup \Gamma_{\text{out}}$, that are defined according to the continuous, divergence-free advection velocity field  $\bm{c}$ and outward normal $\bm{n}$,
\begin{align*}
    \Gamma_{\text{in}} = \left\{x\in \Gamma \colon \bm{c}\cdot\bm{n} \leq 0\right\},\\
    \Gamma_{\text{out}} = \left\{x\in \Gamma \colon \bm{c}\cdot\bm{n} > 0\right\} \, .
\end{align*}
The linear advection problem defined in equation (\ref{eq:linear_advection_PDE}) admits the following semi-discrete discontinuous Galerkin discretization \cite{cockburn2003discontinuous,Kronbichler2021}: we seek $u_h \in W_h^\porder$ such that
\begin{equation*}
    \left(w, \frac{\partial u_h}{\partial t}\right)_{\mathcal{T}_h}
    -\left(\nabla w,\, \bm{c}u_h \right)_{\mathcal{T}_h} 
    + \left\langle \jmp{w},\, u^*\, (\bm{c}\cdot\bm{n}) \right\rangle_{\varepsilon^\circ}
    + \left\langle w,\, u_h (\bm{c}\cdot\bm{n}) \right\rangle_{\Gamma_{\text{out}}}
    = (w, f)_{\mathcal{T}_h}
    - \left\langle w,\, g_D (\bm{c}\cdot\bm{n}) \right\rangle_{\Gamma_{\text{in}}},
    \label{eq:DG-advection-discretization}
\end{equation*}
for all $w\in W_h^\porder$.
There are many choices for the single-valued, inter-element numerical flux $u^*$. 
To mimic the physics of the problem, we use the common upwinded flux
\begin{equation*}
  u^* = 
  \begin{cases}
    u_h^{+}, & \bm{c}\cdot\bm{n} < 0 \\
    u_h^{-}, & \bm{c}\cdot\bm{n} \geq 0.
  \end{cases}
\end{equation*}
In the steady case, the time derivative term is set to zero. In the unsteady case, a standard explicit time-integration scheme can be used for temporal discretization using a method of lines approach \cite{hesthaven2007nodal}; we use LSERK-45 \cite{carpenter1994fourth}.

\subsection{Advection-diffusion equation}%
\label{sec:advection-diffusion-eqn}

To demonstrate that the methodology generalizes well across different problem physics as well as across different finite element discretizations, we will consider the advection-diffusion equation, which includes as a subset second-order Poisson-like problems.
Namely, we consider the set of problems described by the PDE
\begin{equation}
  \begin{aligned}
    \frac{\partial u}{\partial t}
    + \nabla \cdot \left(\bm{c} u\right) 
    - \nabla \cdot \left(\kappa \nabla u\right) &=  f, \qquad \text{ in } \Omega \\
    \left(-\kappa \nabla u + \bm{c}u \right)\cdot\bm{n} &= g_N, \qquad \text{ on } \Gamma_N, \\
    u &= g_D, \qquad \text{ on } \Gamma_D,
  \end{aligned}
  \label{eq:advection-diffusion-PDE}
\end{equation}
where $\Gamma_D$ and $\Gamma_N$ denote the Dirichlet and Neumann segments of the boundary, respectively.
The problem in equation (\ref{eq:advection-diffusion-PDE}) admits a hybridizable discontinuous Galerkin (HDG) formulation, developed in \cite{nguyen2009implicit}: we seek $(\bm{q}_h,u_h,\hat{u}_h)\in V_h^\porder\times W_h^\porder \times M_h^\porder$ such that
\begin{equation}
  \begin{aligned}
    \left(\bm{v},\, \kappa^{-1} \bm{q}_h\right)_{\Th}
    - \left(\nabla \cdot \bm{v},\, u_h \right)_{\Th}
    + \left\langle \bm{v}\cdot\bm{n},\, \hat{u}_h\right\rangle_{\partial \Th}
    &= 0 \\
   \left(w,\, \frac{\partial u_h}{\partial t}\right)_{\Th}
   -\left(\nabla w,\, \bm{c}u_h\right)_{\Th}
   + \left(w,\,  \nabla \cdot \bm{q}_h \right)_{\Th}
   + \left(w,\,  (\bm{c}\cdot\bm{n}) \hat{u}_h\right)_{\partial\Th}
   + \left\langle w,\, \tau(u_h - \hat{u}_h) \right\rangle_{\partial\Th}
   &= \left(w,\,  f\right)_{\Th}\\
   \left\langle \mu,\, \bm{q}_h\cdot\bm{n} 
       + (\bm{c}\cdot\bm{n})\hat{u}_h 
       + \tau(u_h - \hat{u}_h)
       \right\rangle_{\partial\Th}
    &=
    \left\langle \mu,\, g_N \right\rangle_{\partial\Th}
  \end{aligned}
  \label{eq:HDG-advection-diffusion}
\end{equation}
for all $(\bm{v}, w, \mu) \in V_h^\porder\times W_h^\porder \times M_h^\porder$.
We take the diffusion coefficient $\kappa = 1$, choose the stabilization parameter $\tau = \kappa/\ell + |\bm{c}\cdot\bm{n}|$, where we use as a diffusion length scale $\ell = 1/5$.
In order to solve Poisson-like problems, we set the convective velocity field $\bm{c}$ to zero.
In the unsteady case, an implicit time-integration scheme can be applied for the temporal discretization; we use the third-order backward difference formulae (BDF3).
We refer the reader to \cite{nguyen2009implicit} for a thorough discussion of these choices.

\subsection{Error indicators and AMR heuristics}
\label{sec:error_indicators}

For the test cases in \S\ref{sec:numerical_experiments}, we benchmark the RL policy against two common AMR heuristics, the Kelly Error indicator and a gradient indicator. The indicators are shown in Table \ref{tab:AMR_error_indicators}.

\begin{table}[H]
\centering
\begin{tabular}{ c c c } 
\hline
 Error indicator & faces $F = K \cap K^\prime$ & boundary face $\Gamma_N$ \\ 
 \hline
 Kelly  & $\displaystyle \sum_{F\in\partial K} c_F\int_{F} \left\llbracket a \frac{\partial u_h}{\partial \bm{n} } \right\rrbracket^2 \, dF$  & $ n_F \int_F \left \lvert g_N - a \frac{\partial u_h}{\partial \bm{n}} \right \rvert^2 dF $ \\ 
 Gradient-based & $\displaystyle Y^{-1}\sum_{K^\prime} \frac{y_{K^\prime}}{\left \lVert y_{K^\prime} \right \rVert} \frac{u_h(x_{K^\prime}) - u_h(x_K)}{ \left\lVert y_{K^\prime} \right \rVert }$ & - \\  
 \hline
\end{tabular}
\vspace{1em}
\caption{\small Cell-wise error indicators, contributions by face.}
\label{tab:AMR_error_indicators}
\end{table}
The Kelly indicator uses parameters $c_F = n_F = h_K / 24$ for element width $h_K$. 
The approximate gradient is formed using the distance vectors $y_K^\prime = x_K^\prime - x_K$ between the cell centers of element $K$ and its neighbors $K^\prime$, as well as the matrix
\begin{equation*}
    Y = \sum_{K^\prime} \left(
 \frac{y_{K^\prime}}{\left \lVert y_{K^\prime} \right \rVert} 
 \frac{y_{K^\prime}^T}{\left \lVert y_{K^\prime} \right \rVert} 
    \right).
\end{equation*}
We scale the approximate gradient $\tilde{\nabla} u_h$ in Table \ref{tab:AMR_error_indicators} by a power of the mesh width, using $h_K^{1 + d/2} \left\lVert \tilde{\nabla} u_h\right\rVert$.
To execute an AMR strategy based on either of these estimators, we use either a bulk or fixed fraction rule.
As described in \S\ref{sec:introduction}, a bulk strategy refines and coarsens the cells in which the top and bottom percentages of the estimated error occurs; therefore the actual number of cells coarsened or refined is entirely dependent on the estimate. 
On the other hand, a fixed-fraction refinement strategy sorts the cells according to their error estimates and refines and coarsens a top and bottom percentage of the total number of cells.
We denote these strategies as \bulk{refine percentage}{coarsen percentage} and, similarly
\fixedfraction{$\cdot$}{$\cdot$};\, \eg, \fixedfraction{0.4}{0.6} refers to an AMR strategy where the top 40 percent of cells in terms of their estimated error are refined and the bottom 60 percent are coarsened.
For additional details on both these indicators and refinement strategies, see \cite{dealII93}.

\subsection{Numerical implementation}%
\label{sec:implementation_details}

The finite element forward models were implemented in \texttt{C++} and make use of the finite element library \texttt{deal.ii} \cite{dealII93}.
The POMDP
characterizing the RL environment was implemented using the OpenAI Gym framework \cite{brockman2016openai}, and the custom deep RL architectures were implemented in the open-source RL framework Stable-Baselines 3 \cite{raffin2019stable}. 

Unless otherwise specified, the policy architecture used is A2C, and the hyperparameter $\gamma_c = 25$ (see \S\ref{sec:reward_fn}, \S\ref{sec:static_1D_LA}), which we empirically found to give good performance over a wide range of test cases. 
Additional values of training parameters are given in Appendix C.
As is typical in deep RL, the particular number of training time steps used is not particularly meaningful as compared to the order of magnitude of the total number of steps (see discussion in \S\ref{sec:lin_adv_training_and_deployment}), as long as the policy network is given sufficient time to increase its mean episodic reward significantly from its starting performance.
For all experiments, we used on the order of $10^5$ RL training time steps, consisting of 100-200 time step episodes, and training took on the order of 1-3 hours on a desktop computer without GPU acceleration. 
More detailed benchmarking would be extraneous, as performance in terms of training time is dominated by the cost of running the numerical solver rather than updating the policy network, for all but trivial problems.

All linear systems arising from the finite element discretizations are solved using direct solvers (UMFPACK) to avoid the complicating factors of iterative solver tolerances and stopping criteria. 
All surface and volume integral operators are discretized with Gaussian quadrature using $p_{\text{order}}+1$ one-dimensional (1D) quadrature points in each spatial direction, where $p_{\text{order}}$ is the polynomial order of the finite element space $W_h^\porder$.

$L^2$-errors, when shown, are computed as a post-processing step using a known exact solution and numerically integrated using $\porder + 3$ Gaussian quadrature points in each spatial direction to ensure that the calculation of errors is not adversely affected by integration error.
We re-emphasize that no exact solutions are used at any point during training or deployment of the RL model itself.

\section{Numerical experiments}
\label{sec:numerical_experiments}

The numerical experiments in this section are designed to demonstrate the features and performance of the deep RL AMR (DRL-AMR) approach.
We start with
a 1D, steady linear advection problem in \S\ref{sec:static_1D_LA} as an illustrative example to exhibit the feasibility of the approach and to demonstratively explore the characteristics of the RL policy.
We provide a discussion of details related to training and model evaluation, as well as performance comparisons between different RL algorithms, to which the simplicity and interpretability of the test case lend itself well.

Subsequent sections focus on the performance and generalizability of the approach.
We use the same trained RL model in \S\ref{sec:static_1D_LA}, but apply it to a different linear advection problem in \S\ref{sec:static_1D_gen} to demonstrate generalization of the policy to different boundary conditions and forcing functions.
To show the generalization of the method to unsteady dynamics, we train a model on a 1D time-dependent advection problem in \S\ref{sec:time-dependent-1D-adv}.
Lastly, to support the claim that the approach is not specific to a particular PDE or finite element scheme, we apply the DRL-AMR framework to the Poisson equation discretized with an HDG scheme (\S\ref{sec:advection-diffusion-eqn}).
As the Poisson equation is a second-order, elliptic PDE, its solutions are characterized by different physics than the hyperbolic, first-order advection problems solved in \S(\ref{sec:static_1D_LA}-\ref{sec:time-dependent-1D-adv}).
Similarly, as HDG schemes are mixed methods, their formulation (\ref{eq:HDG-advection-diffusion}) involves vector-valued finite element spaces and use of a traced finite element space \cite{nguyen2009implicit} on the mesh skeleton; that is, the underlying numerical method is significantly different than those used for the solution of the advection equation (\cf (\ref{eq:DG-advection-discretization})). 

To show that the methods are not relegated to small or one-dimensional problems and that the RL approach generalizes to higher spatial dimensions and larger, more complex problems, we examine the performance of deep RL policies trained on 2D problems in \S(\ref{sec:steady-2D-advection}-\ref{sec:steady-2D-advection-diffusion}).
The steady advection problem in \S\ref{sec:steady-2D-advection} allows evaluation of the methodology in the context of the two-dimensional generalization of the problem in \S\ref{sec:static_1D_LA}.
The steady advection-diffusion problem in \S\ref{sec:steady-2D-advection-diffusion} examines the effectiveness of the deep RL policy when both advection and diffusion processes occur, and highlights its ability to detect and resolve non-trivial features in an automated way by using solution smoothness.
The unsteady advection problem in \S\ref{sec:ring_test_case} shows the ability of the method to preserve salient physical features of a numerical solution over long-time integration horizons.
Throughout the entire set of numerical experiments, we focus on the goal stated in the introduction of providing a solution that captures all physically relevant features of a numerical solution efficiently in terms of problem degrees of freedom, avoiding spurious diffusion and dispersion effects due to under-resolution.

\subsection{Steady 1D linear advection}%
\label{sec:static_1D_LA}

For proof-of-concept, we consider the linear advection equation described in \S\ref{sec:linear_advection_equation} on the 1D spatial domain $\Omega = [0,1]$. We choose the boundary conditions $g_D$ at the inlet and forcing function $f$ such that the exact solution takes the form
\begin{equation*}
  u(x) = 1 - \tanh\left[ \alpha(1- 4(x - 1/4)) \right],
\end{equation*}
where the steepness parameter $\alpha=10$. As the exact solution has the form of a smooth step function, the resultant meshes and numerical solutions are easily interpretable---resolution should be concentrated in the steep gradient region around the \say{step}.
In light of this, we will use this illustrative example to demonstrate the features of the trained deep RL policy, as well as the details of training and model deployment.
Subsequent sections will focus on generalizability. 

Using a numerical solution of polynomial order $p_{\text{order}}=3$, we train the RL-agent for 20$\cdot10^3$ episodes on a computational \say{budget} of 25 cells (\S\ref{sec:implementation_details}).
However, at the time of model deployment, we are free to give the trained RL policy whatever budget we wish; we emphasize that this is because in general, we would like to train our model on much cheaper problems than those we intend to solve.

During deployment, starting with a very coarse mesh consisting of 4 elements, we perform 6 AMR cycles comparing the DRL-AMR model to an AMR heuristic that attempts to refine the elements responsible for the top 50 percent of the total error and attempts to coarsen the elements responsible for the bottom 50 percent of total error as measured by the approximate gradient error indicator (Table \ref{tab:AMR_error_indicators}). That is, a \bulk{0.5}{0.5} strategy (see \S\ref{sec:error_indicators}). 
We expect this heuristic indicator to perform well, as this test case has only one feature, the steep gradient in the center of the domain.

\begin{figure}[ht]
    \centering
    \begin{subfigure}{0.4\textwidth}
        \includegraphics[height=4.25cm]{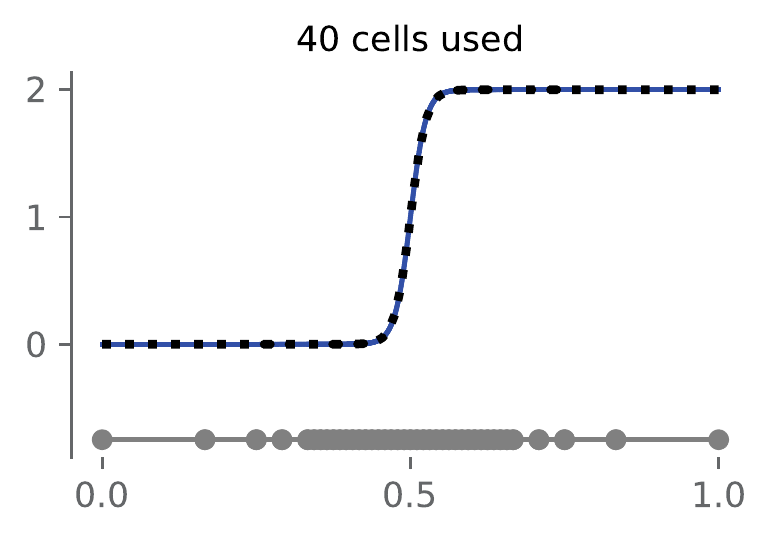}
        \caption{\small AMR heuristic}
        \label{fig:case0:AMR}
    \end{subfigure}
    \begin{subfigure}{0.45\textwidth}
        \includegraphics[height=4.25cm]{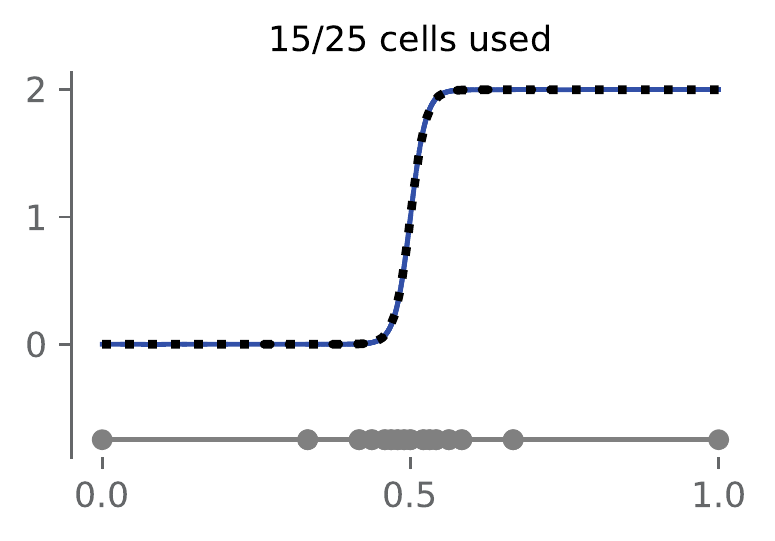}
        \caption{\small RL Agent}
        \label{fig:case0:RL}
    \end{subfigure}
    \caption{\small Steady 1D linear advection (\S\ref{sec:static_1D_LA}). Numerical solution with the mesh resulting from 6 cycles of refinement, using the approximate gradient error indicator as an AMR heuristic and the deep RL policy resulting from training. The exact solution is overlaid in both cases for comparison.}
    \label{fig:case0_results}
\end{figure}

\begin{figure}[ht]
    \centering
    \begin{subfigure}{0.45\textwidth}
        \includegraphics[height=6cm]{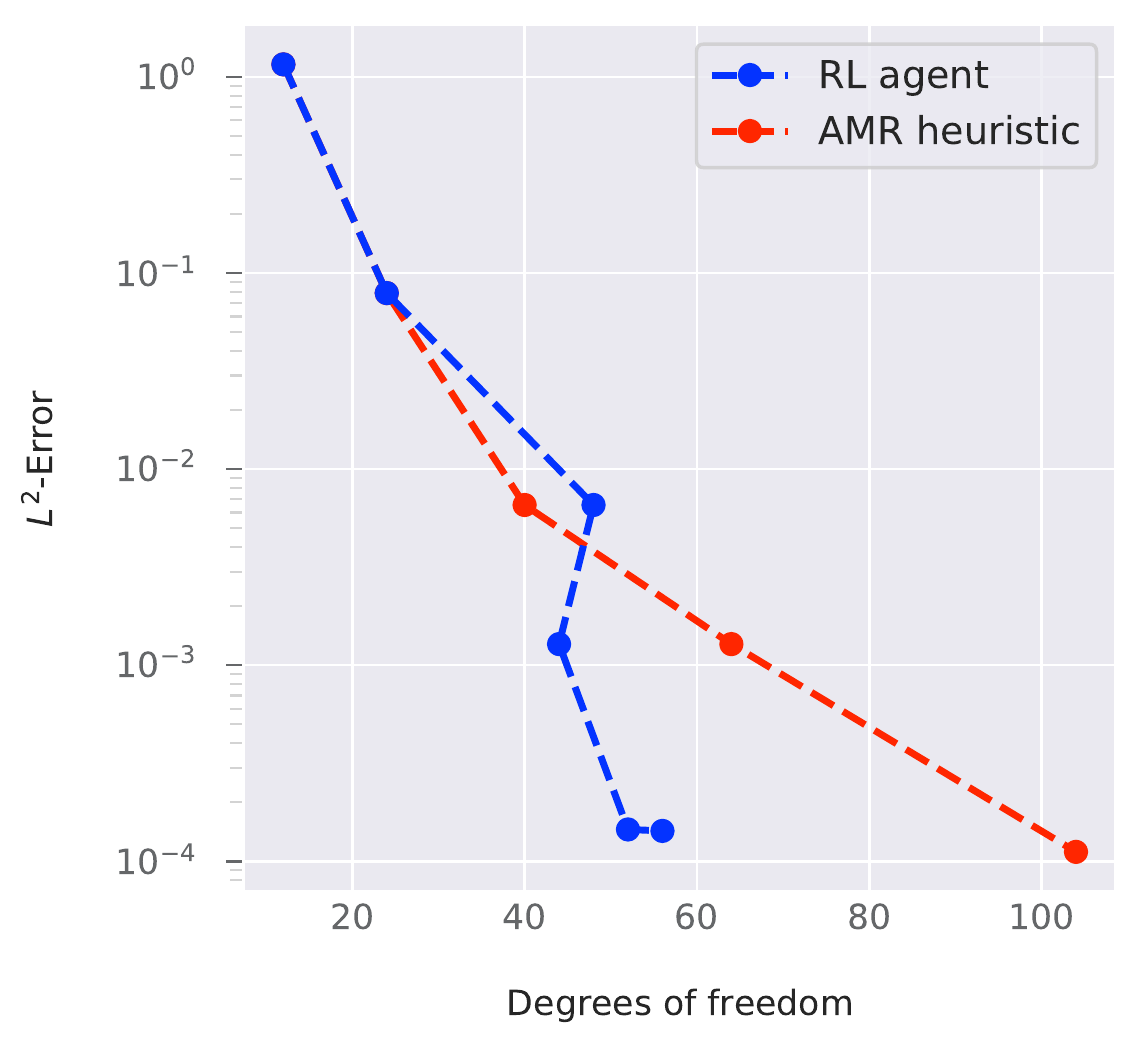}
        \caption{\small 25 cell budget}
        \label{fig:case2_compare_25}
    \end{subfigure}
    \begin{subfigure}{0.45\textwidth}
        \includegraphics[height=6cm]{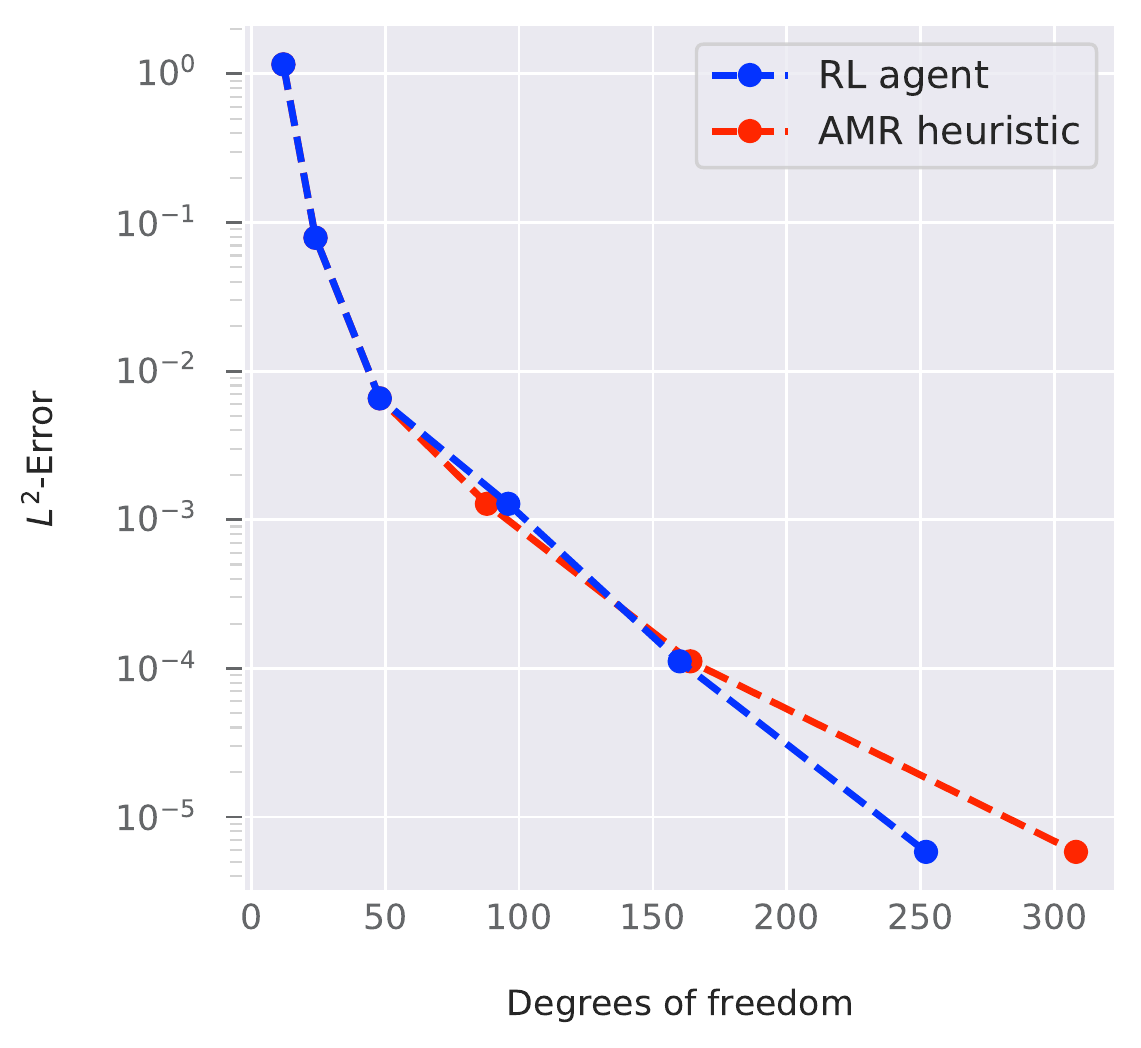}
        \caption{\small 500 cell budget}
        \label{fig:case2_compare_500}
    \end{subfigure}
    \caption{\small Steady 1D linear advection (\S\ref{sec:static_1D_LA}).
    $L^2$-error plotted against problem degrees of freedom over 6 cycles of refinement, using the gradient-based error indicator as an AMR heuristic and the deep RL policy: (\ref{fig:case2_compare_25}) deployed with a computational budget of 25 elements; (\ref{fig:case2_compare_500}) with a computational budget of 500 elements.}
    \label{fig:case2-err-per-dof}
\end{figure}

Figure \ref{fig:case0_results} shows the numerical solution and meshes proposed by the two approaches.
We see that the deep RL agent deployed with a 25-cell budget is able to find a high-quality solution with fewer elements than that of the AMR heuristic recommendation.
This is corroborated by the more precise comparison in Figure \ref{fig:case2-err-per-dof}, which shows that at model deployments with budgets of 25 and 500 cells, the RL agent outperforms the AMR heuristic over six refinement cycles in terms of $L^2$ accuracy per degree of freedom.
That is, the RL policy provides a numerical solution of approximately the same accuracy as that of the AMR heuristic, but does so with fewer degrees of freedom after the refinement cycles, both on the problem size it was trained on, as well as at a much larger problem size.

Although for 1D problems, differences in problem degrees of freedom are small, this example illustrates the central difference between the learned policy and the classical AMR approaches. 
AMR algorithms mark cells for refinement either in terms of the estimated volume fraction of total error in the case of bulk refinement or by a certain number of cells in the case of fixed-number refinement. 
In either case, the AMR algorithms rely on estimators purely to decide which cells to refine, but the actual refinement behavior is in some sense, specified \textit{a priori}. 
Similarly, the commonly-used way to prevent AMR heuristics from continuing to refine beyond a computational limit is to manually specify a max refinement depth, again an \text{a priori} choice uninformed by the particulars of the solver or PDE, and specified everywhere.
The DRL-AMR approach, in contrast, not only estimates which cells are likely responsible for a disproportionately large or small share of the total error but also estimates a stopping point, beyond which the solution smoothness indicates that additional decreases in error are likely to be marginal from the perspective of adding new unresolved features to the solution.
This is because, during training, the RL agent is given global information related to available computational resources, which it balances against the expectation of change in the solution upon local refinement, with respect to computational cost.

Where the stopping point occurs, in terms of the solution error, is controllable through the hyperparameter $\gamma_c$. 
A larger value of $\gamma_c$ means a cheaper numerical solution with a greater aversion to incurring a cost.
On the other hand, a small value of $\gamma_c$ will result in a numerical solution with a lower error and less tolerance for interface jumps.
Informally, $\gamma_c$ is a user-specified setting indicating the trade-off between cost and accuracy, and is a way to inform the agent ``how good'' versus ``how cheap'' to make the solution.
In appendix A, we provide an approach to remove $\gamma_c$ as a training hyperparameter entirely, and instead make it a user-chosen value during deployment by modifying the policy architecture; however, this discussion is beyond the scope of the main contributions of this paper, and we train using a single value of $\gamma_c$ for all of the numerical experiments in this section.
Next, we explore the training and deployment behavior of the RL policy for this simple case to better understand the model.

\subsubsection{Training and deployment considerations}
\label{sec:lin_adv_training_and_deployment}

\textbf{Initialization.} At the start of each training episode, we can either allow the RL agent to begin on the coarsest possible mesh (coarse initialization) or initialize the mesh to a random state by performing a random number of refinements.

\begin{figure}[ht]
    \centering
    \begin{subfigure}{0.55\textwidth}
        \includegraphics[height=5.5cm]{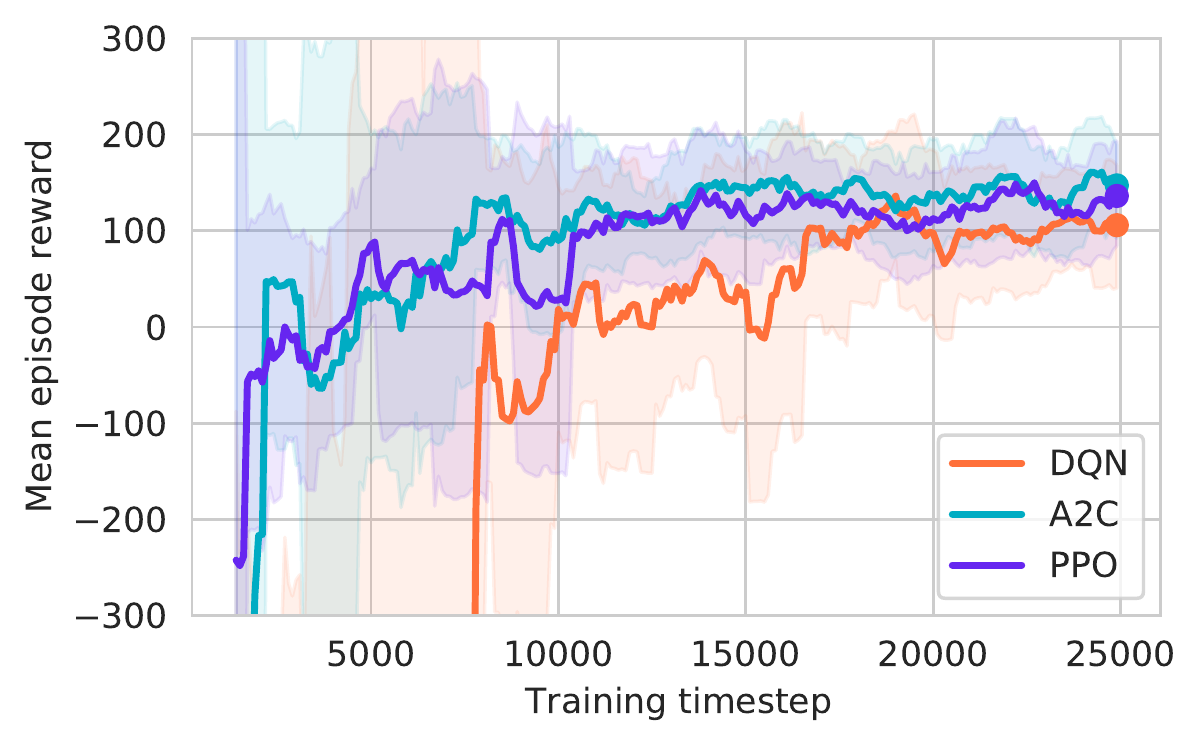}
        \caption{\small DRL-AMR algorithm performance over training}
        \label{fig:RL_algorithm_training_performance}
    \end{subfigure}
    \begin{subfigure}{0.35\textwidth}
        \includegraphics[height=5.5cm]{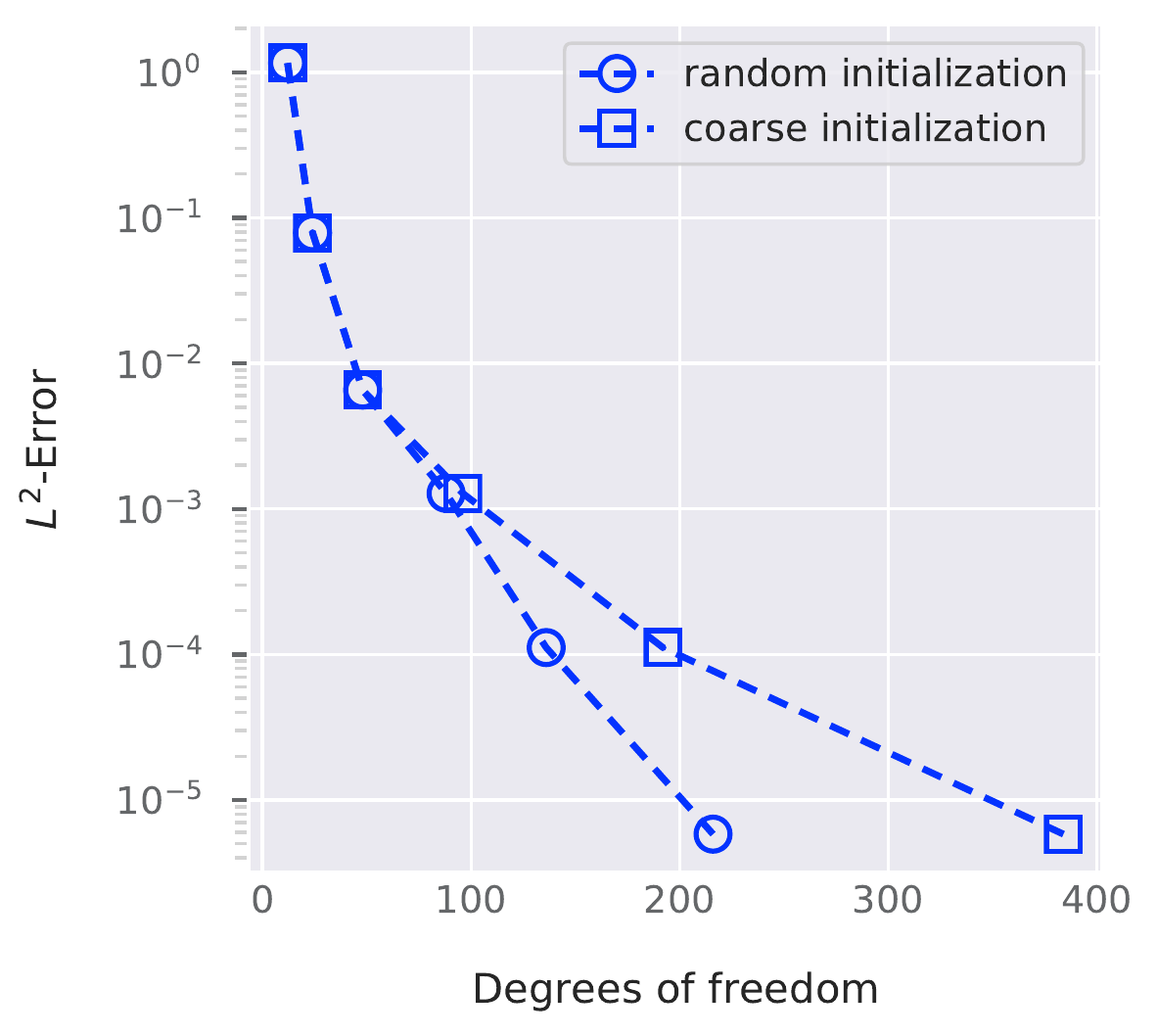}
        \caption{\small Mesh initialization}
        \label{fig:random_vs_coarse_init}
    \end{subfigure}
    \caption{\small
    For the steady linear advection test in \S\ref{sec:static_1D_LA}:
    (a) Performance achieved during training for different deep RL training algorithms, as measured by average episode reward (solid lines). %
    Shaded bands indicate the sample variance of the reward over the episode.
    (b) Performance of DRL-AMR models trained using random state initialization and coarse initialization, as measured by $L^2$-Error per degree of freedom over 6 refinement cycles
    (on a 25-cell budget example and deployed with a 500-cell budget).
    }
\end{figure}
Random initialization and coarse initialization perform similarly for the small problem sizes on which they are trained; however, per unit training time, we find that random initialization often yields better policies when deployed on substantially larger problems.
Figure \ref{fig:random_vs_coarse_init} compares the performance of two DRL-AMR models trained for $10^5$ training time steps, one with the mesh in a coarse state at the beginning of each episode, and the other with a mesh in a random state at the beginning of each episode.
Both models were trained on the linear advection problem in \S\ref{sec:static_1D_LA}  with a maximum budget of 25 cells but were deployed
with a budget of 500 cells over six refinement cycles. 
Comparing the accuracy with the problem degrees of freedom, it is clear that the model trained with random initialization significantly outperforms the model with coarse initialization.
Both models begin the refinement cycle on the same coarse mesh during deployment. The model trained with random initialization finds a solution of the same accuracy but uses only half the elements as the model trained with coarse initialization. 
We obtained similar results for other test cases we ran.

We hypothesize that random initialization during training leads to a more aggressive exploration of the action space and produces better results because initializing the mesh to a random state produces regions that are over-refined and under-refined.
On the other hand, initialization on a coarse mesh tends to produce an under-resolved solution everywhere, leading the agent to become biased towards refinement during deployment. 
As might be expected, in the limit of long training times, performance becomes comparable between the two initialization strategies.

In light of these performance advantages, all results shown correspond to models trained using random initialization.

\textbf{Learning algorithms.} The performance---as measured by mean episodic reward---for the three RL policy architectures (DQN, A2C, and PPO, see \S\ref{sec:policy_architectures}) are shown in Figure  \ref{fig:RL_algorithm_training_performance}. The sample variance bands are found to tighten non-monotonically over the training duration, indicating more consistent performance.
All three algorithms achieve similar mean reward and variance after roughly 25,000 training time steps, successfully solving the problem.
However, as is common in RL problems, the model performance during training is non-monotonic, and it is advantageous to periodically save the most performant model state for deployment, rather than deploying the policy occurring at the end of the training window.
In general, we have not found any particular architecture to be consistently better in terms of performance over the set of test cases considered in this paper.
We note however that we have not yet attempted to accelerate the training process through optimization of training parameters such as batch size and episode length.

\textbf{Model deployment.}
Unlike the AMR heuristic, the deployed DRL-AMR model finds solutions by apportioning computational resources over the mesh in a way that handles the smoothness/efficiency trade-off, then changes the mesh topology over the subsequent refinement cycles in order to increase accuracy.
This is the sense in which the algorithm considers a richer set of strategies than a purely greedy strategy. 
\begin{figure}[h!]
  \centering
  \includegraphics[width=0.9\linewidth]{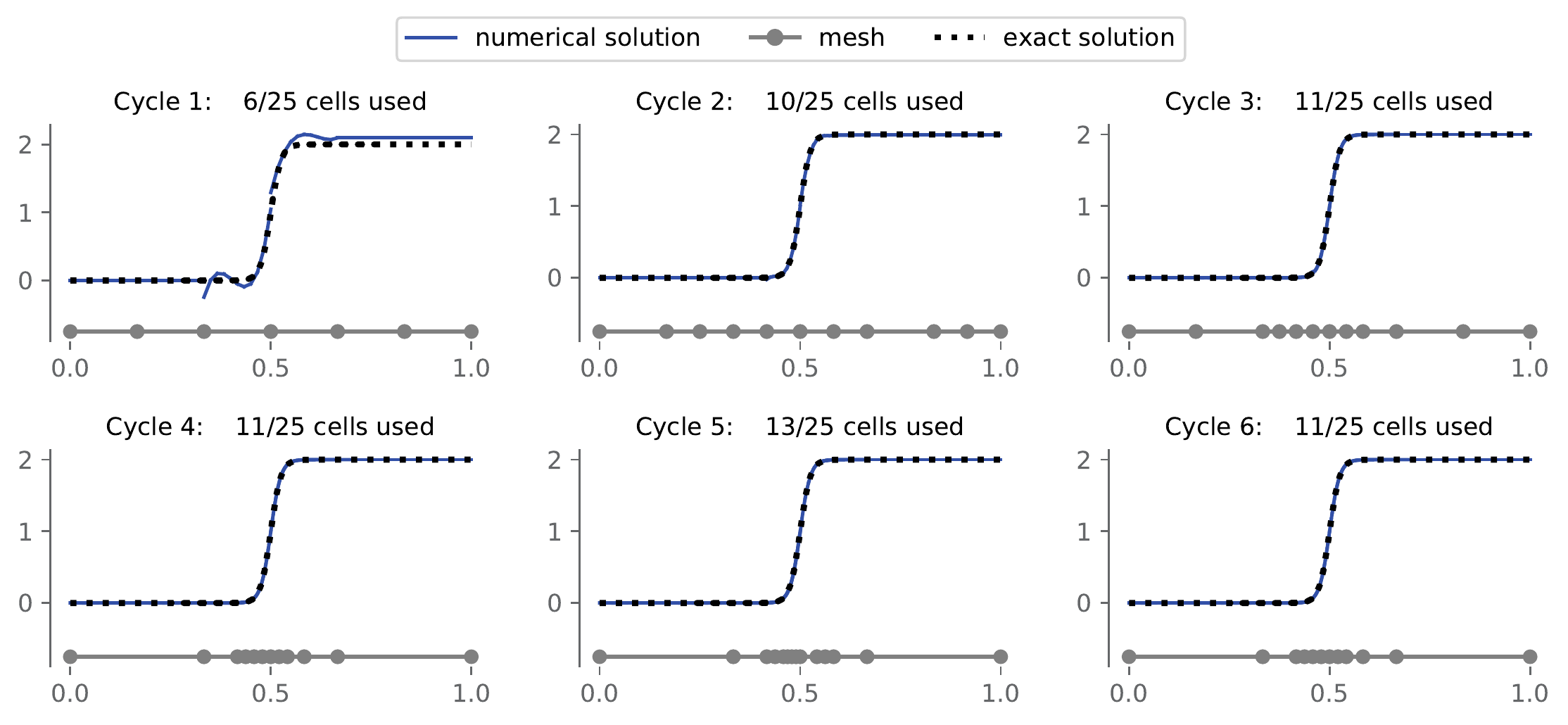}
  \caption{\small Steady 1D linear advection (\S\ref{sec:static_1D_LA}). State of the numerical solution and mesh during the deployment of the deep RL agent AMR policy, over six refinement cycles.}
  \label{fig:case2_cycle_plots}
\end{figure}

Figure \ref{fig:case2_cycle_plots} shows the mesh and resultant numerical solution over a set of six refinement cycles.
Although cycles 2-6 all contain the order of 10 elements, the topology of their allocation over the domain changes to concentrate around the steep gradient region. 
This showcases a particular strength of the method: namely, that the decision of the number of elements to use during the exploratory cycles while preserving solution smoothness is delegated to the machine learning model, rather than manually needing to be specified. 
This allows the number of elements to increase and decrease exploratively, as opposed to AMR heuristics, which must increase or decrease the number of elements according to the error estimator or by a fixed number.

In practice, the cost of each cycle depends on the problem size but we find that the trained RL policies reach a converged mesh state in a few cycles, typically around 5-10, independent of problem size.
This is in part because the number of cells can potentially double or halve every cycle, so locally under-resolved regions either become resolved very quickly, or the computational budget is approached and refinement recommendations become more conservative.

\textbf{Introspection of the neural network model.}
As an RL policy is ultimately a trained neural network, we can query the neural network in order to introspect the policy suggestions.
In practice, this space is high-dimensional---the input to the network is a vector of dimension equal to that of the observation space.
However, insight can be gained by considering a model trained on a reduced observation space.

\begin{figure}[h!]
  \centering
  \includegraphics[width=0.8\linewidth]{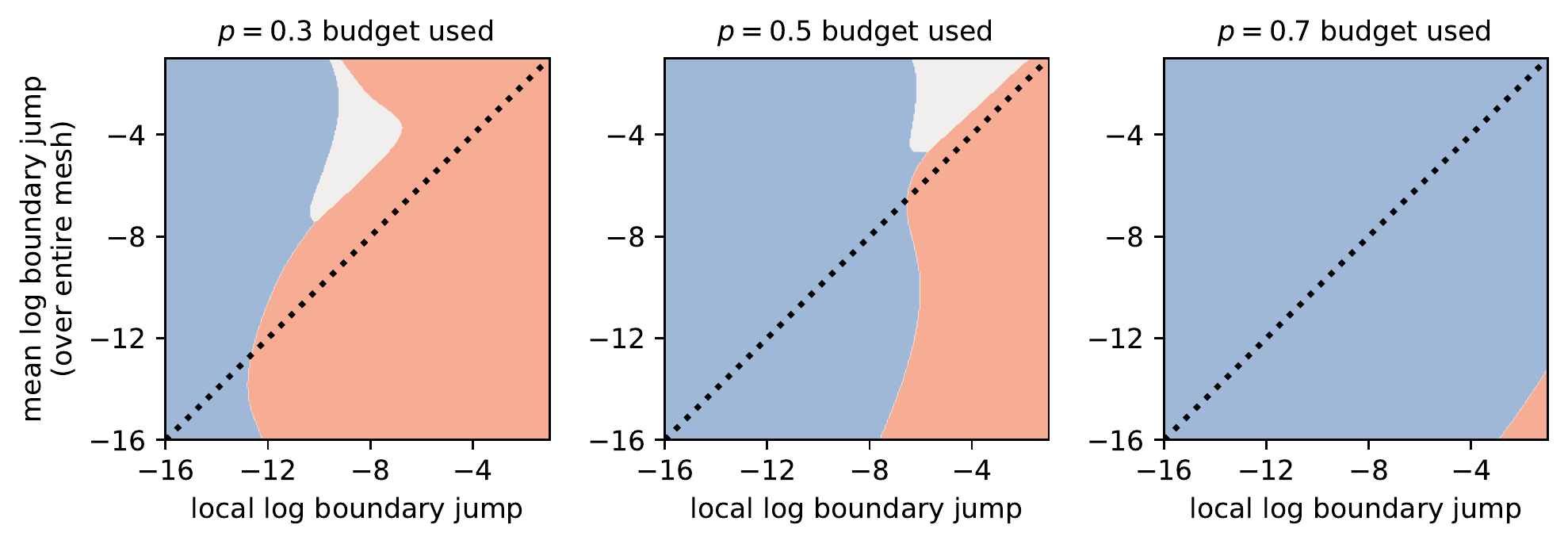}
  \caption{\small 
  Policy network recommendation as a function of a simplified observation space. Decision boundaries are shown between network recommendations to coarsen (blue), do nothing (grey), and refine (salmon). The dashed line indicates where the local boundary jump is the same as the mean cell-wise boundary jump over the entire mesh.}
  \label{fig:network_introspection}
\end{figure}

Figure \ref{fig:network_introspection} shows the recommendation of a model trained on the problem in \S\ref{sec:static_1D_LA} using a value $\gamma=25$ and a simplified observation space consisting only of the cell boundary jump, the mean boundary jump over all cells, and the current use of computational resources $p$.
We consider a range of [-16, -1] in log space, as these are the range of jump values encountered by the network for this test case; recommendations outside of this region are uninformed by data encountered during training.
Each subfigure shows the decision boundary regions corresponding to the action recommended by the policy.
Although the observation space has been simplified, the overall model is qualitatively discernible and interpretable.
At low use of computational resources ($p=0.3$), the model suggests mostly refinement, even in regions where the observed boundary jump is significantly lower than the average over the entire mesh. We hypothesize that this corresponds to exploratory refinement in some sense.
At moderate use of computational resources ($p=0.5$), the model provides more conservative recommendations, suggesting refinement where the local boundary jump exceeds the mesh mean and is not less than $10^{-8}$ in terms of magnitude.
When computational resources become scarce ($p=0.7)$, the model suggests coarsening except in regions where the local log boundary jump is much greater than the mean.
The decision boundaries are highly nonlinear, even for this simplified observation space.
This demonstrates the potential power and flexibility of the DRL-AMR approach, as it learns complex relationships between any physical relevant information included in the observation space and the utility of coarsening or refining.

\subsection{Generalization to different boundary conditions and forcing functions}%
\label{sec:static_1D_gen}

We deploy the same trained DRL-AMR model in \S\ref{sec:static_1D_LA} to the same PDE, but on a different domain $\Omega = (-4,4)$, with a different set of boundary conditions and forcing.
We choose the boundary conditions $g_D$ at the inlet and forcing function $f$ such that the exact solution takes the form
\begin{equation*}
  u(x) = \sin(n x) \exp\left(-\frac{1}{2} x^2\right).
  \label{eq:1D_LA_generalization}
\end{equation*}

At the time of model deployment, we give the trained RL policy a budget of 100 elements.  We evaluate the model over 6 AMR cycles this time using an AMR heuristic which makes use of the Kelly error indicator (Table \ref{tab:AMR_error_indicators}, \cite{kelly1983posteriori}) and implements a \fixedfraction{0.5}{0.1} strategy, as described in \S\ref{sec:error_indicators}.
\begin{figure}[ht]
    \centering
    \begin{subfigure}{0.49\textwidth}
        \includegraphics[width=\linewidth]{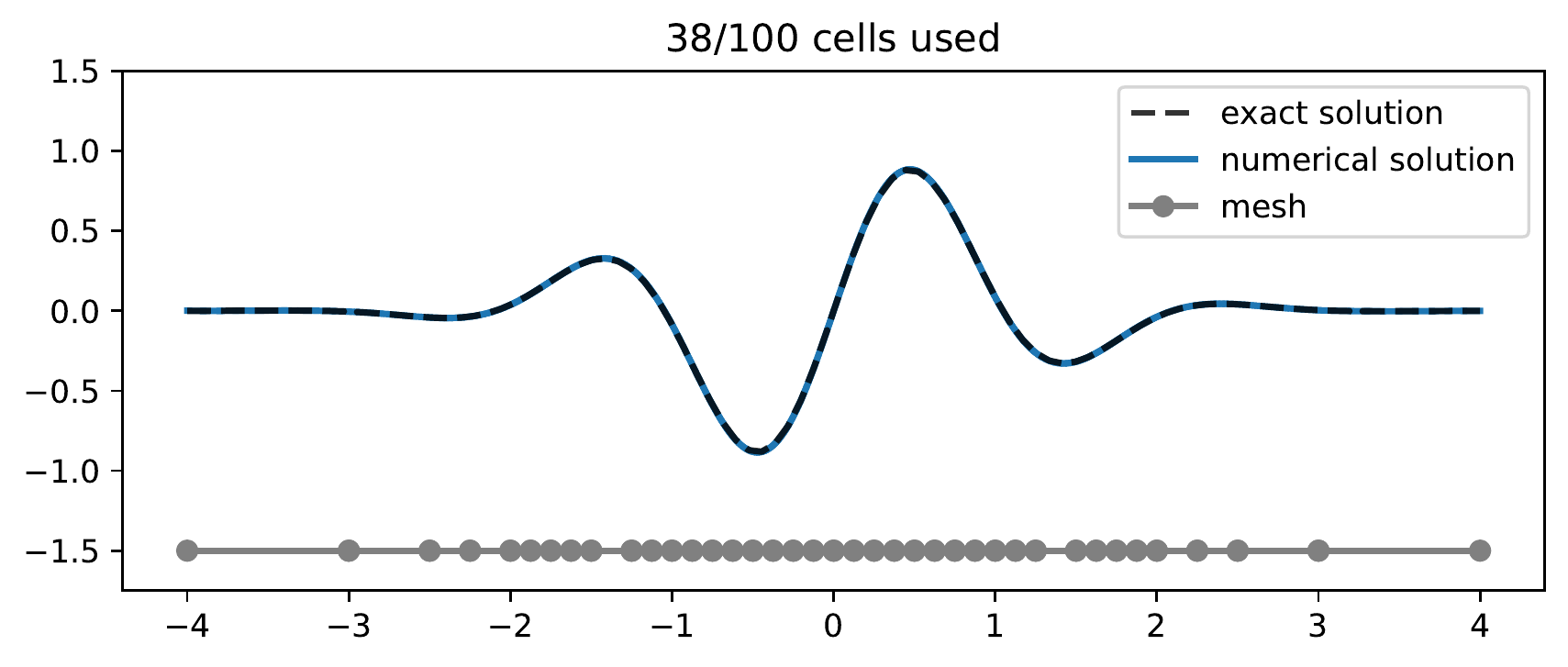}
        \caption{\small AMR heuristic}
    \end{subfigure}
    \begin{subfigure}{0.49\textwidth}
        \includegraphics[width=\linewidth]{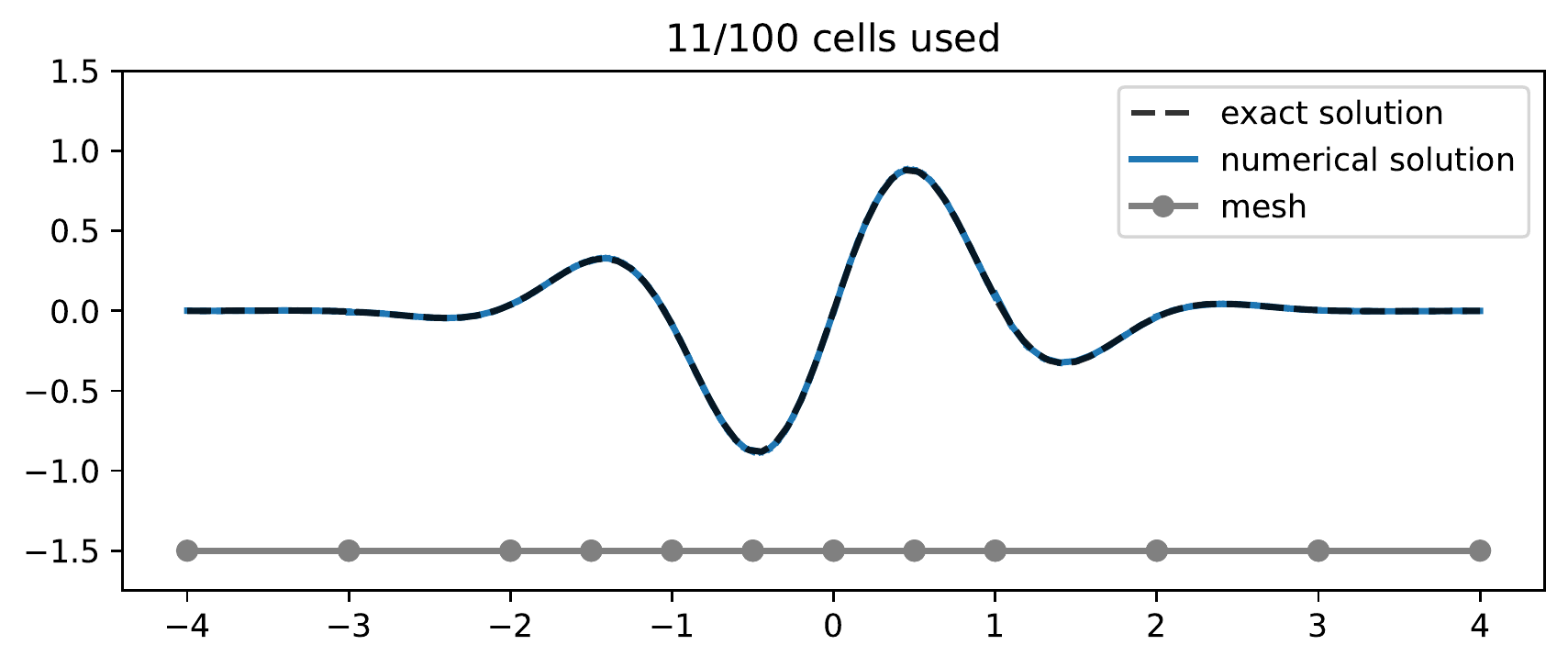}
        \caption{\small RL Agent}
    \end{subfigure}
    \caption{\small DRL-AMR model trained on the test case in \S\ref{sec:static_1D_LA} but deployed on the problem in equation (\ref{eq:1D_LA_generalization}). Numerical solution with the mesh resulting from 6 cycles of refinement, using the Kelly error indicator as an AMR heuristic and the RL policy resulting from training. The exact solution is overlaid in both cases for comparison.}
    \label{fig:1D_lin_adv_generalization}
\end{figure}

The final mesh and numerical solution for both is shown in Figure \ref{fig:1D_lin_adv_generalization}. 
Our findings are similar to the test case in the previous section.
Both methods find a mesh and corresponding numerical solution that very closely matches the exact solution, however, the RL policy finds a coarse mesh on which the solution is well represented and stops refinement past the fourth refinement cycle.
Similar to Figure \ref{fig:case2-err-per-dof}, after the six refinement cycles, the RL policy returns a mesh that provides strictly better accuracy per problem degree of freedom (not shown). 

This highlights the generality of the method; the RL policy did not suggest refining close to the center of the domain as it did for the example in \S\ref{sec:static_1D_LA}.
This is to be expected, as the RL agent is never given global location information during training, only local cell information along with the surrounding interface jumps; therefore it's impossible to over-fit the RL agent to a specific training example during training.
However, we remark that if the agent were trained against a pathological example such as the Weierstrass function, where the correct action could always be to refine, this will be reflected in the trained model, although this can hardly be considered overfitting.
Because the agent learns to relate features of the PDE and local jumps to the local smoothness of the solution, we find that the agent generalizes well across different test cases from the same PDE.
This is an important characteristic of the approach, as we wish to deploy the trained model on problems of interest other than the subset on which the model was trained.

\subsection{Unsteady 1D linear advection}%
\label{sec:time-dependent-1D-adv}

As an extension to the steady experiments, we consider the time-dependent Sommerfeld wave equation 
\begin{equation*}
    \begin{aligned}
        \frac{\partial u}{\partial t} + \frac{\partial u}{\partial x} &= 0 \qquad \text{ in } \Omega \times [0,T], \\
        u &= g_D \qquad \text{ on } \Gamma_{\text{in}}, \\
    u_0 &= \exp\left(\frac{1}{2\sigma^2} (x-\mu)^2\right),
    \end{aligned}
\end{equation*}
on the computational domain $\Omega = (-4,4)$. 
The Dirichlet boundary condition $g_D(t)$ at the inlet is chosen according to the exact solution to the time-dependent wave equation $u_{\text{exact}} = u_0(x-t)$.
The scalar constants are $\mu=-4$ and $\sigma^2 = 0.25$. 
The outlet boundary of the domain is taken to be an outflow condition.
We consider numerical solutions at a polynomial order $\porder = 3$.

\begin{figure}[ht]
    \centering
    \begin{subfigure}{0.45\textwidth}
        \includegraphics[width=\linewidth]{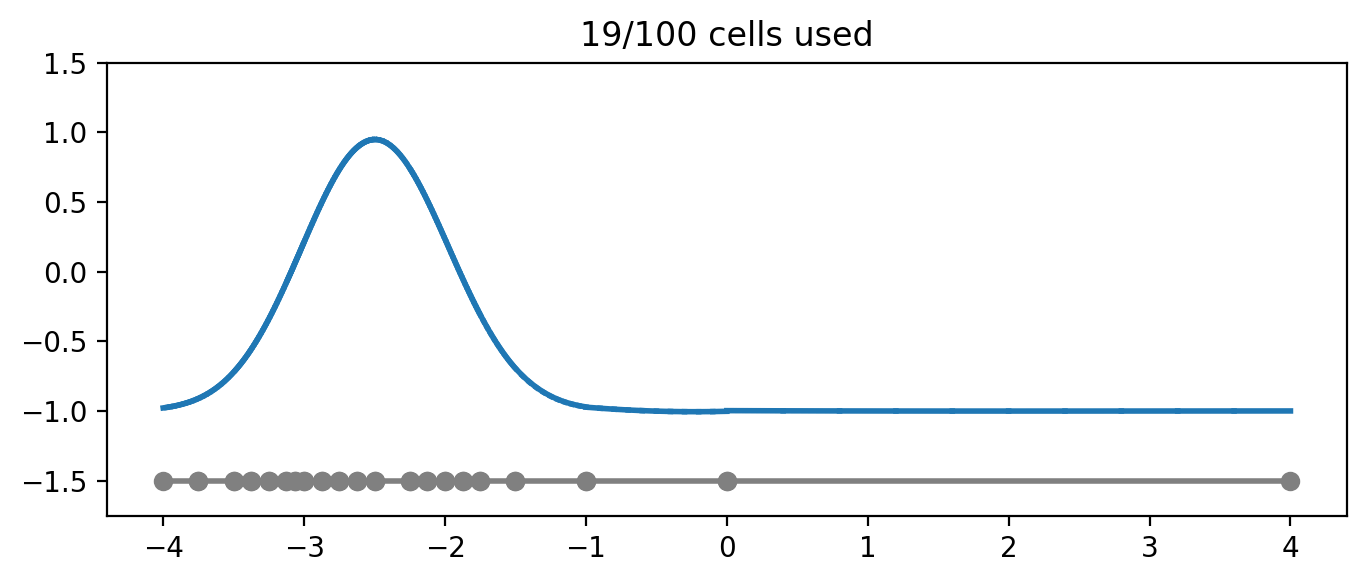}
    \end{subfigure}
    \begin{subfigure}{0.45\textwidth}
        \includegraphics[width=\linewidth]{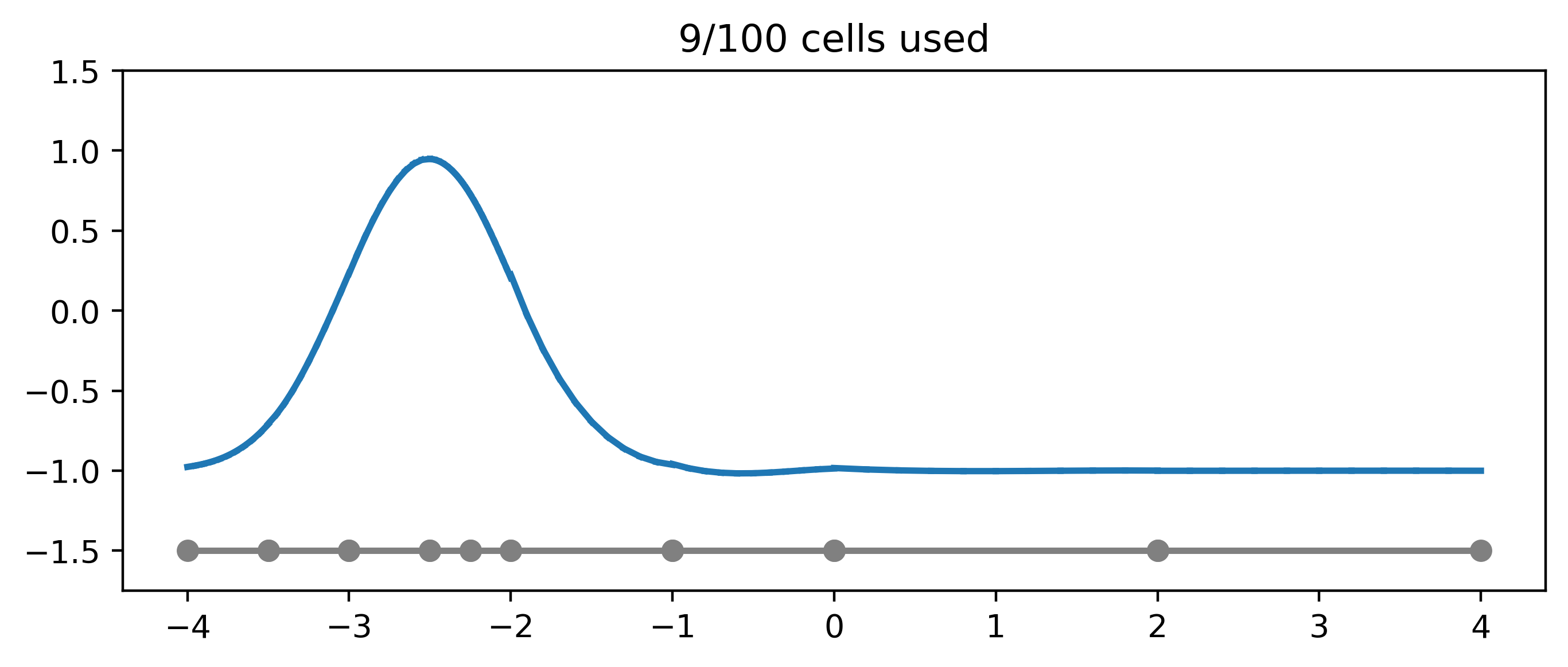}
    \end{subfigure}
    \begin{subfigure}{0.45\textwidth}
        \includegraphics[width=\linewidth]{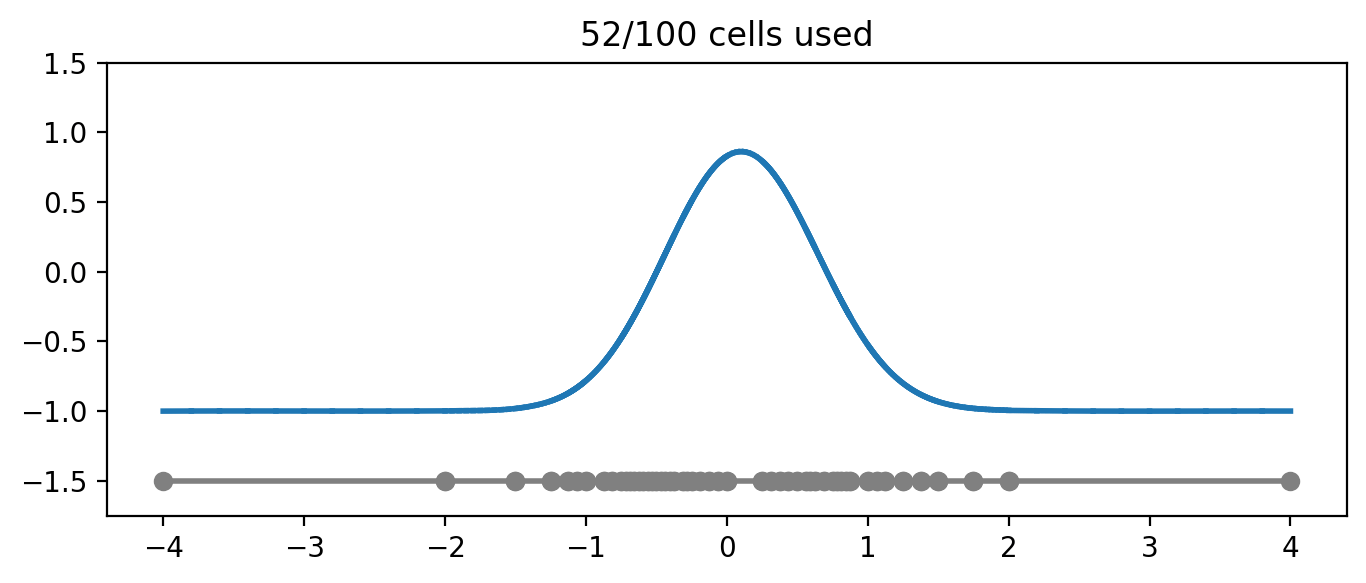}
    \end{subfigure}
    \begin{subfigure}{0.45\textwidth}
        \includegraphics[width=\linewidth]{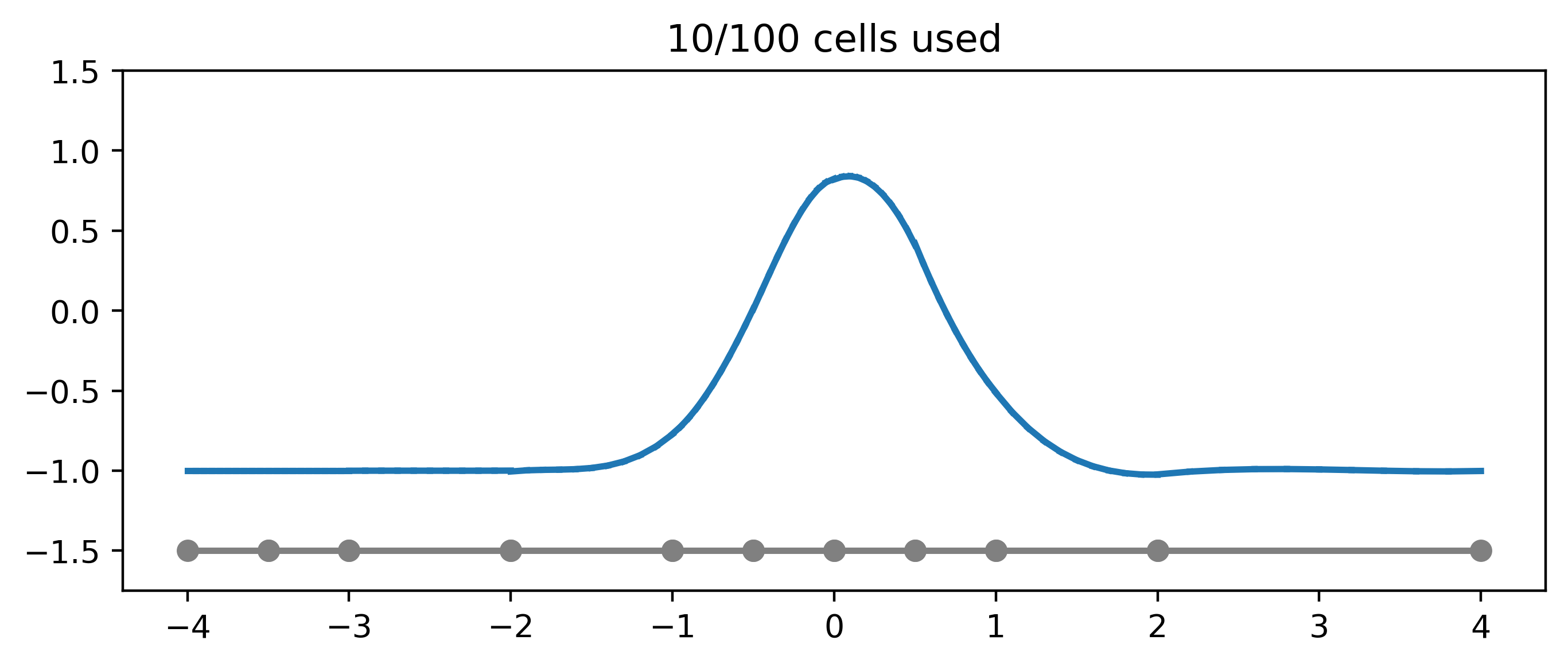}
    \end{subfigure}
    \begin{subfigure}{0.45\textwidth}
        \includegraphics[width=\linewidth]{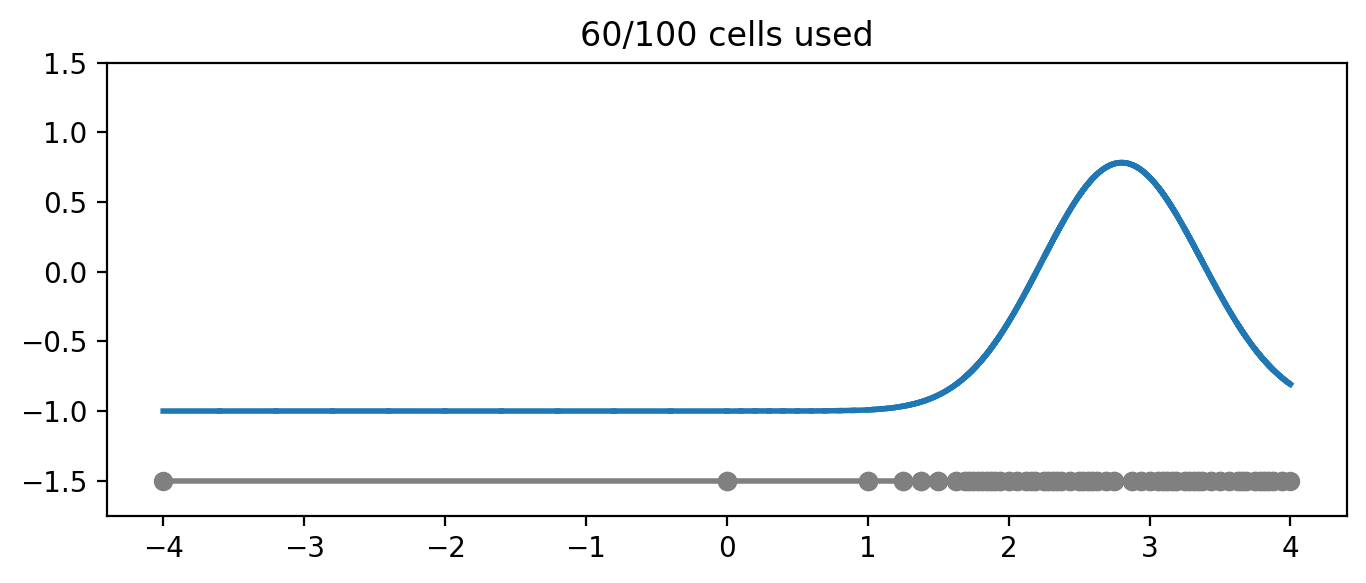}
        \caption{\small AMR heuristic}
    \end{subfigure}
    \begin{subfigure}{0.45\textwidth}
        \includegraphics[width=\linewidth]{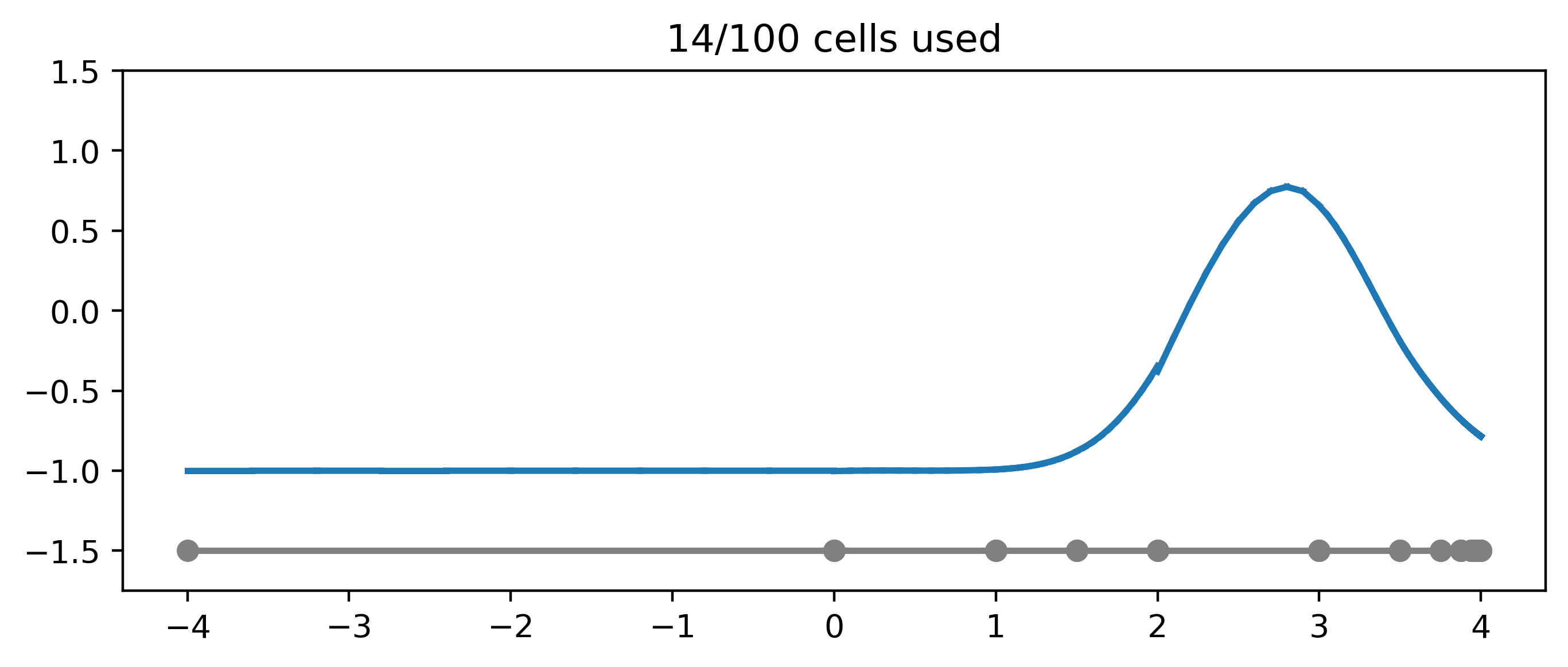}
        \caption{\small RL policy}
    \end{subfigure}
    \caption{\small Unsteady 1D linear advection. Numerical solution with time-dependent AMR policies using a gradient-based heuristic, and the trained RL policy. Results shown at times $T=1.6$ (top), $T=4.2$ (middle), and $T=6.9$ (bottom).}
    \label{fig:case2_results}
\end{figure}

We train the DRL-AMR model for $50\cdot 10^3$ RL time steps using the time-dependent solver and training procedure as detailed in \S\ref{sec:training} with a computational budget of 25 elements and using a scaling factor value $\gamma_c=100$ to render a relatively coarse numerical solution (\ref{eq:reward_fn_general}), as the computational cost is more heavily penalized.
All other training parameters assume the default values (\S\ref{sec:implementation_details}).
For the AMR heuristic, we employ a gradient indicator (Table \ref{tab:AMR_error_indicators}), that uses an approximation to the gradient of the numerical solution to estimate the error. Given the fast-decaying tails of the Gaussian pulse being advected, the gradient-based refinement indicator can be expected to accurately recommend coarsening outside of the pulse.

During deployment, we allow the AMR heuristic and the RL policy to perform one cycle of refinement/coarsening at every time step before advancing the numerical solution in time.
We consider an increased budget of 100 elements for both, although we don't explicitly enforce this requirement for the AMR heuristic; rather, we perform a bulk-refinement approach where we refine the cells responsible for the top 50 percent of the numerical error as measured by the gradient indicator, and coarsen the bottom 50 percent of the numerical error as measured by the same, \ie, a \bulk{0.5}{0.5} strategy.
Using a time step $\Delta t = 0.01$, we simulate to a final time of $T=7$. 

The results are shown in Figure \ref{fig:case2_results}.
Similar to the other experiments, we find that both algorithms are able to capture the advection of the initial condition; the RL policy does so using a computational mesh with far fewer elements than the AMR heuristic.

Namely, the DRL-AMR model is able to preserve the features of the solution while using fewer than 25 percent of the computational cost incurred by the AMR heuristic.
Due to the increased value of $\gamma_c$ as compared to the other numerical experiments considered thus far, the RL agent is more tolerant of a lack of smoothness in the solution. This is reflected in the small discontinuities of the numerical solution at the location $x=0$ at $t=1.6$ and at $x=2$ at $t=6.9$.

\subsection{Generalization to different PDEs}
\label{sec:1D_HDG}

To demonstrate that the DRL-AMR approach is not specific to the advection equation considered above, we apply it to the second-order Poisson equation, a subset of advection-diffusion PDEs (\S\ref{sec:advection-diffusion-eqn}).
\begin{equation}
  \begin{aligned}
    -\frac{\partial^2 u}{\partial x^2} = f \qquad \text{ in } \Omega = (-1,1), 
  \end{aligned}
  \label{eq:steady_1D_poisson_problem}
\end{equation}
with mixed boundary conditions consisting of a Dirichlet condition $u = g_D$ on boundary $\Gamma_D = \{-1\}$ and Neumann condition $\nabla u \cdot \bm{n} = g_N$ on boundary $\Gamma_N = \{1\}$.
The boundary conditions and forcing function are deduced from the exact solution, chosen to be a superposition of Gaussian functions
\begin{equation}
  u = \sum_{i=1}^{3} \frac{1}{\sigma \sqrt{2\pi}} 
  \exp\left(-\frac{(x - r_i)^2}{\sigma^2} \right)
  \label{eq:1d_poisson_exact_soln}
\end{equation}
with parameters $r_1 = -1/3$,\, $r_2 = 0$, \,$r_3 = 1/3$ and $\sigma = 1/10$.
We discretize the problem using HDG finite elements as described in \S\ref{sec:advection-diffusion-eqn}, with polynomial order $p_\text{order}=3$. We set the diffusivity coefficient $\kappa=1$ and the convective velocity $c = 0$.

\begin{figure}[ht]
    \centering
    \begin{subfigure}{0.55\textwidth}
        \includegraphics[width=\linewidth]{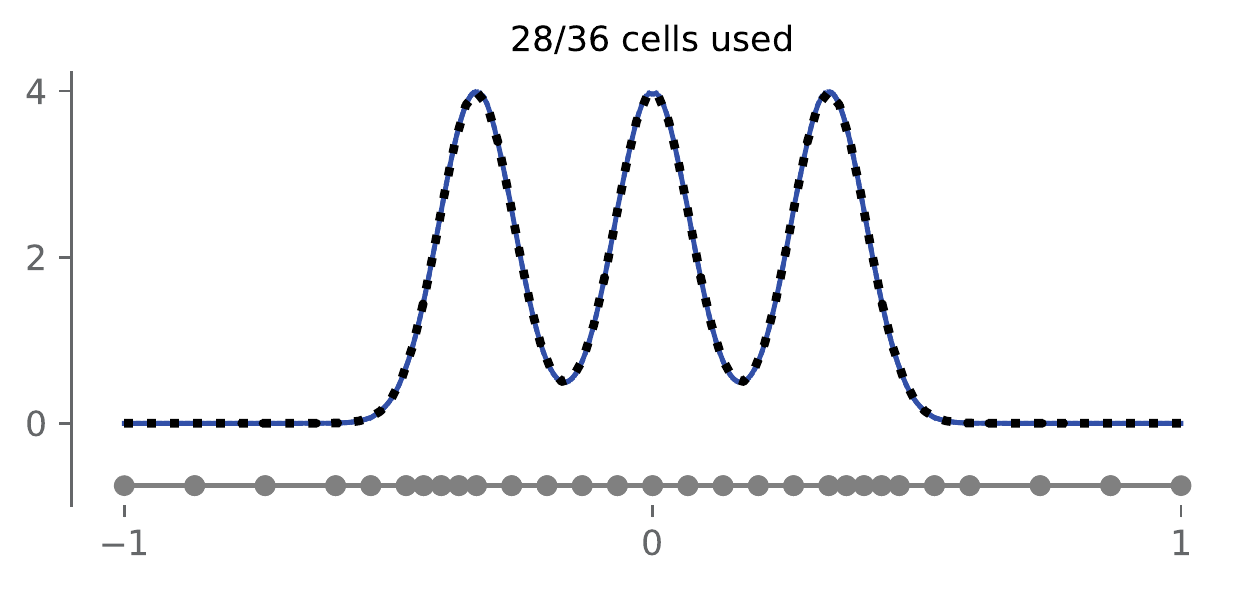}
    \end{subfigure}
    \begin{subfigure}{0.36\textwidth}
        \includegraphics[height=5.75cm]{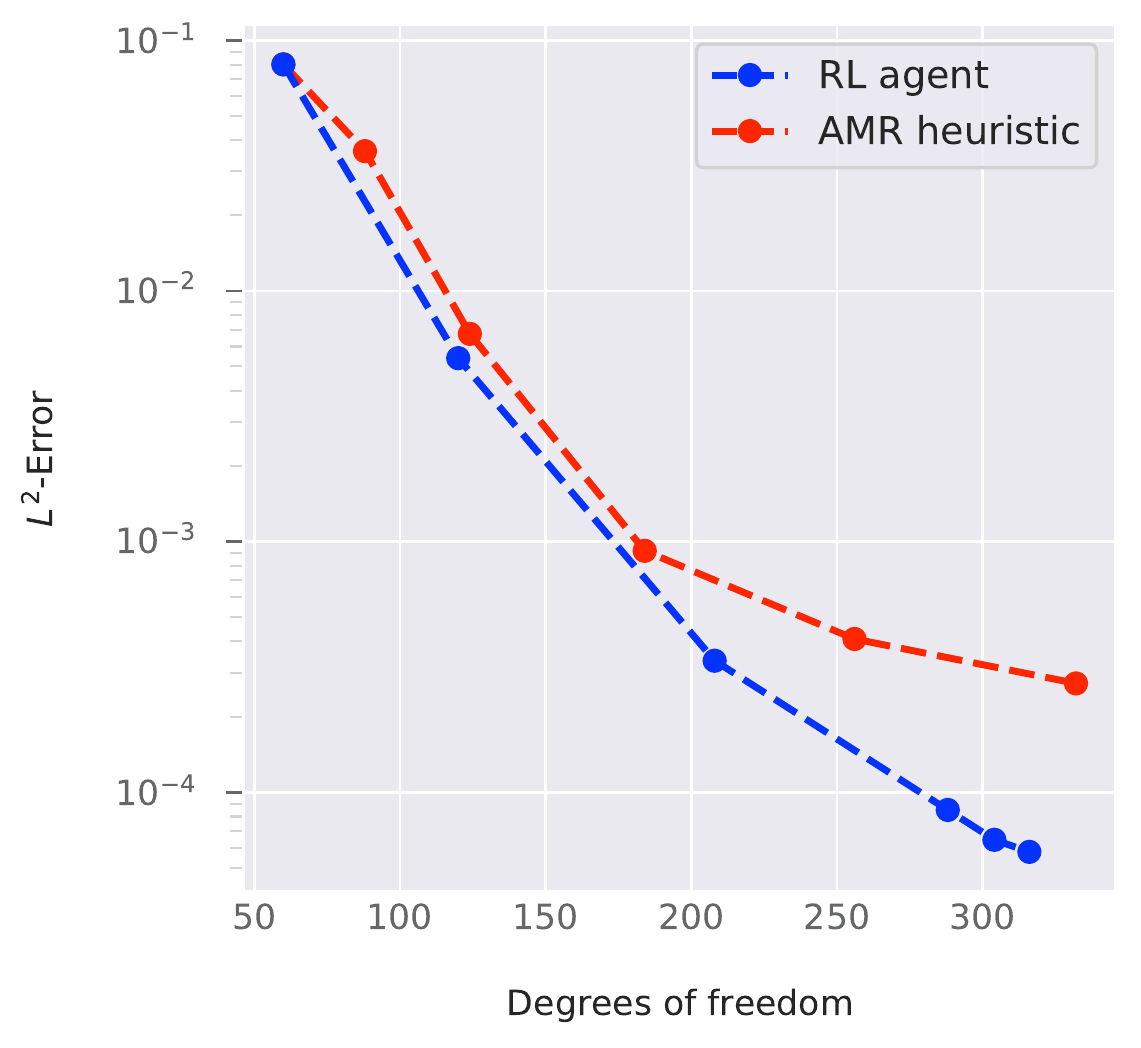}
        \label{fig:1D_HDG_L2_err}
    \end{subfigure}
    \caption{\small Steady 1D Poisson problem (\ref{eq:steady_1D_poisson_problem}). (Left) Mesh distribution and corresponding numerical solution resulting from 6 cycles of refinement using the trained RL policy with a budget of 36 cells; the exact solution (\ref{eq:1d_poisson_exact_soln}) (black, dashed line) is overlaid for comparison. (Right) $L^2$-Error of the numerical solution vs problem degrees of freedom over 6 mesh refinement cycles for the RL agent and the Kelly error estimator as AMR heuristic.}
    \label{fig:case3_results}
\end{figure}
Results of the trained RL policy are shown in Figure \ref{fig:case3_results}. 
Here we train for $2\cdot 10^5$ training time steps, and all non-specified training parameters assume the default values (\S\ref{sec:implementation_details}).
We see that the features of the solution are well resolved, and match the exact solution.

Furthermore, the trained policy identifies regions on the outer edges of the Gaussian mixture as in need of refinement; we hypothesize that this region is specifically sensitive to Gibbs phenomena caused by the change from a near-zero solution to a non-trivial one.
The more precise comparison between $L^2$-error and problem degrees of freedom in Figure \ref{fig:case3_results} shows that the DRL-AMR model outperforms the AMR heuristic.
As the Kelly error indicator was specifically designed for this type of Poisson problem and can be referred to as an error estimator in this context, the fact that the RL policy is competitive is an encouraging result.

Lastly, the PDE in (\ref{eq:steady_1D_poisson_problem}) is elliptic. 
Errors in the solution are therefore well-known to be globally propagated over the domain, in contrast to the advection equation, where errors are advected along with the solution \cite{lermusiaux_2.29_notes,strauss2007partial}.
The fact that the RL methodology was nonetheless able to satisfactorily solve the problem lends credence to the POMDP formulation and use of a local observation space during training.

\subsection{Steady 2D linear advection}%
\label{sec:steady-2D-advection}

To examine the generalizability and performance in the context of larger, higher-dimensional problems, we solve the steady version of the linear advection equation (\S\ref{sec:linear_advection_equation}) on the two-dimensional (2D) spatial domain $\Omega =(0,1)^2$ with a circular, counter-clockwise convective velocity field $\bm{c}(x) = 1/\left\lVert x \right\rVert_2\left(-x_2,\, x_1\right)$.
On the inflow boundary $\Gamma_{\text{in}}$, we specify the boundary condition $g_D$ according to the chosen solution,
\begin{equation}
  u(r) = 1- \tanh\left( \alpha(1 - 4(r - x_0 ))\right),
  \label{eq:exact_solution_steady_shock}
\end{equation}
where $r$ denotes the radius in cylindrical coordinates. We use the parameters $\alpha=10$, $x_0 = 1/4$ and forcing function $f=0$.
The exact solution is a two-dimensional, cylindrical coordinate generalization of the test case considered in \S\ref{sec:static_1D_LA}; that is, the solution contains a smooth, steep gradient.
However, unlike the test case in \S\ref{sec:static_1D_LA}, due to the higher problem dimensionality, the steep gradient is present both in the domain interior as well as at the domain boundary.
\begin{figure}[h]
    \centering
    \begin{subfigure}{0.4\textwidth}
        \includegraphics[width=\linewidth]{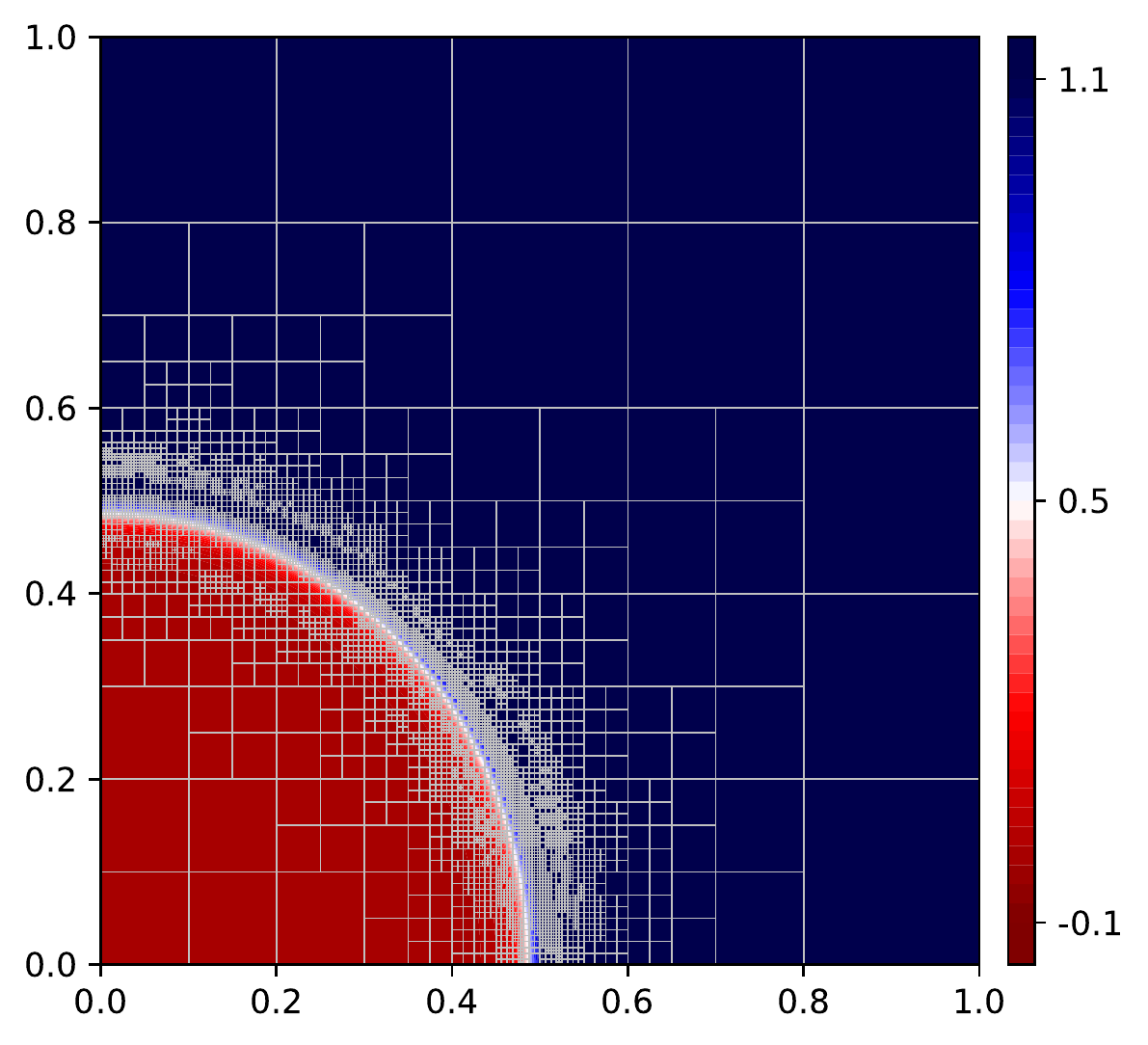}
        \caption{\small AMR gradient heuristic}
        \label{fig:shockMR:AMR}
    \end{subfigure}
    \begin{subfigure}{0.4\textwidth}
        \includegraphics[width=\linewidth]{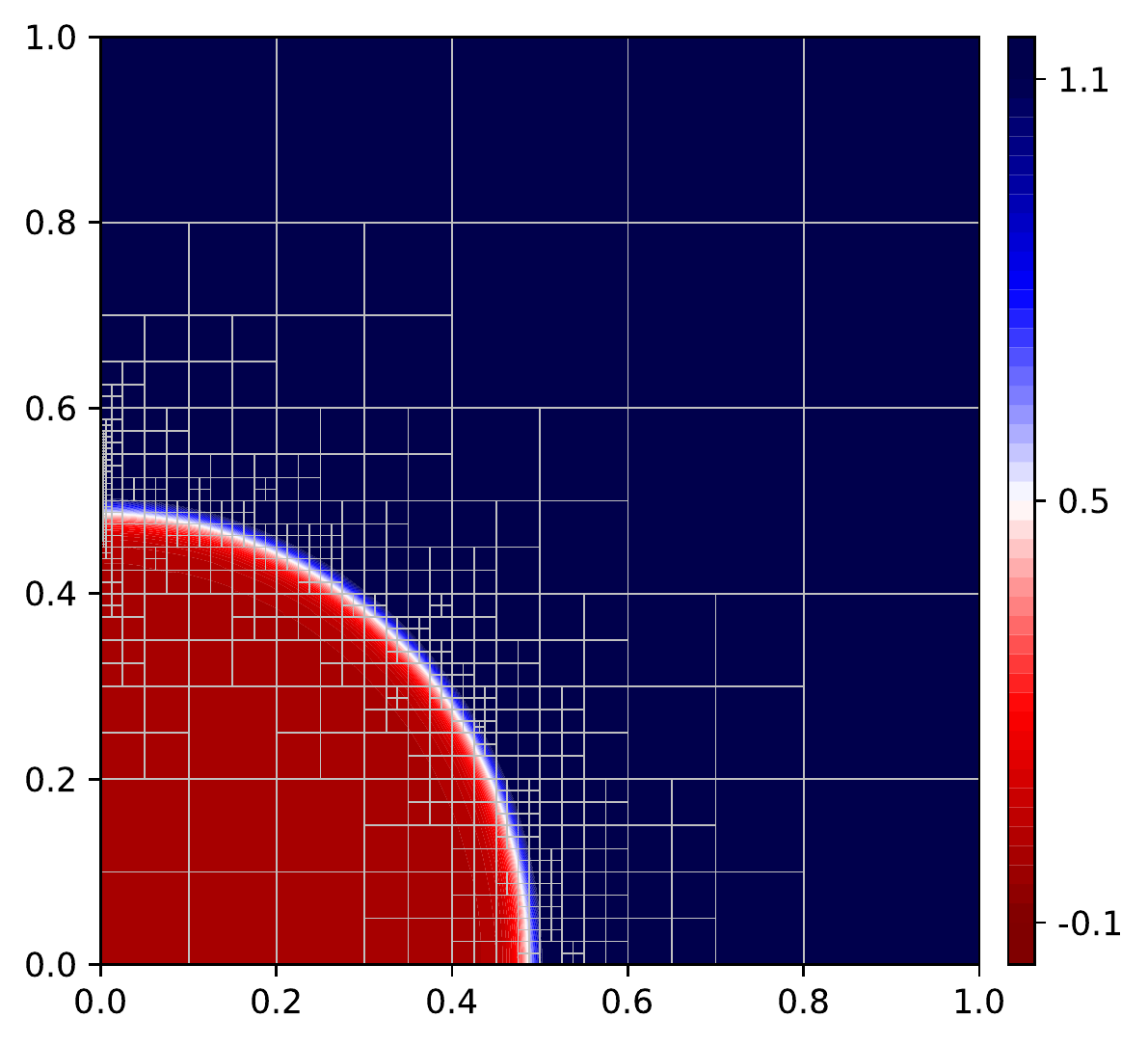}
        \caption{\small RL Agent}
        \label{fig:shockMR:RL}
    \end{subfigure}
    \begin{subfigure}{0.4\textwidth}
        \includegraphics[width=\linewidth]{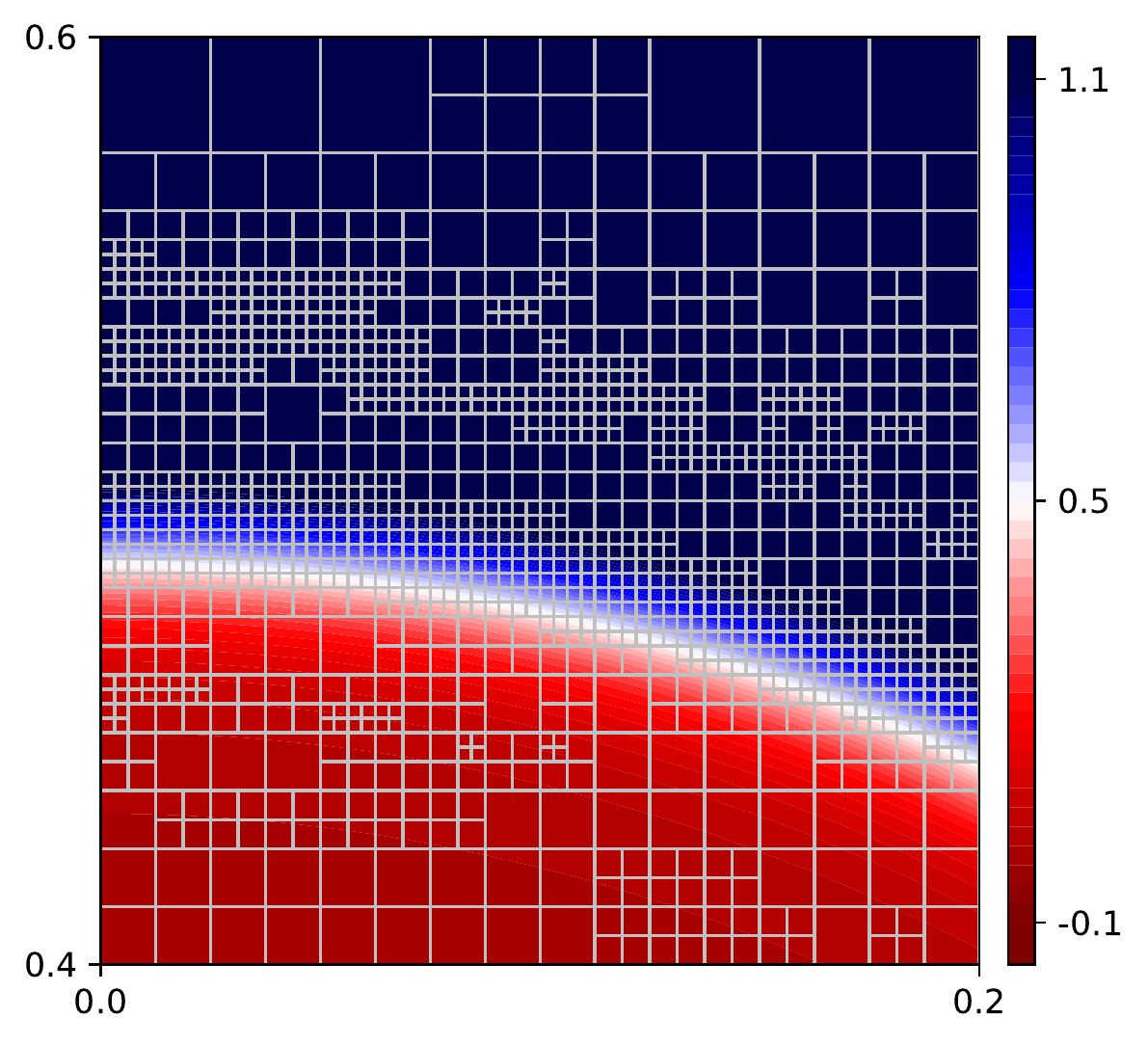}
        \caption{\small AMR gradient heuristic (zoom, outflow boundary)}
        \label{fig:shockMR:AMR:outflow_zom}
    \end{subfigure}
    \begin{subfigure}{0.4\textwidth}
        \includegraphics[width=\linewidth]{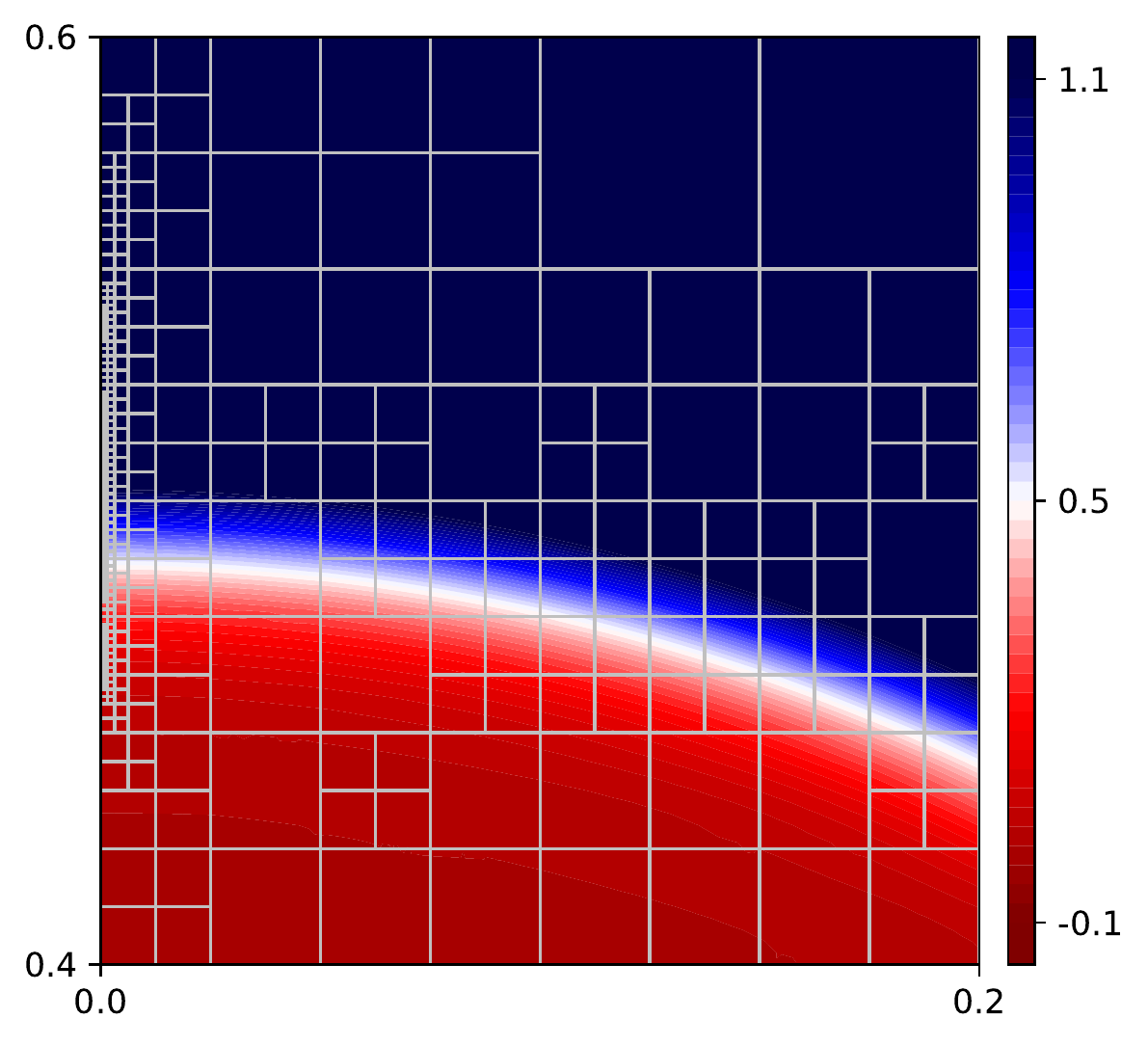}
        \caption{\small RL Agent (zoom, outflow boundary)}
        \label{fig:shockMR:RL:outflow_zoom}
    \end{subfigure}
    \caption{ \small Steady 2D linear advection. Numerical solution overlaid with the mesh refined using a gradient-based error indicator and the RL policy (1000 cell budget). (Top row) Entire problem domain. (Bottom row) Zoomed-in portion of the domain where the steep gradient meets the left outflow boundary.}
    \label{fig:shockMR}
\end{figure}
\begin{figure}[h]
    \centering
    \includegraphics[width=0.45\linewidth]{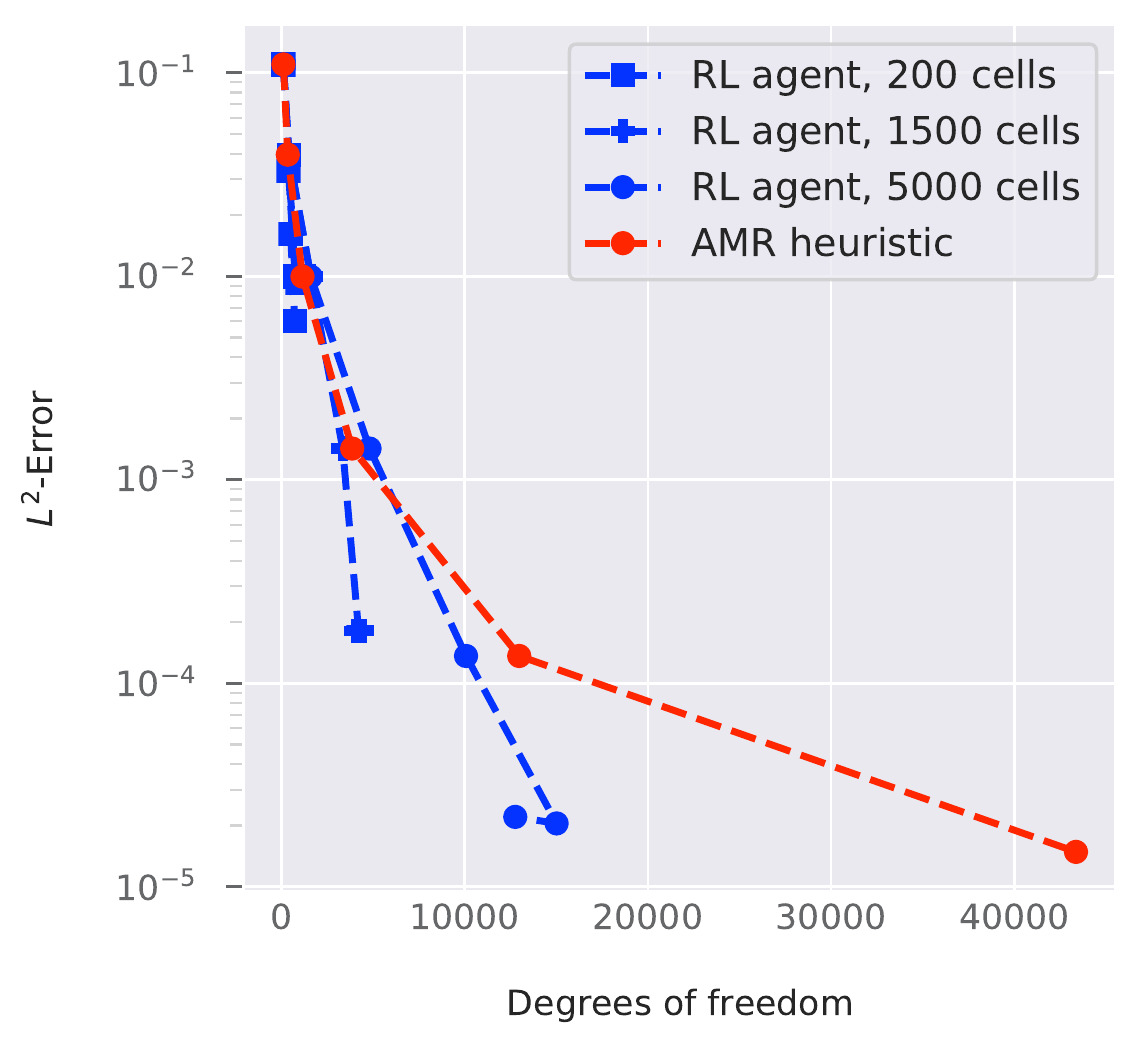}
    \caption{\small  Steady 2D linear advection. $L^2$ error as a function of the number of degrees of freedom for the refinement cycles resulting from a gradient-based error indicator and the RL policy.}
    \label{fig:shock-err-per-dof}
\end{figure}

To train the network, we solve the linear advection problem 
\begin{equation*}
\begin{aligned}
     \nabla \cdot (\bm{c} u) = 0, &\qquad \text{in } \Omega,  \\
     u = u(r), &\qquad \text{in } \Gamma_\text{in}
\end{aligned}
\end{equation*}
using the DG discretization given in \S\ref{sec:linear_advection_equation} with elements of polynomial order $\porder=2$, and deploy the model on the same problem. 
However, we train the RL policy network with a resource budget of only 110 cells, limiting the learning process to a relatively small number of elements compared to the deployment budgets; all other training parameters are the default values (\S\ref{sec:implementation_details}).
After $2\cdot 10^5$ training, time steps, we deploy the trained RL policy over a series of 6 refinement cycles, starting with a coarse mesh of 25 elements with 5 elements in each direction. 

We compare the performance of the RL agent to the gradient-based error indicator (Table \ref{tab:AMR_error_indicators}) that uses a \bulk{0.5}{0.5} refinement strategy.
Similar to the test case considered in \S\ref{sec:static_1D_LA}, this AMR heuristic can be expected to perform well, given the steep gradient as the dominant feature of the exact solution.

We compare the overall $L^2$ error of the resulting solution with the number of degrees of freedom resulting from the numerical discretization at each level of the refinement cycle in Figure \ref{fig:shock-err-per-dof},
showing results for the RL policy deployed at different cell budgets. 
We consider deployment budgets of 200, 1500, and 5000 elements.
By the final refinement cycle, the RL policy finds a solution of comparable or better accuracy than the AMR heuristic for the same number of degrees of freedom. In particular, with a budget of 5000 cells, the RL agent substantially outperforms the AMR heuristic. 

To shed light on why the RL policy is able to outperform the heuristic, numerical solutions with the overlaid meshes proposed by the two approaches are shown in Figure \ref{fig:shockMR}. 
In this figure, the RL policy has been deployed with a budget of 1000 cells; it makes use of roughly 2/3 of its budget allocation by the final refinement cycle.
Qualitatively, both approaches are able to resolve the steep gradient; however, compared to the AMR heuristic, the RL agent has much more conservatively refined the mesh around the slope.
Instead, the RL agent refines preferentially in the region where the steep gradient encounters the outflow boundary (\cf Figure \ref{fig:shockMR:AMR:outflow_zom} and Figure \ref{fig:shockMR:RL:outflow_zoom}).
That is, in training, the RL agent learned that the numerical solution is sensitive to the discontinuities arising where the gradient meets the outflow boundary. 
Although the agent has no information on where cells are located, the cells in this problem region exhibited behaviors in their observation space that indicated the numerical solution would improve by refining them.
That is to say, the RL policy learned a non-trivial, spatially-heterogeneous strategy for regional refinement.
In contrast, the gradient-based indicator detected the steep discontinuity but was not able to exploit the sensitivity of the error to the resolution close to the outflow boundary.

\subsection{Steady 2D advection-diffusion equation}%
\label{sec:steady-2D-advection-diffusion}

Section \S\ref{sec:steady-2D-advection} involves a relatively simple numerical solution, in the sense that the only interesting solution feature is the steep gradient region. 
It is natural to ask whether the methodology can capture solutions with more complicated features.
In this section, we turn our attention to the advection-diffusion problem (\ref{eq:advection-diffusion-PDE}) with mixed boundary conditions, and the less straightforward hybridizable finite element discretization (\ref{eq:HDG-advection-diffusion}) of the same. 

We choose a circular velocity field $\bm{c} = (x_2, -x_1)$, and take the top and bottom boundaries of the domain $\Omega = (-1,1)^2$ as a Dirichlet boundary $\Gamma_D $, whereas the left and right boundaries of $\Omega$ constitute a Neumann boundary $\Gamma_N$.
The boundary values and forcing function are deduced from the exact solution, a superposition of Gaussian functions chosen to create non-trivial features.
The details of the exact solution are given in Appendix B.
The problem is discretized at polynomial order $p_{\text{order}} = 4$ and makes use of element-local post processing, resulting in an observed optimal convergence order of $p_{\text{order}} + 2$ upon uniform mesh refinement (see \cite{nguyen2009implicit}).
\begin{figure}[h]
    \centering
    \begin{subfigure}{0.4\textwidth}
        \includegraphics[width=\linewidth]{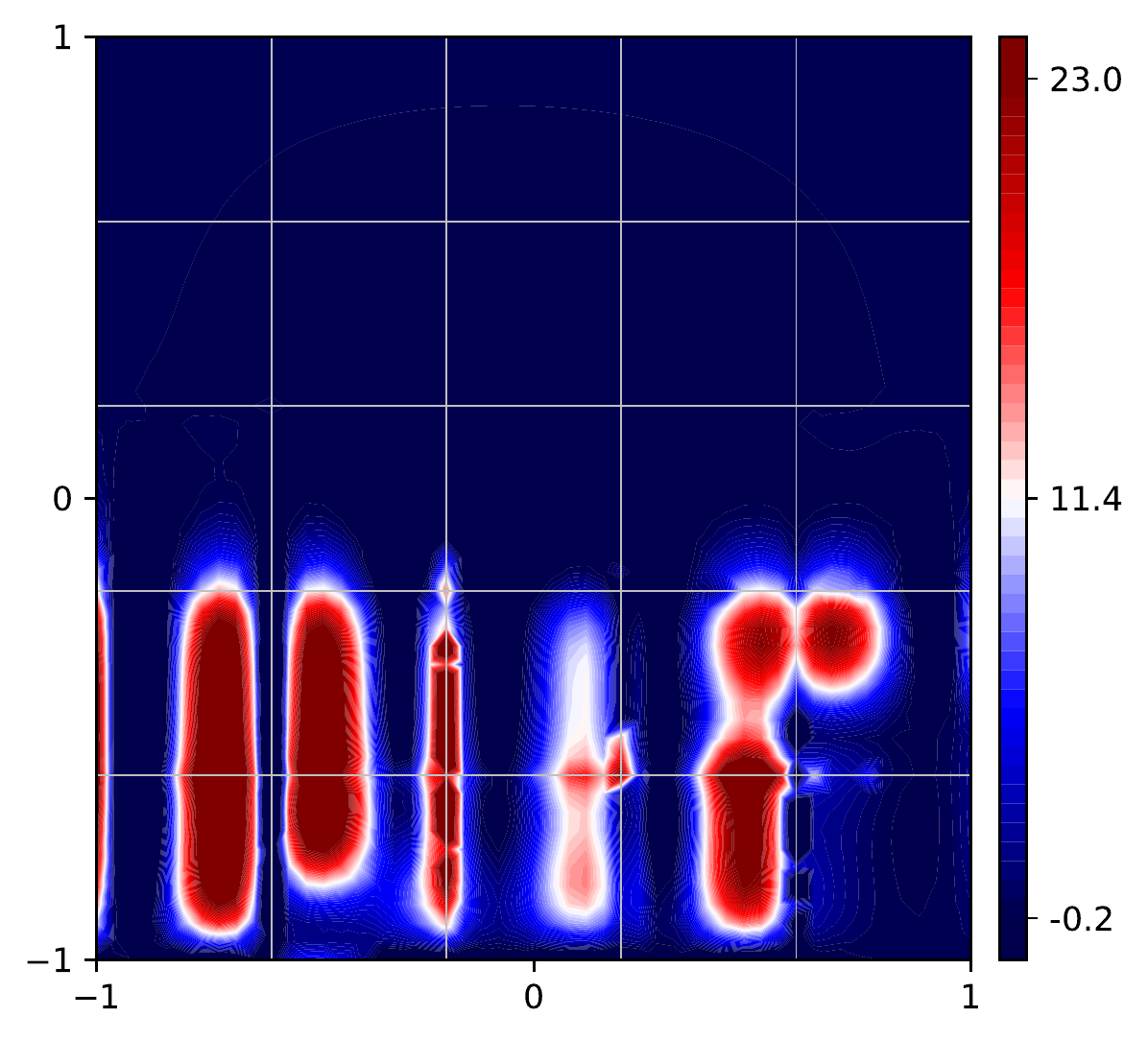}
        \caption{\small Numerical solution, coarse mesh}
        \label{fig:HDG_MIT:base}
    \end{subfigure}
    \begin{subfigure}{0.4\textwidth}
        \includegraphics[width=\linewidth]{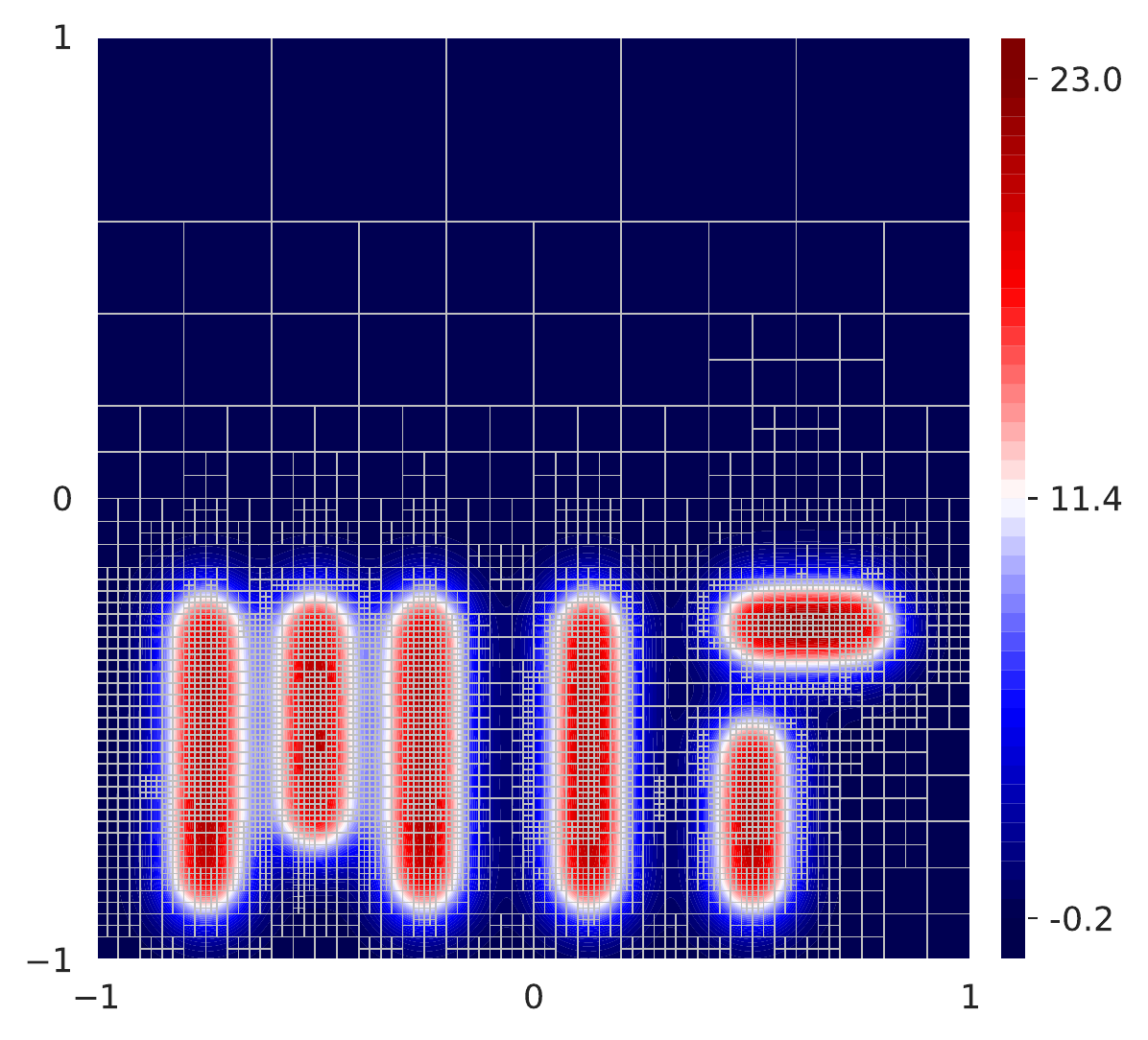}
        \caption{\small AMR gradient heuristic}
        \label{fig:HDG_MIT:AMR}
    \end{subfigure}
    \begin{subfigure}{0.4\textwidth}
        \includegraphics[width=\linewidth]{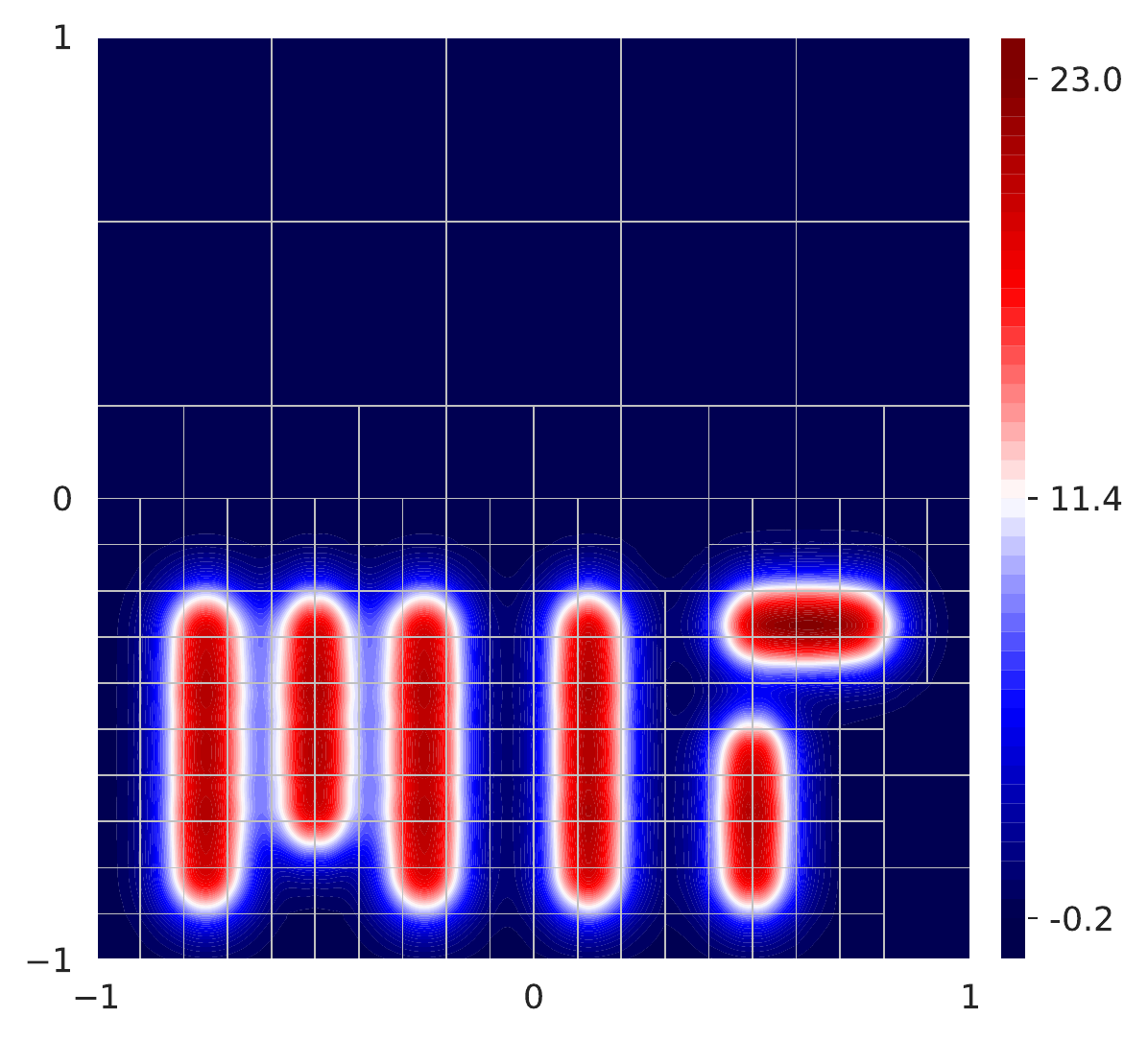}
        \caption{\small RL Agent (1500 cell budget)}
        \label{fig:HDG_MIT:RL_1500}
    \end{subfigure}
    \begin{subfigure}{0.4\textwidth}
        \includegraphics[width=\linewidth]{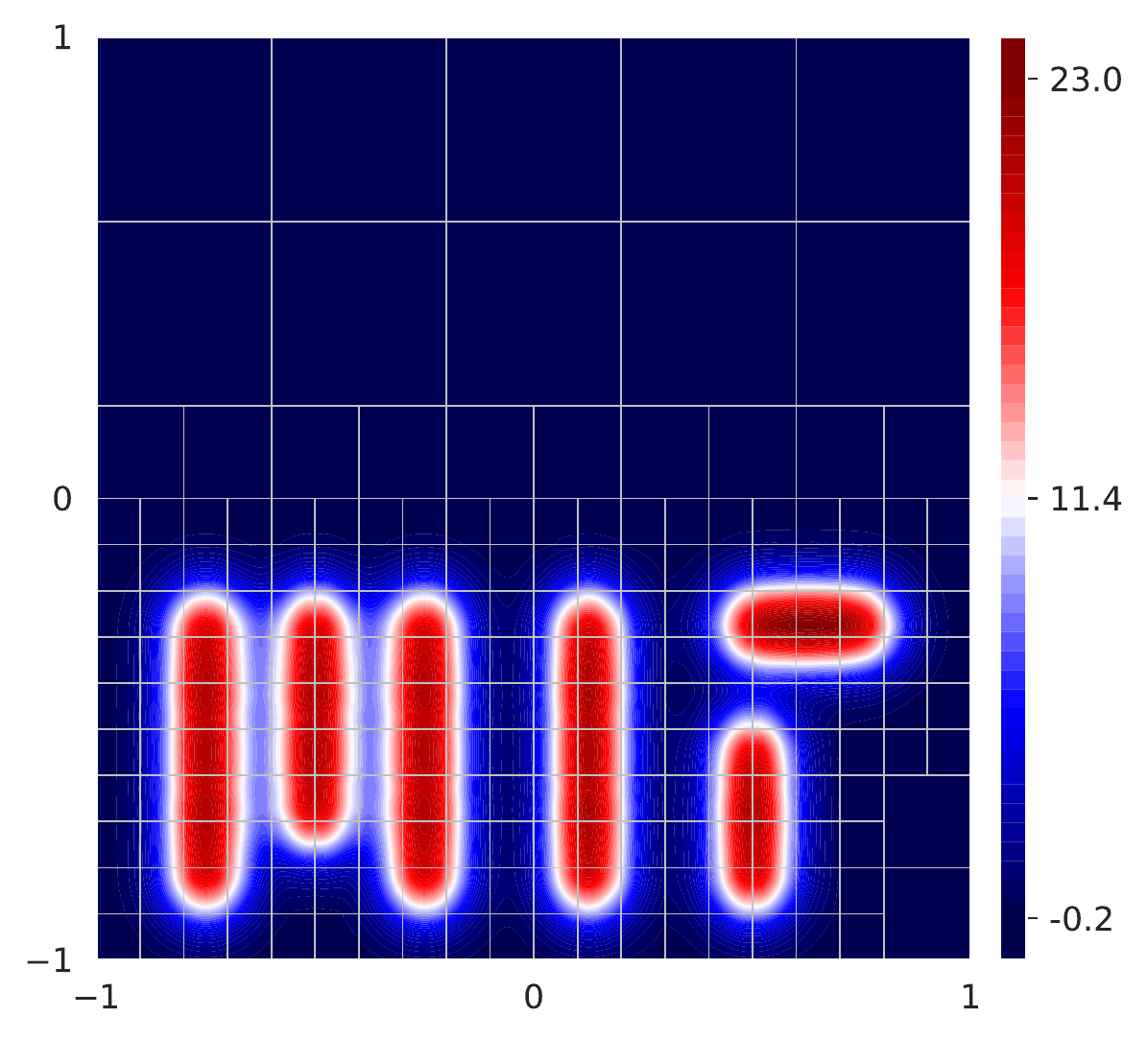}
        \caption{\small RL Agent (5000 cell budget)}
        \label{fig:HDG_MIT:RL_5000}
    \end{subfigure}
    \caption{\small Steady 2D advection diffusion. Numerical solution with overlaid mesh for the this problem in \S\ref{sec:steady-2D-advection-diffusion}.}
    \label{fig:HDG_MIT}
\end{figure}

We train the RL policy for $3\cdot 10^5$ training time steps, using a budget of 200 cells; all other training parameters are the default values (\S\ref{sec:implementation_details}).
As in the other numerical experiments, the training budget was chosen to be significantly cheaper than the deployment budget.
For comparison, we consider an AMR heuristic that makes use of the Kelly error indicator with a \bulk{50}{50} refinement strategy, refining and coarsening cells responsible for the top 50 percent and bottom 50 percent of the total estimated error, respectively.
Both strategies are employed over 5 refinement cycles, beginning from the coarse mesh shown in Fig.\ \ref{fig:HDG_MIT:base}.

The results of the comparison between the two algorithms are given in Figure \ref{fig:HDG_MIT}.
Examination of the $L^2$-error per degree of freedom (not shown) again demonstrates that the RL agent is competitive with any of the AMR heuristics considered in this paper.
However, in this experiment, we focus on the relative ability of the heuristic and RL AMR strategies to resolve relevant features of the solution, pursuant to the discussion in \S\ref{sec:objectives_and_novel_contributions}.
We see from Figure \ref{fig:HDG_MIT:base} that on the coarse starting mesh, the numerical solution is acutely under-resolved.
Both the AMR heuristic (Figure \ref{fig:HDG_MIT:AMR}) and the trained RL agent (Figures \ref{fig:HDG_MIT:RL_1500}, \ref{fig:HDG_MIT:RL_5000}) are able to satisfactorily resolve the features of the numerical solution, but the RL agent does so with a more parsimonious allocation of elements over the mesh.
Even when the computational budget is increased from 1500 cells to 5000 cells, the RL agent makes only minor adjustments to the mesh.

The explanation for this is that the trained DRL-AMR model makes decisions based on the learned relationship between the features of the observation space local to each element, and the change in the numerical solution upon refinement.
The HDG finite element schemes (\ref{eq:HDG-advection-diffusion}) enjoy a faster convergence rate and, typically, better overall accuracy than the advection DG scheme (\ref{eq:DG-advection-discretization}) due to its dual formulation and element-wise post-processing, as discussed above.
Additionally, we use a relatively high-order (sixth-order effective convergence rate) finite element scheme as compared to the smooth step example in \S\ref{sec:steady-2D-advection}, 
which exhibited second-order convergence.
The RL policy is able to detect the fast convergence to the exact solution (according to the local numerical solution as well as the jump discontinuities over the boundary of each element) and elects to use only a small portion of its computational budget in order to resolve the solution, approximately 8 percent and 2 percent for budgets of 1500 cells and 5000 cells, respectively.
We hypothesize that the inclusion of additional HDG-specific features into the observation space, such as the numerical gradient $\bm{q}_h$ or the approximate numerical flux on the mesh skeleton $\hat{u}_h$ (\ref{eq:HDG-advection-diffusion}) could additionally improve performance.

In summary, the RL policy is able to judiciously allocate resources in order to capture non-trivial features of a numerical solution. 
Furthermore, the generality of the RL methodology extends to more complicated finite element schemes and is able to take advantage of their properties---in this case, high-order convergence due to a post-processed solution.

\subsection{Unsteady 2D linear advection}%
\label{sec:ring_test_case}

Inspired by the novel advection-diffusion test cases in \cite{kulkarni_lermusiaux_JCP2019}, we consider the two-dimensional, unsteady linear advection problem described in \S \ref{sec:linear_advection_equation} 
with the velocity field given by a reversible swirl flow 
\begin{equation}
  \bm{c}(x,y,t) = \left(
    \frac{3}{2} a(t) \sin^2(\pi x) \sin(2\pi y),\,
    -\frac{3}{2} a(t) \sin^2(\pi y) \sin(2\pi x)
  \right)
  \label{eq:reversible_swirl_flow}
\end{equation}
over the square domain $\Omega=(0,1)^2$. 
Here $0 \leq t \leq 1$ and $a(t) = 1$ if $t < 0.5$ and $a(t) = -1$ if $t \geq 0.5$, which reverses the direction of the flow field at time $t=0.5$.
At the time $T=1$, the analytical initial tracer distribution returns to its initial location due to the flow symmetry and can be compared to the initial condition to measure the error in the numerical solution.
The boundary conditions and forcing function are taken to be zero over the duration of the simulation.

\begin{figure}[ht]
    \centering
    \begin{subfigure}{0.24\textwidth}
        \includegraphics[width=\linewidth]{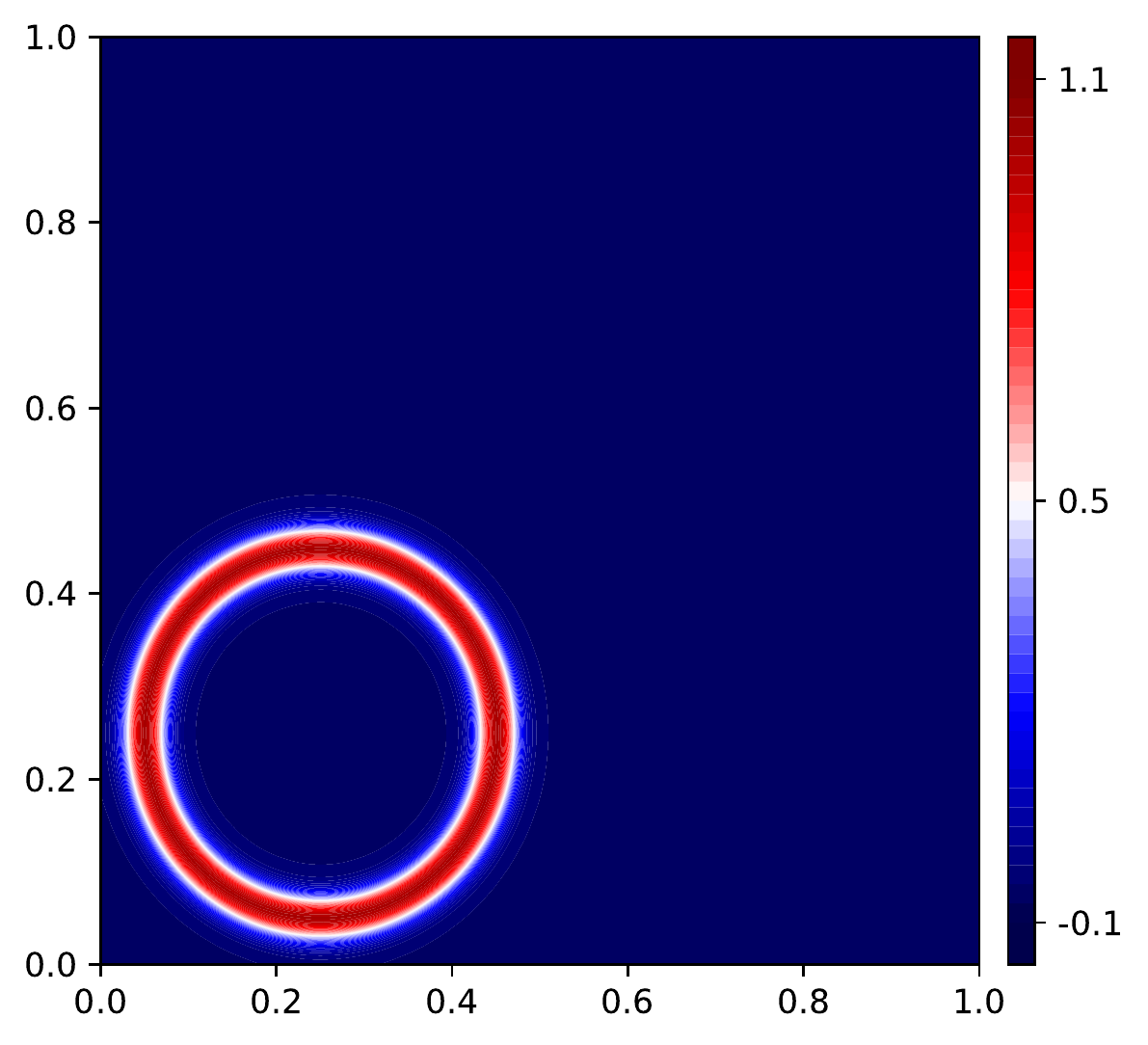}
        \caption{\small $t=0$}
        \label{fig:ring:IC}
    \end{subfigure}
    \begin{subfigure}{0.24\textwidth}
        \includegraphics[width=\linewidth]{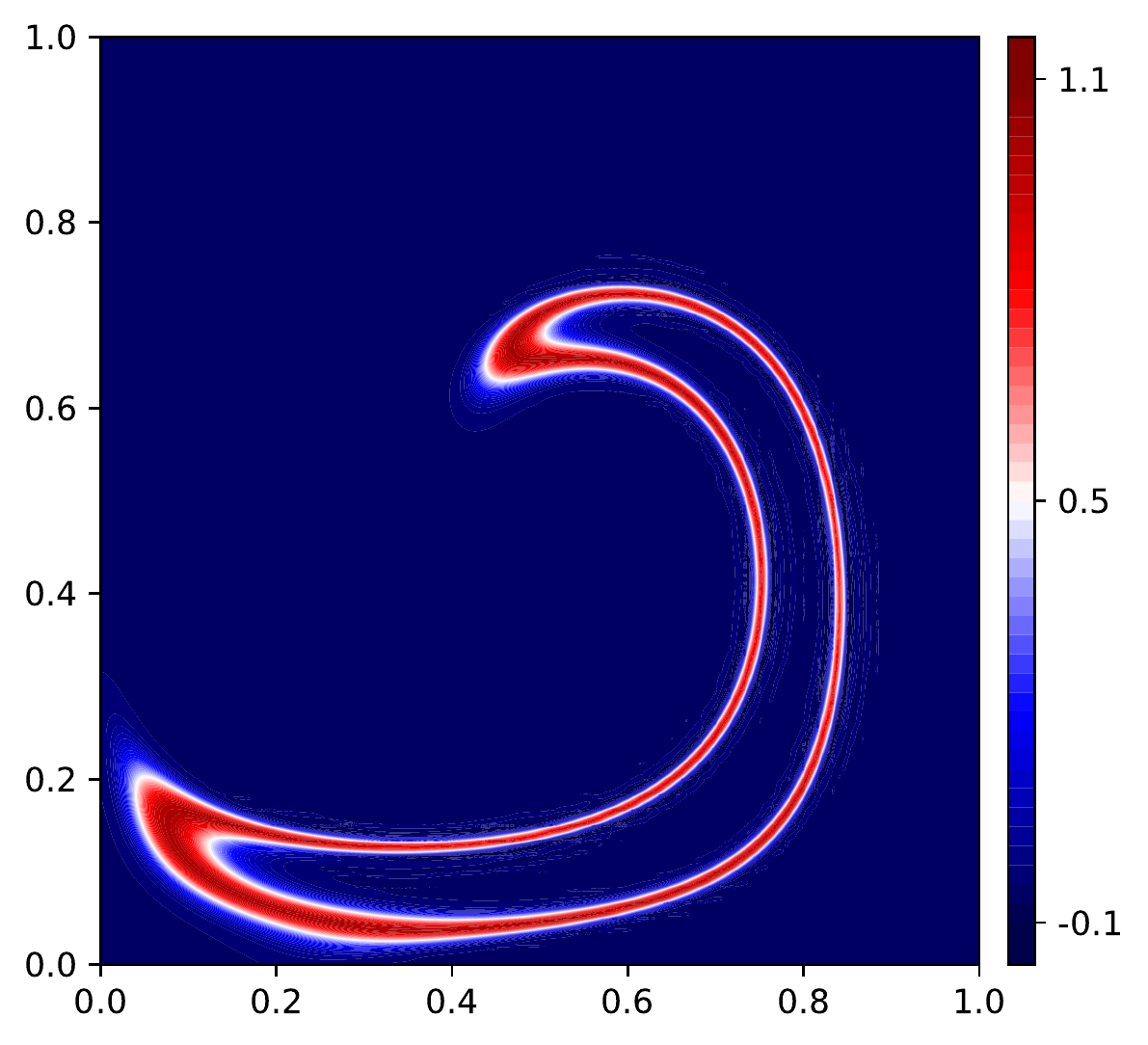}
        \caption{\small $t=0.5$}
        \label{fig:ring:half_T}
    \end{subfigure}
    \begin{subfigure}{0.24\textwidth}
        \includegraphics[width=\linewidth]{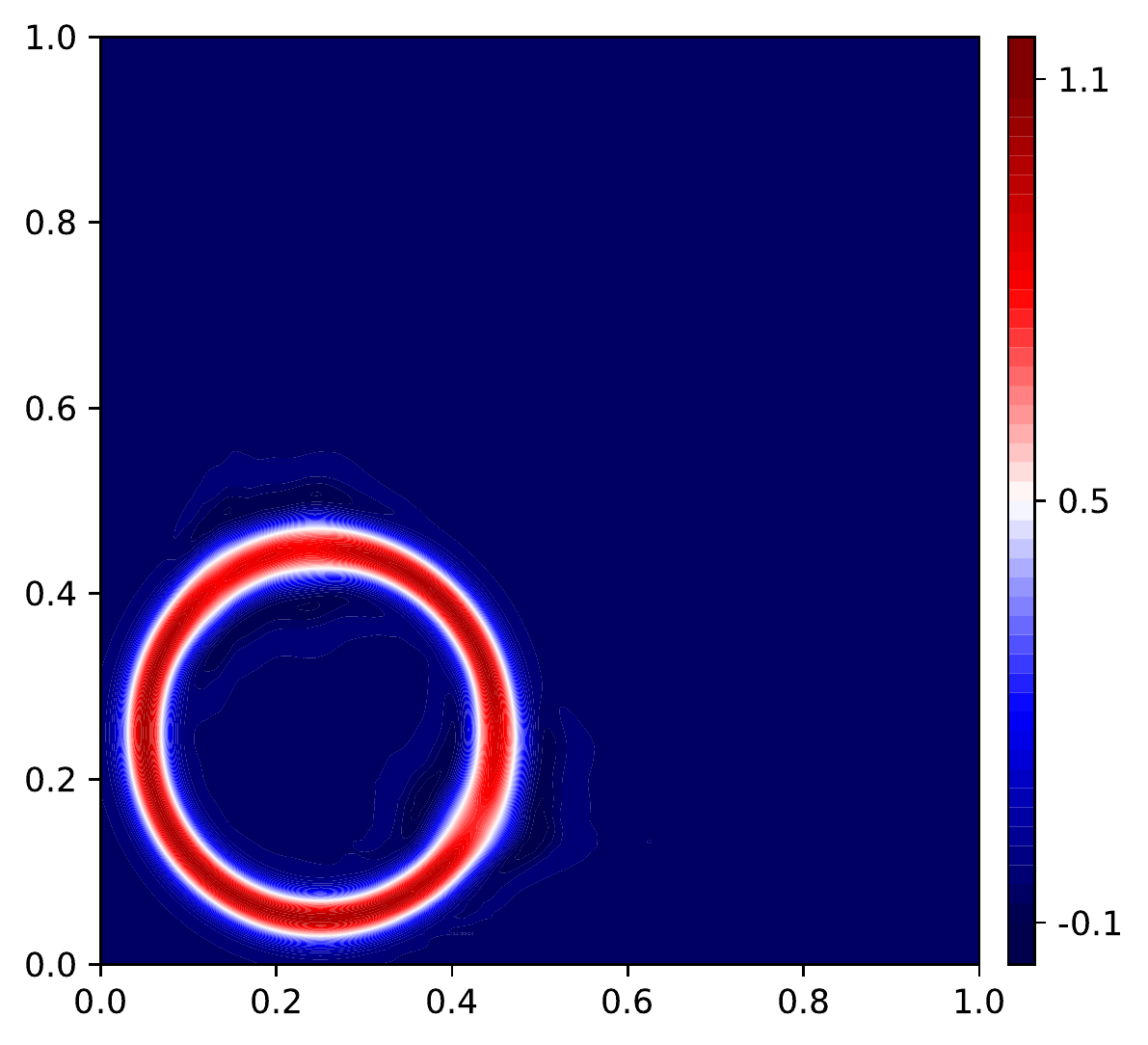}
        \caption{\small $t=1$}
        \label{fig:ring:T}
    \end{subfigure}
    \begin{subfigure}{0.24\textwidth}
        \includegraphics[width=\linewidth]{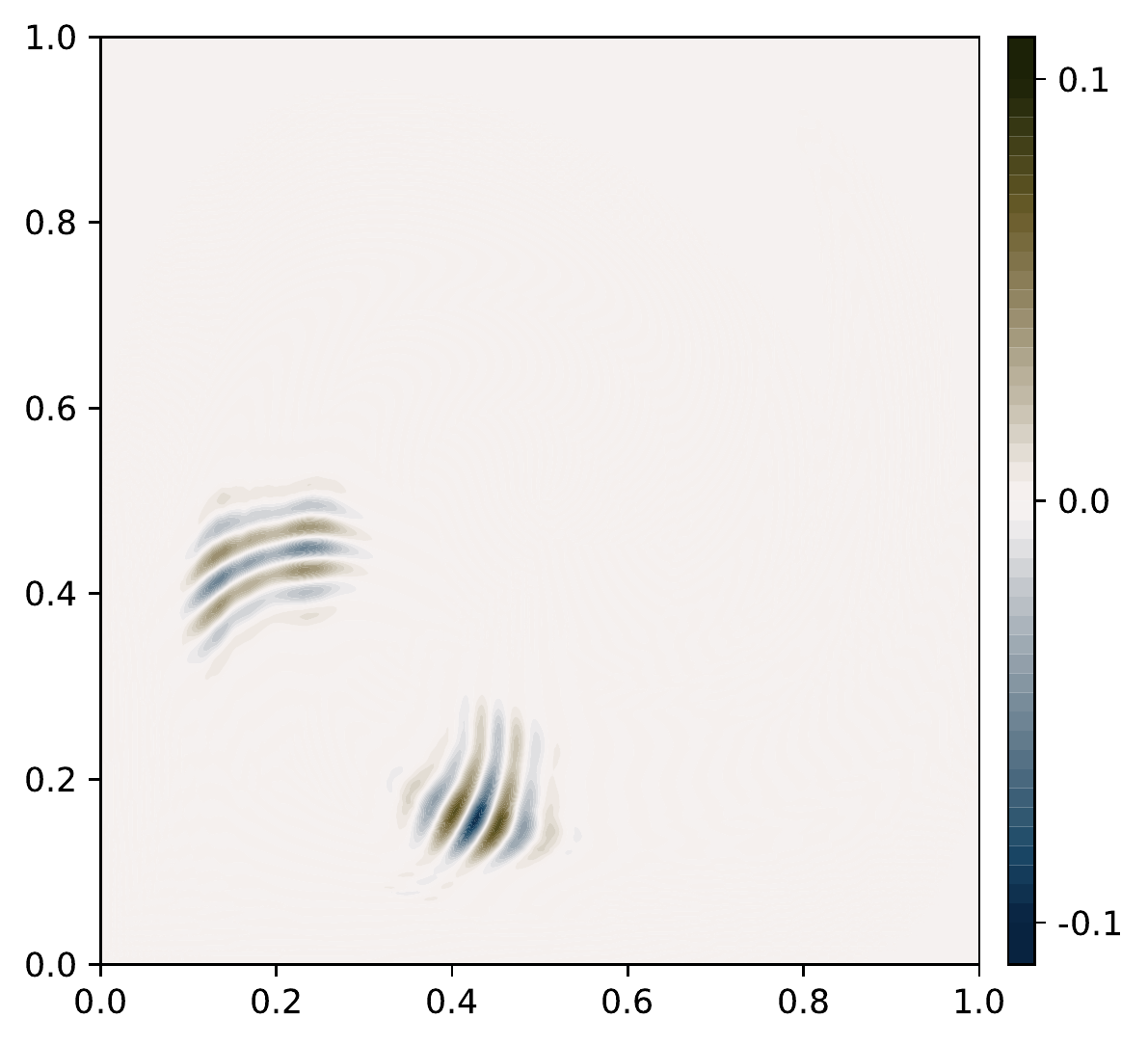}
        \caption{\small $u_h - u_{\text{exact}}$, $t=1$}
        \label{fig:ring:uniform_error}
    \end{subfigure}
    \caption{\small Unsteady 2D Ring advection. Numerical solution on a $64^2$ uniform grid.}
    \label{fig:ring_numerical_solution_uniform}
\end{figure}

The initial condition is given by
\begin{equation}
  u_0 = \exp\left(- \frac{1}{2\sigma^2} \left(\sqrt{(x-x_0)^2 + (y-y_0)^2} - r_0 \right)^2 \right),
\end{equation}
specifying a ring with inner radius $r_0$ and approximate thickness $3\sigma$ centered at the point $(x_0,\, y_0)$.
For this test case, we choose the parameters $r_0 =0.2$, $(x_0,\, y_0) = (0.25, 0.25)$, and $3\sigma = 0.05$, describing a thin ring centered in the bottom right-hand corner of the domain, depicted in Figure \ref{fig:ring:IC}.
The presence of steep gradients in the solution along the inner and outer circumferences of the ring during both the forward and backward swirl advection process makes this problem challenging, as well as a suitable test case to assess the ability of a numerical scheme to avoid spurious diffusion and dispersion, and to preserve features of the analytical solution \cite{kulkarni_lermusiaux_JCP2019}. 
All numerical simulations are run at polynomial order $p_{\text{order}} = 3$ and make use of a small time step $\Delta t = 10^{-3}$ over the LSERK45 time integration so that the overall error is determined by the error due to the spatial discretization.

We train the reinforcement learning agent for $3\cdot10^5$ training time steps on a budget of 200 cells; all other training parameters are the default values (\S\ref{sec:implementation_details}).
During deployment, we compare the trained DRL-AMR model to an AMR heuristic which uses the Kelly error indicator with a conservative \bulk{60}{40} refinement strategy. 
Both strategies start on a coarse mesh of 4x4 elements and are allowed 6 refinement cycles during initialization, after which, they are allowed one refinement cycle per time step for the duration of the simulation.
In order to prevent the number of degrees of freedom from growing unbounded over time integration for the AMR heuristic, we specify a maximum refinement depth of 4, corresponding to the finest possible grid of 64x64 cells.
To provide a fair comparison, the RL policy is deployed with an equal maximum cell budget of $64^2=4096$ elements.

The numerical solution which results from running the simulation on a uniform grid of 64x64 elements, corresponding to the resolution limit for the AMR heuristic, is shown for comparison in Figure \ref{fig:ring_numerical_solution_uniform}.
The uniform grid simulation is qualitatively able to well-preserve the features of the ring over the forward and reverse advection, %
at the final integration time $t=1$ (Figure \ref{fig:ring:T}). 
This result serves as a baseline comparison for the AMR heuristic and the RL policy.

\begin{figure}[h!]
    \centering
    \begin{subfigure}{0.32\textwidth}
        \includegraphics[width=\linewidth]{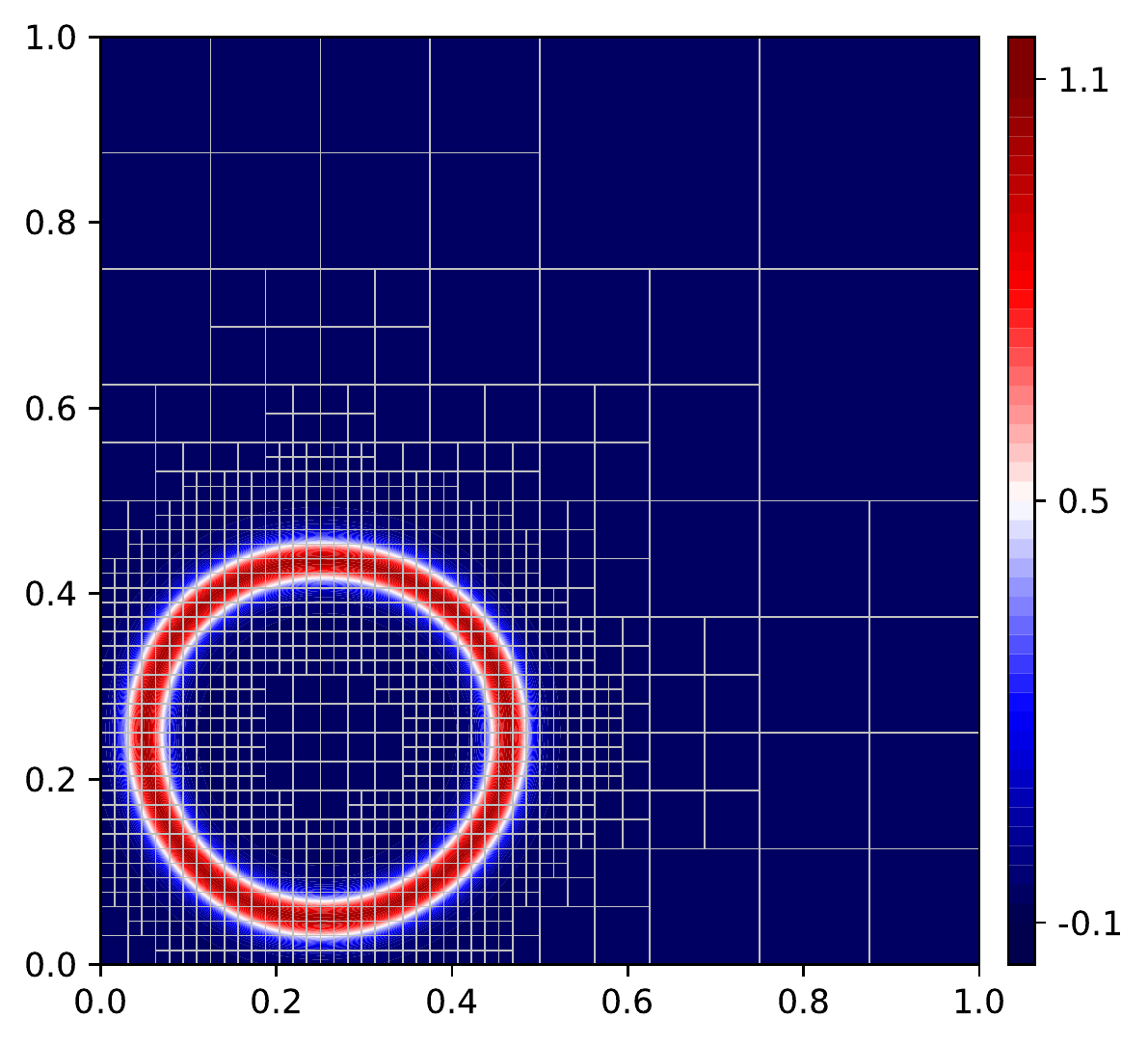}
        \caption{\small $t=0.1$}
        \label{fig:ring:AMR_t0.1}
    \end{subfigure}
    \begin{subfigure}{0.32\textwidth}
        \includegraphics[width=\linewidth]{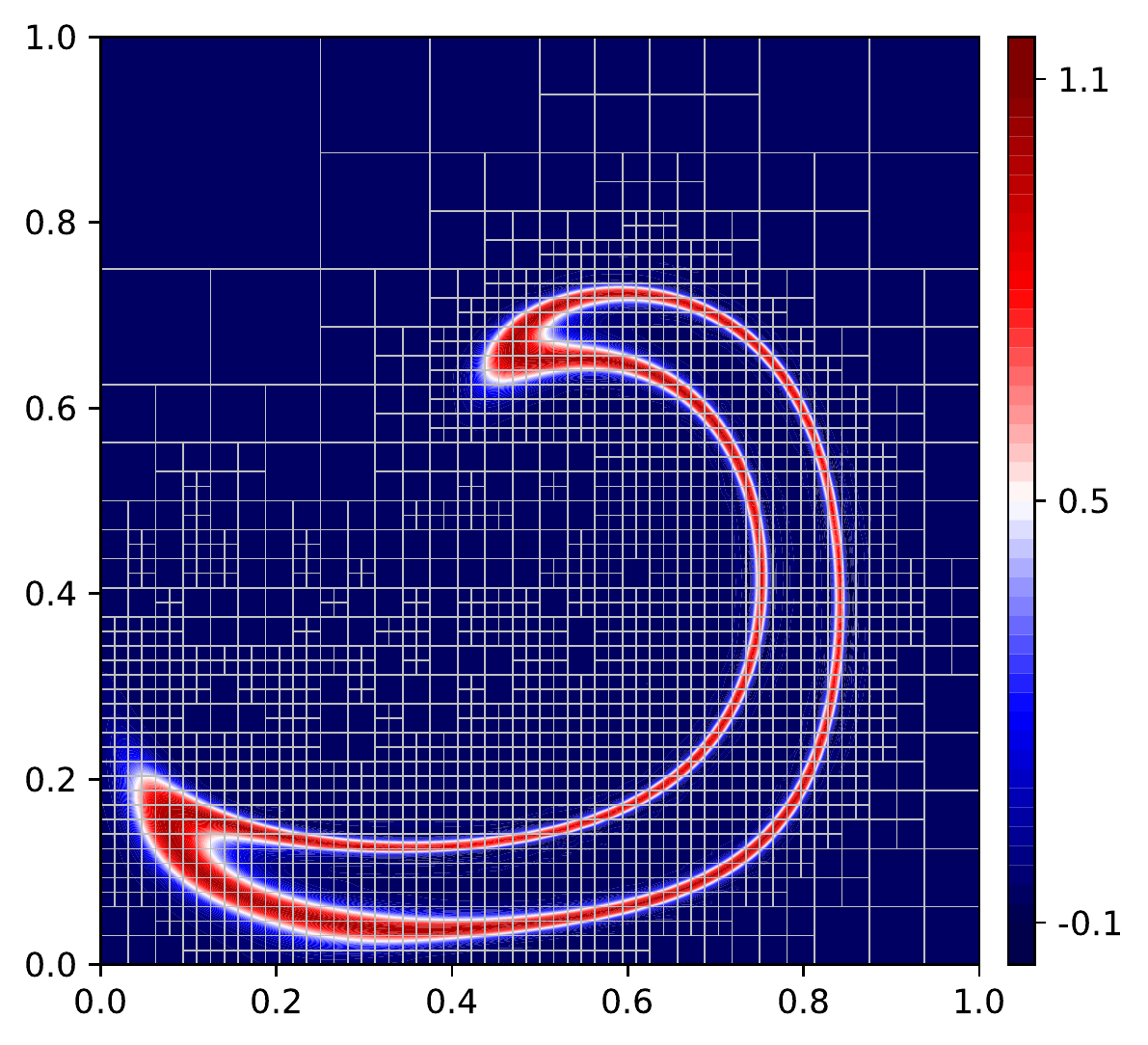}
        \caption{\small $t=0.5$}
        \label{fig:ring:AMR_t0.5}
    \end{subfigure}
    \begin{subfigure}{0.32\textwidth}
        \includegraphics[width=\linewidth]{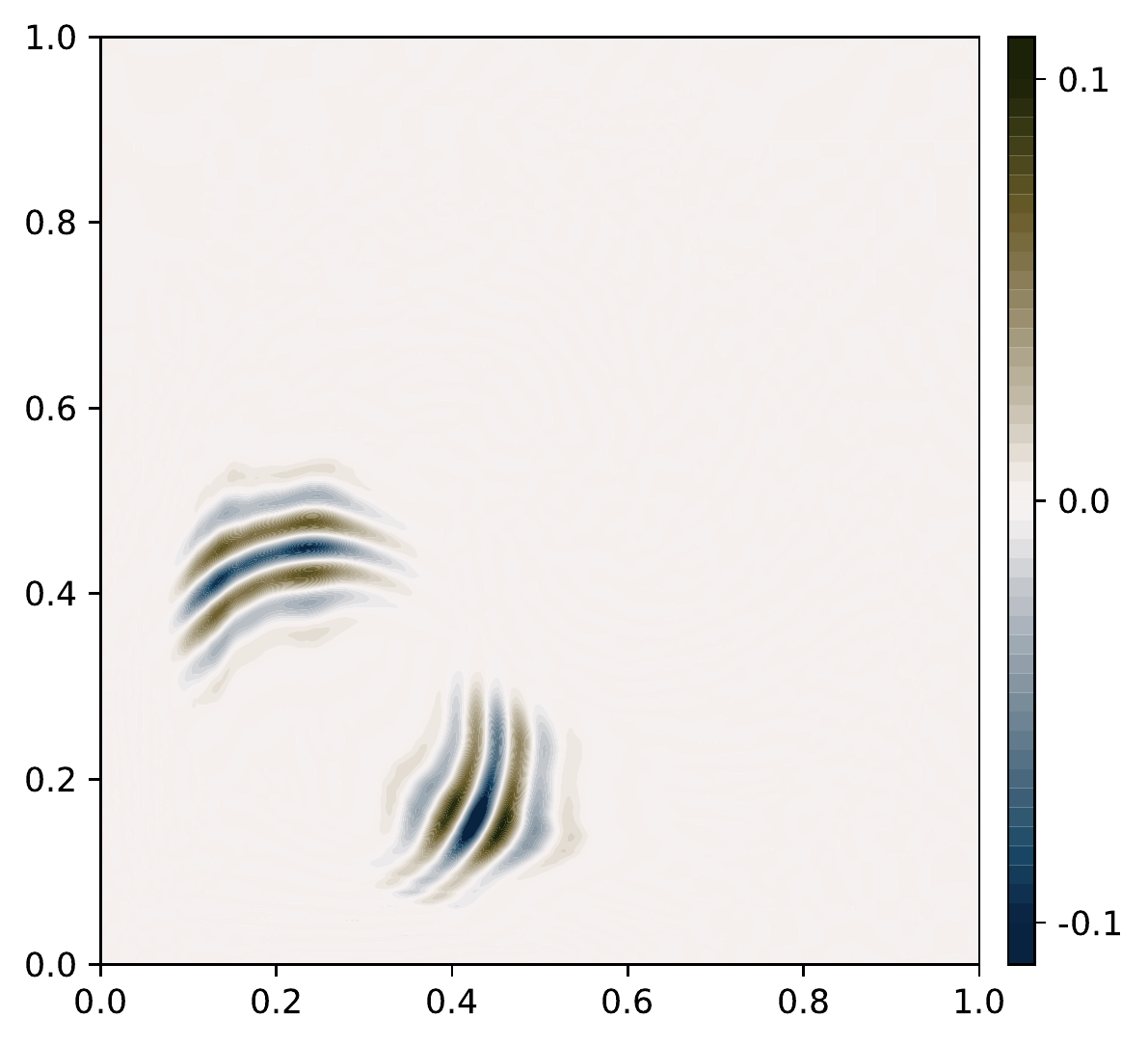}
        \caption{\small $u_h - u_{\text{exact}}$, $t=1$}
        \label{fig:ring:AMR_final_error}
    \end{subfigure}
    \begin{subfigure}{0.32\textwidth}
        \includegraphics[width=\linewidth]{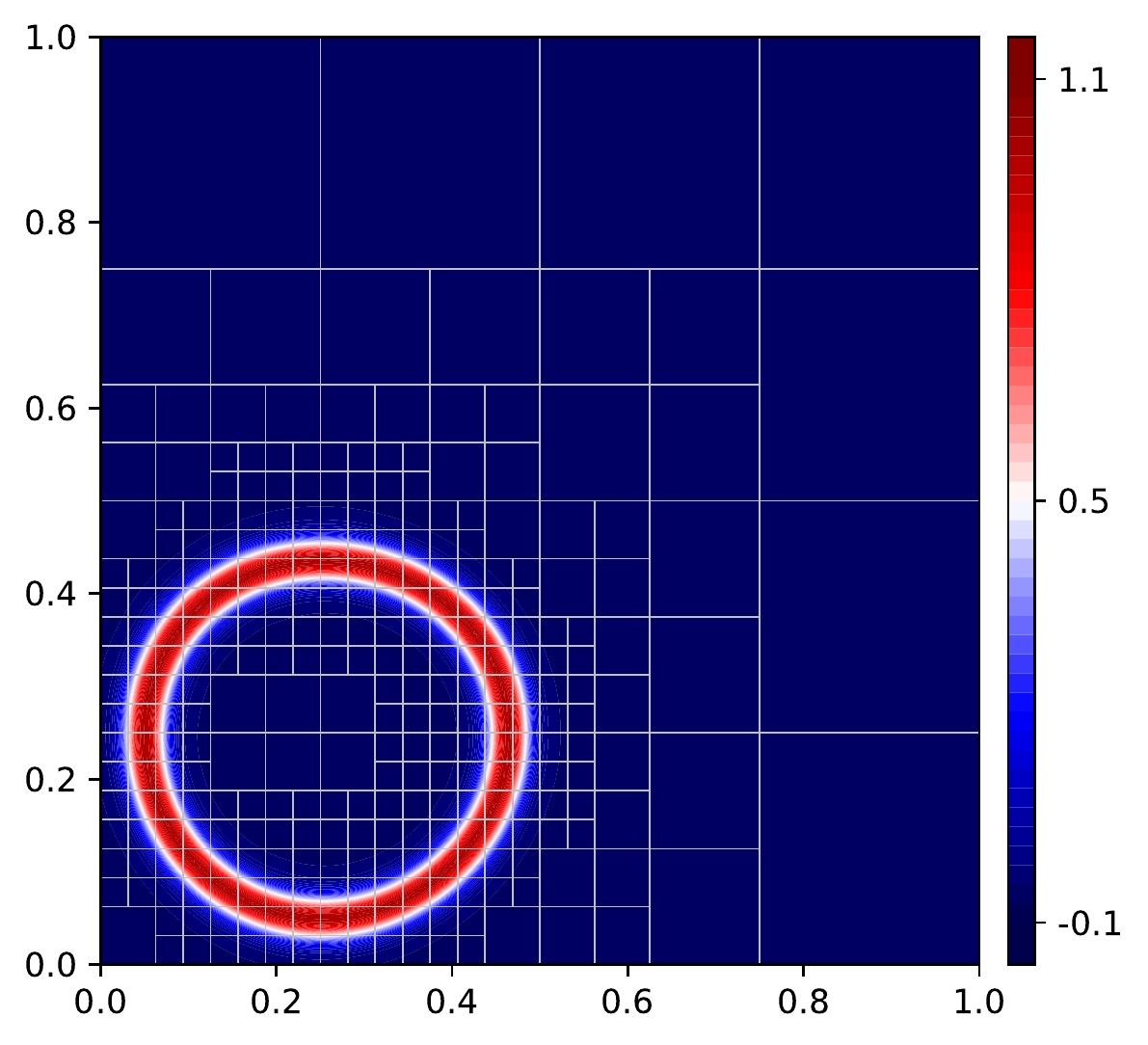}
        \caption{\small $t=0.1$}
        \label{fig:ring:RL_t0.1}
    \end{subfigure}
    \begin{subfigure}{0.32\textwidth}
        \includegraphics[width=\linewidth]{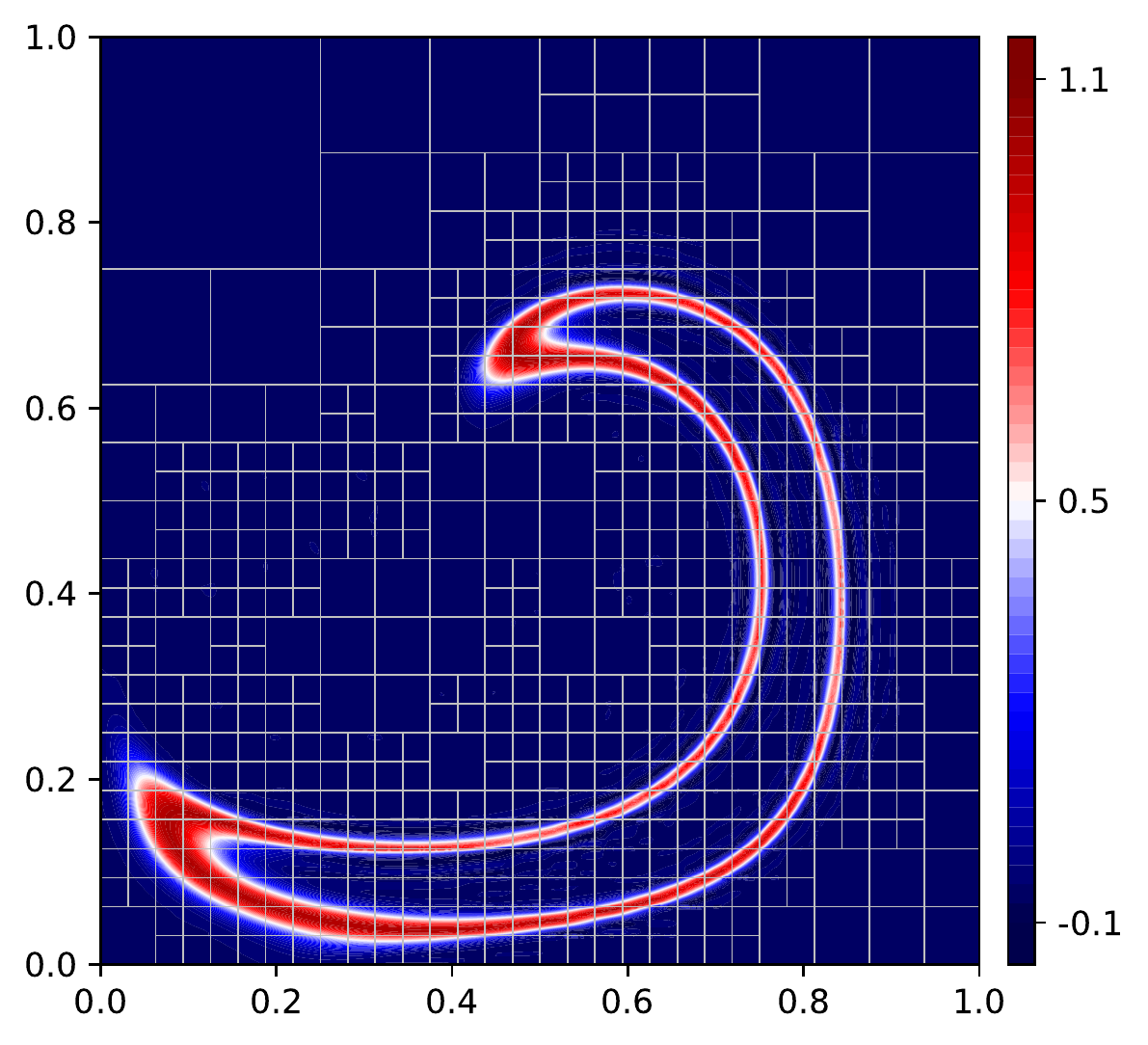}
        \caption{\small $t=0.5$}
        \label{fig:ring:RL_t0.5}
    \end{subfigure}
    \begin{subfigure}{0.32\textwidth}
        \includegraphics[width=\linewidth]{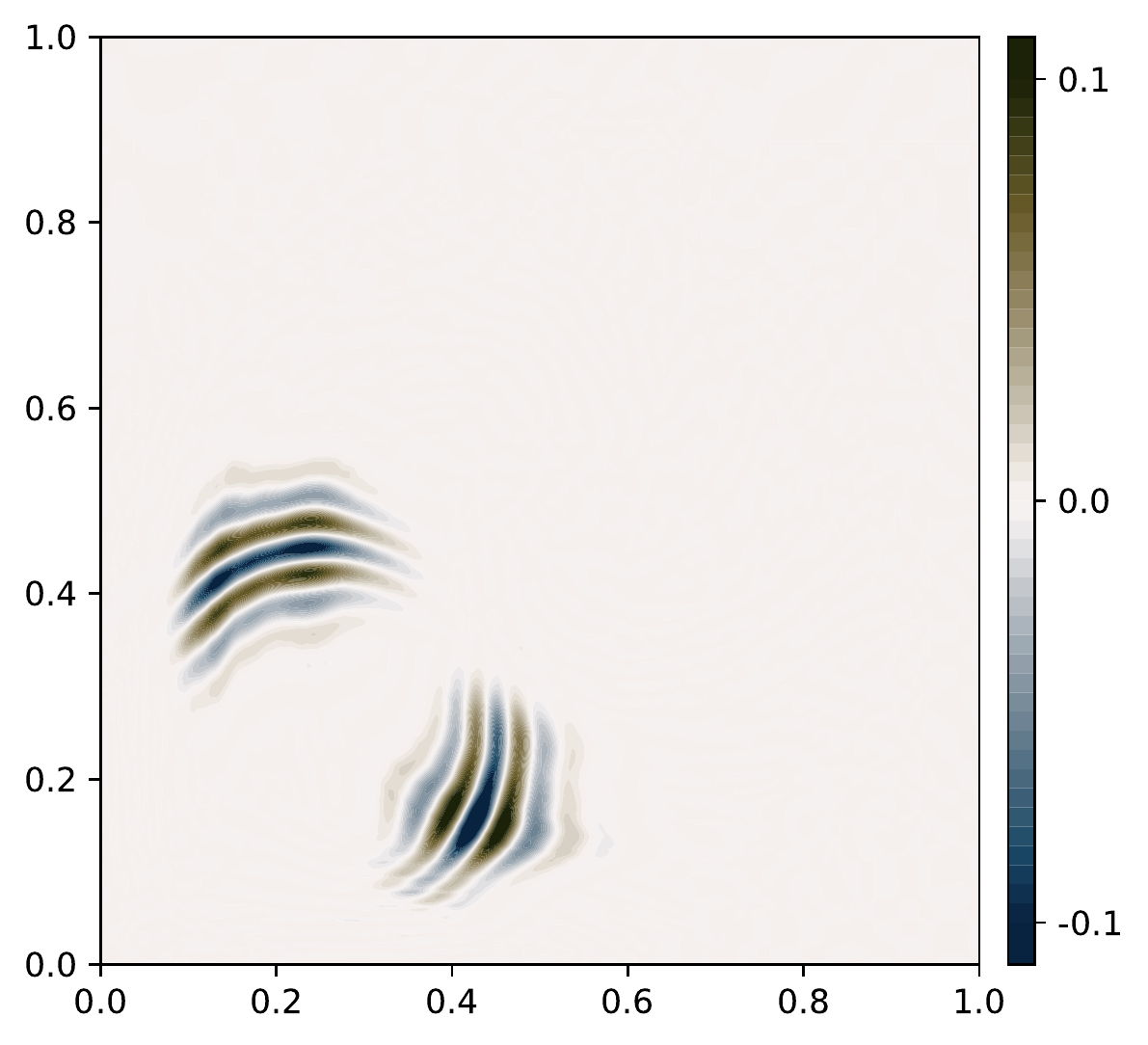}
        \caption{\small $u_h - u_{\text{exact}}$, $t=1$}
        \label{fig:ring:RL_final_error}
    \end{subfigure}
    \caption{\small Unsteady 2D Ring advection. AMR heuristic (top row) vs RL (bottom row).}
    \label{fig:ring:AMR_RL_comparison}
\end{figure}

The proposed meshes and resultant numerical solutions are shown for the AMR heuristic and the RL policy in Figure \ref{fig:ring:AMR_RL_comparison}. 
As can be seen in
panels \ref{fig:ring:AMR_t0.1}, \ref{fig:ring:AMR_t0.5}, \ref{fig:ring:RL_t0.1}, and \ref{fig:ring:RL_t0.5},
the qualitative features of the numerical solution resulting from both approaches are very similar to that of the uniform grid simulation; that is, both approaches are able to successfully resolve the features of the solution.
Similar to the other experiments, the RL agent is able to achieve comparable accuracy to the AMR heuristic, but with a comparatively coarser mesh, as is corroborated in Figures \ref{fig:ring:AMR_final_error} and \ref{fig:ring:RL_final_error}; the numerical errors for each approach are comparable and within 10 percent of the numerical error resulting from the uniform mesh.
However, in this particular test case, at each time step, the AMR heuristic used roughly a factor of 10 more elements than the RL policy.

\subsection{Discussion of numerical experiments}%
\label{sec:numerical_experiment_summary}

The numerical experiments in \S\ref{sec:static_1D_LA}-\S\ref{sec:static_1D_gen} demonstrated a proof of concept for the DRL-AMR approach; we were able to train a model on a small budget, and apply it effectively to larger, different problems governed by the same PDE.
Our experiments indicated that random initialization %
led to a better exploration of the decision space and that all learning algorithms considered were able to solve the problem.
We saw that for a simplified observation space, the neural network representing the trained RL policy was interpretable in terms of local solution conformity.
The numerical experiments in \S\ref{sec:time-dependent-1D-adv}-\S\ref{sec:1D_HDG} showed that the method generalizes to time-dependent problems, as well as to different PDEs with mixed boundary conditions and significantly more complicated numerical schemes.

The test case in \S\ref{sec:steady-2D-advection} demonstrated that the RL agent was able to learn a spatially heterogeneous, non-trivial strategy to increase accuracy per cost by preferentially refining around a problem region on the boundary, a strategy on which the AMR heuristic was not able to capitalize. 
The advection-diffusion problem in \S\ref{sec:steady-2D-advection-diffusion} extended the findings in \S\ref{sec:1D_HDG} to higher dimensions and exhibited the ability of the RL policy to take advantage of the fast convergence of a post-processed solution as well as to resolve non-trivial features of a numerical solution cheaply. Lastly, the unsteady 2D advection problem in \S\ref{sec:ring_test_case} displayed the ability of the trained RL policy to preserve similar non-trivial features over the course of time integration in problems with dynamic, sharp gradients.

Overall, our DRL-AMR methodology is flexible and effective at delivering efficient, high-quality solutions for both static and time-dependent problems over a wide range of different PDEs, boundary conditions, dimensions, and problem sizes.
\vskip -0.2truecm

\vskip -0.2truecm
\section{Conclusions and future work}%
\label{sec:conclusions_and_future_work}

We introduced a novel deep reinforcement learning formulation for adaptive mesh refinement based on a partially observable Markov decision process designed to balance numerical accuracy with computational cost, with the goal of providing a learned, high-quality strategy for resolving solution features efficiently.
The underlying idea is that rather than hand-designing a heuristic error indicator \textit{a priori}, the reinforcement learning agent will instead learn one through trial and error during training by solving many inexpensive problems. 
This is advantageous because the learned policy network can be arbitrarily complex and make use of any feature included in the observation space, including information about computational cost.
Furthermore, training the RL policy requires no domain-specific knowledge, as the relevant information is encoded in the numerical solver, and, more abstractly, in the underlying PDE. 
Conversely, specific features of a numerical scheme such as polynomial order and problem dimension are implicitly considered in training through use of the numerical solver, as opposed to having to be analyzed and explicitly specified in the case of a manually-defined error indicator.

Our implementation shows that the resultant trained policies are able to execute adaptive refinement strategies which are competitive with, and in many cases, better than common AMR heuristics in terms of accuracy of the final solution per problem degree of freedom.
Our methodology is not specific to any particular PDE, spatial dimension, or numerical scheme, and can flexibly incorporate physical or temporal history data into the observation space.
The local nature of the RL problems allows for training the RL agent on small problems and deploying the policy on much larger problems, ensuring scalability.
Lastly, at no point during training nor model deployment do we ever make use of an exact solution or a ``ground truth.''

Future work could incorporate $p$-refinement into the learning process to allow the RL agent access to a richer set of finite element representations of the numerical solution, 
incorporate a more sophisticated belief distribution as to the regions of under- or over-resolution in the numerical solution,
or integrate transfer learning using existing error estimators to accelerate training. Finally, the application of the new DRL-AMR schemes to the Navier--Stokes and geophysical equations such as those used in storm surge \cite{mandli2014adaptive,thomas2021improving}, ocean \cite{haley_lermusiaux_OD2010,haley_et_al_OM2015,ueckermann_lermusiaux_JCP2016,foucart_et_al_Oceans2018,foucart_et_al_Oceans2021}, and atmospheric \cite{kopera2014analysis} applications, as well as studies of related numerical topics such as slope-limiting, preconditioning, and adaptive time integration, are the subject of ongoing research.

\section*{Acknowledgements}%
\label{sec:acknowledgements}

We thank the members of the MSEAS group, especially Dr. C. Mirabito for review of the manuscript. We are grateful to the Office of Naval Research (ONR) for partial research support under grants  N00014-18-1-2781 (DRI-CALYPSO) and N00014-20-1-2023 (MURI ML-SCOPE) to the Massachusetts Institute of Technology.
We also thank the Spring 2022 course staff of 2.C51 for helpful discussions related to machine learning for physical systems. 
C.\ Foucart would additionally like to thank the active \texttt{deal.ii} software development community for helpful discussion related to cross-language software development.

{ \footnotesize
\bibliographystyle{abbrv}
\bibliography{references,mseas}  
}

\section*{Appendix A. Tunable policies}
Computational resources are user- and time-dependent. For different applications, or even in the middle of a simulation, we may seek to adjust how aggressive our reinforcement learning agent is in refining or coarsening our mesh. We directly encode the trade-off between accuracy and computational cost in hyperparameter $\gamma_c$, which the user may tune freely. However, naively changing $\gamma_c$ would require re-training our reinforcement learning agent. Next, we show how we can avoid re-training while simultaneously allowing for policy tuning.

The more general setting to this problem is multi-objective reinforcement learning (MORL). We seek an agent that is optimized over a continuum of objective functions depending upon $\gamma_c$. The value function $V_t^\pi(s_t)$ is the expected reward following policy $\pi$ given state $s_t$. Traditionally, $V_t^\pi : S \rightarrow \mathds{R}$ maps an observation to a scalar, but in MORL, the value function $V_t^\pi : S \rightarrow \mathds{R}^d$ maps an observation to a $d$-dimensional vector indexed by the different objective functions. In our case, the value function $V_t^\pi(s_t;\gamma_c) : S \rightarrow \mathcal{C}_b(\mathds{R})$ maps an observation to a bounded continuous function. Most approaches in the MORL literature attempt to find either a single policy by scalarizing the vector-valued $V_t^\pi$ or multiple policies by repeatedly training over different objective functions (see \cite{hayes2022practical} for a literature review). However, due to a particular feature of our expected reward function $Q_t$, we are able to simply learn over a continuum of objective functions.

\begin{figure}[ht]
    \centering
    \includegraphics[width=0.8\linewidth]{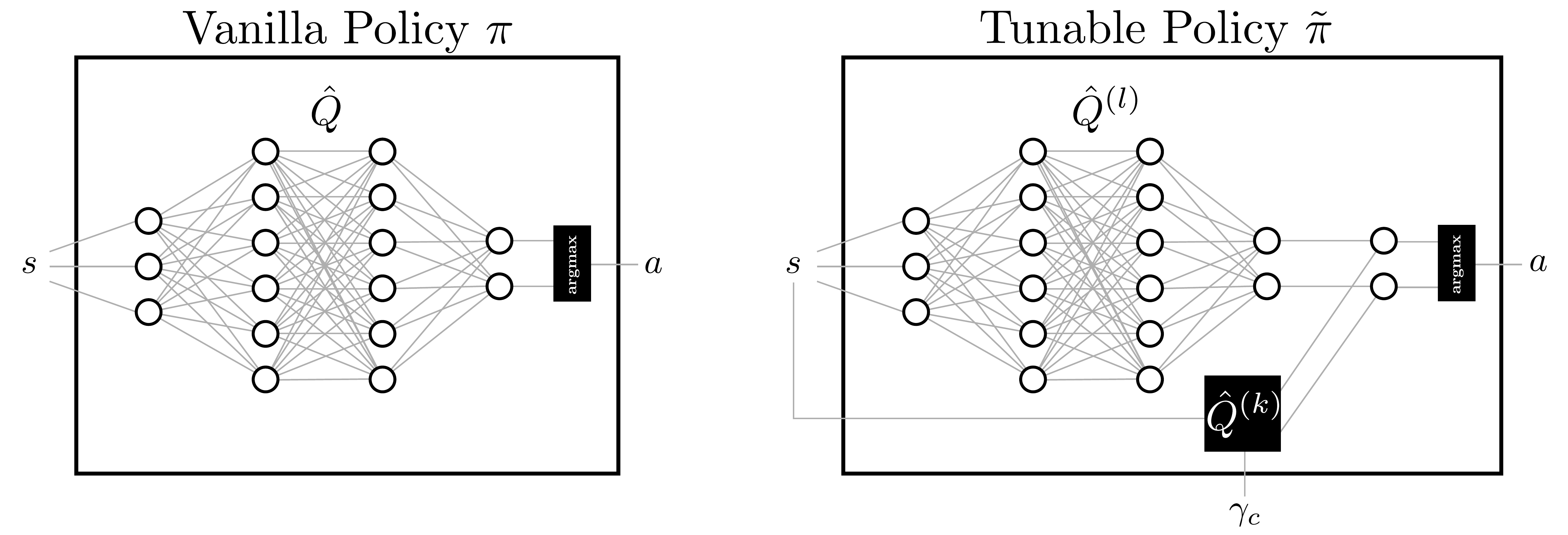}
    \caption{\small We depict a vanilla policy based on Deep Q-Learning and a tunable policy. The tunable policy splits the $\hat{Q}$ function into known and learned functions. The learned function is represented by a neural network, while the known policy can be evaluated explicitly given the observations. In the tunable policy, the hyperparameter $\gamma_c$ is an input to the known $\hat{Q}^{(k)}$ function which does not have to be re-learned.}
    \label{fig:tunable_policy}
\end{figure}

Consider an expected reward function $Q_t : \mathcal{O}\times \mathcal{A} \rightarrow \mathds{R}$ that can be split into two parts: a function that can be explicitly computed given the observation, i.e.\ the \say{known} function $Q^{(k)}$, and a function that needs to be learned, $Q^{(l)}$.
\begin{gather}
	Q_t(s,a;\gamma_c) = Q^{(k)}_t(s,a;\gamma_c) + Q^{(l)}_t(s,a)
\end{gather}
There is no reason to learn $Q^{(k)}$ if it can be explicitly computed; instead, we should just represent $Q^{(l)}$ with a deep neural network as in deep Q-learning. Then, the outputs of the two $Q$ functions can be combined at the end, and the argmax of the resulting sum will be the action that we take. Importantly, the hyperparameter $\gamma_c$ is only an argument of the known function that we can explicitly compute. This allows us to learn one function $Q^{l}$ but then change our policy adaptively as a function $\gamma_c$. Computationally, we really learn the function $\hat{Q}_t : S \rightarrow \mathds{R}^{|\mathcal{A}|}$ that maps an observation to the expected reward for every action in the action space, and we denote this modified and approximate expected reward function with a hat. In figure \ref{fig:tunable_policy}, we delineate a traditional deep Q-learning policy $\pi$ from a tunable policy $\tilde{\pi}$.

In our particular case, the reward function (\ref{eq:reward_fn_general}) is naturally decomposed into a part that needs to be learned and one that can be explicitly computed: the accuracy needs to be learned but the computational cost and barrier function are explicitly known. Call these partial rewards $R^{(l)}$ and $R^{(k)}$, respectively. We can write our $Q$ function as follows.
\begin{gather}
Q_t(s,a) = \mathds{E}\left[\left.\sum_{i=0}^\infty \gamma_k^i R_{t+i+1}^{(k)} + \sum_{i=0}^\infty \gamma^i R_{t+i+1}^{(l)}  \right| S_t = s, A_t = a\right]
\end{gather}

Above, we have used different discount factors $\gamma_k$ and $\gamma$ for $R^{(k)}$ and $R^{(l)}$, respectively. We consider the case where $\gamma_k = 0$. In our example, setting the discount factor on the computational cost to zero is justified in that we are usually only concerned with the computational cost at a given instant. Furthermore, the computational load on a machine is often highly unpredictable; for example, other users may submit jobs in the middle of our simulation. As such, a greedy approach to discounting the computational penalty is desirable, and we may rewrite our expected reward as follows.
\begin{gather}
Q_t(s,a) = \underbrace{\mathds{E}\left[\left. R_{t+1}^{(k)}  \right| S_t = s, A_t = a      \right]}_{Q_t^{(k)}} + \underbrace{\mathds{E}\left[\left.\sum_{i=0}^\infty \gamma^i R_{t+i+1}^{(l)}  \right| S_t = s, A_t = a \right]}_{Q_t^{(l)}}
\end{gather}

With that, we have defined our known and learned $Q$ functions, and to train we simply omit $R^{(k)}$ from our reward entirely by setting $\gamma_c = 0$. Once trained, we append $Q^{(k)}$ to the end of our $Q^{(k)}$ network in order to make predictions, and $\gamma_c$ may be set arbitrarily without retraining. We note that separable rewards are not limited to this case where we balance computational cost and simulation accuracy, and we posit that there are many other areas where this framework may be beneficial.

While we have described how to perform MORL in instances where $Q$ is separable and the action space is discrete, this framework should be extendable to the continuous case. In actor-critic policies where the critic learns the $Q$ function (e.g.\ deep deterministic policy gradient (DDPG) \cite{ddpg}, twin delayed DDPG (TD3) \cite{td3}, and soft actor critic (SAC) \cite{sac} methods), one should be able to apply a similar methodology: after learning $Q^{(l)}$, one should be able to append $Q^{(k)}$. In the continuous case, however, the actor must be re-trained since we cannot just take the argmax of a continuous action space. Fortunately, re-training the actor should be computationally inexpensive since the policy networks tend to be small, and we can just randomly sample from the action space. That is, we do not need to re-sample from our environment; instead, because we already know our $Q$ function, we simply need to re-train the actor to maximize the estimated reward given an action from our action space. This procedure is the subject of ongoing research.

\section*{Appendix B. Exact solution, \S\ref{sec:steady-2D-advection-diffusion}}
\label{app:MIT_exact_solution}

The exact solution is given by
\begin{equation*}
  u(\bm{x}) = \sum_{i=1}^{25} \frac{1}{2\pi \sigma^2} \exp\left(-\frac{1}{\sigma^2} \left\lVert \bm{x} - \bm{x}_i\right\rVert^2  \right)
\end{equation*}
with parameter $\sigma=1/10$ and source centers $\bm{x}_i$ described in Table \ref{tab:MIT_source_ctrs}.

\begin{table}[H]
\centering
\footnotesize
\begin{tabular}{lclclclclc}
\hline
i & $\bm{x_i} = (x_1, x_2/\alpha)$ &&&&&&&& \\
\hline
1  & (-0.75,  -0.25) & 6  & (-0.25,  -0.25) & 11 & (0.125, -0.75) & 16 & (-0.75, -0.625) & 21 & (-0.25, -0.375) \\ 
2  & (-0.75,  -0.75) & 7  & (-0.25,  -0.50) & 12 & (0.5,   -0.75) & 17 & (-0.50, -0.625) & 22 & (0.125, -0.625) \\ 
3  & (-0.75,  -0.50) & 8  & (-0.25,  -0.75) & 13 & (0.5,   -0.5 ) & 18 & (-0.25, -0.625) & 23 & (0.5  , -0.625) \\ 
4  & (-0.50,  -0.25) & 9  & (0.125,  -0.25) & 14 & (0.5  , -0.25) & 19 & (-0.75, -0.375) & 24 & (0.625  ,-0.25) \\ 
5  & (-0.50,  -0.50) & 10 & (0.125,  -0.50) & 15 & (0.75 , -0.25) & 20 & (-0.50, -0.375) & 25 & (0.125, -0.375) \\ 
\end{tabular}
\vspace{1em}
\caption{\small Description of source centers, where $\alpha=1.1$.}
\label{tab:MIT_source_ctrs}
\end{table}

\section*{Appendix C. Additional training parameters}
\label{app:training_params}

For all numerical experiments, we used a time discount factor $\gamma=0.99$, which is the default for the Stable-Baselines-3 library. 
The large negative reward accrued by the agent upon running out of computational resources was taken to be $R_{\text{exceed}} = -1\cdot 10^3$; this was important as a learning signal in the cases where the RL agent was trained on a small computational budget, as the barrier function $B(p)$ in (\ref{eq:reward_fn_general}) is undefined outside $p=1$. 
Training episodes were terminated after finite number of iterations, we used 200; this choice is arbitrary, as long as the number of iterations is large enough that it is possible for the agent to approach its computational budget and receive the large negative reward for doing so.

At the time of this writing, the authors have made no attempt to accelerate the training process through optimization of training parameters such as batch size and episode length, but point out that doing so constitutes the subject of ongoing research.

\end{document}